%% file: Freedman-11-02.tex
\documentclass[12pt]{article}
\usepackage{amsmath}
\usepackage{amsthm, amscd}
\usepackage{amsfonts}
\usepackage{amssymb}
\usepackage[dvips]{graphicx}
\usepackage{epic}
\usepackage{eepicemu}

\input epsf

\newtheorem{thm}{Theorem}[section]
\newtheorem{con}[thm]{Implication of Ansatz}

\newtheorem{defi}[thm]{Definition}
\newtheorem{lemma}[thm]{Lemma}
\newtheorem{add}[thm]{Addendum}
\newtheorem{pro}[thm]{Proposition}
\newtheorem{obspro}[thm]{Observation}

\newtheorem{rem}[thm]{Remark}
\newtheorem{No}[thm]{Note}

\newcommand{\C}{{\mathbb C}}
\newcommand{\Q}{{\mathbb Q}}

\newcommand{\Z}{{\mathbb Z}}

\def\a{\alpha}
\def\bt{\beta}
\def\c{\circ}

\def\d{\delta}

\def\e{\epsilon}
\def\f{\frac}
\def\Pa{\gamma}
\def\i{\in}
\def\iy{\infty}
\def\l{\lq\lq}
\def\ld {\lambda}
\def\la{\longrightarrow}
\def\ot{\bigotimes}

\def\us{\underset}
\def\s{\sigma}
\def\SS{\Sigma}
\def\m{\mid}

\def\tr{\triangle}

\def\tn{\textnormal}

\renewcommand{\theequation}{\thesection.\arabic{equation}}
\renewcommand{\thefigure}{\thesection.\arabic{figure}}

\title{A magnetic model with a possible Chern-Simons phase}
\author{Michael H. Freedman\footnote{Microsoft Research, Redmond, WA. Email: michaelf@microsoft.com} \\
\small{Appendix by F. Goodman\footnote{Department of Mathematics,
University of Iowa, Iowa City, Iowa. Email:
goodman@math.uiowa.edu} and H. Wenzl\footnote{Department of
Mathematics, University of California, San Diego, California.
Email: wenzl@brauer.ucsd.edu}}}
\date{UPDATED: November 15, 2002}
\begin{document}
\maketitle

\begin{abstract} An elementary family of local Hamiltonians $H_{\c ,
\ell}, \ell = 1,2,3, \ldots$, is described for a $2-$dimensional
quantum mechanical system of spin $=\f{1}{2}$ particles.  On the
torus, the ground state space $G_{\c , \ell}$ is $(\log)$
extensively degenerate but should collapse under $\l$perturbation"
to an anyonic system  with a complete mathematical description:
the quantum double of the $SO(3)-$Chern-Simons modular functor at
$q= e^{2 \pi i/\ell +2}$ which we call $D E \ell$. The Hamiltonian
$H_{\c, \ell}$ defines a \underline{quantum} \underline{loop}
\underline{gas}. We argue that for $\ell = 1$ and $2$, $G_{\c ,
\ell}$ is unstable and the collapse to $G_{\e , \ell} \cong  D E
\ell$ can occur truly by perturbation.  For $\ell \geq 3$,  $G_{\c
, \ell}$ is stable and in this case finding $G_{\e , \ell} \cong D
E \ell$ must require either $\e> \e_\ell> 0$, help from finite
system size, surface roughening (see section 3), or some other
trick, hence the initial use of quotes $\l \quad$".  A
hypothetical phase diagram is included in the introduction.

The effect of perturbation is studied algebraically: the ground
state space $G_{\c , \ell}$ of $H_{\c , \ell}$ is described as a
surface algebra and our ansatz is that perturbation should respect
this structure yielding a perturbed ground state $G_{\e , \ell}$
described by a quotient algebra. By classification, this implies
$G_{\e , \ell} \cong D E \ell$. The fundamental point is that
nonlinear structures may be present on degenerate eigenspaces of
an initial $H_\c$ which constrain the possible effective action of
a perturbation.

There is no reason to expect that a physical implementation of
$G_{\e , \ell} \cong  D E \ell$ as an anyonic system would require
the low temperatures and time asymmetry intrinsic to Fractional
Quantum Hall Effect (FQHE) systems or rotating
Bos$\acute{\tn{e}}$-Einstein condensates $-$ the currently known
physical systems modelled by topological modular functors. A solid
state realization of $D E 3$, perhaps even one at a room
temperature, might be found by building and studying systems,
$\l$quantum loop gases", whose main term is $H_{\c , 3}$.  This is
a challenge for solid state physicists of the present decade. For
$\ell \geq 3, \ell \neq 2 \mod 4$, a physical implementation of
$DE \ell$ would yield an inherently fault-tolerant universal
quantum computer. But a warning must be posted, the theory at
$\ell = 2$ is not computationally universal and the first
universal theory at $\ell =3$ seems somewhat harder to locate
because of the stability of the corresponding loop gas. Does
nature abhor a quantum computer?
\end{abstract}

\medskip


\tableofcontents




\setcounter{section}{-1}
\section{Introduction}
In section 1, we write down a positive semidefinite local Hamiltonian $H_{\c ,
\ell} $ for a system of locally interacting Ising spins on a
$2-$dimensional triangular lattice or surface triangulation, $\ell
= 1, 2, 3, \ldots$. In the presence of topology, e.g. on a torus,
$H_{\c , \ell}$ has a highly degenerate space $G_{\c , \ell}$ of
zero modes. On any closed surface $Y$, different from the
$2-$sphere, the degeneracy is polylog $(2^v ) =$ poly$(v)$, where
$v$ is the number of sites in the triangulation and the $2^v$ is
the dimension of the Hilbert space $h$ of spins. On the torus
$T^2$ the polynomial has degree $=1$, when $Y$ has genus $g>1$ the
polynomial has degree $=3g-3$ (see Proposition 3.8).

We argue for an ansatz (3.4) which
exploits the peculiarly rigid algebraic structure of $G_{\c ,
\ell}\,\, -$ it is a monoidal tensor category with a unique
nontrivial ideal.  The ansatz allows us to model any
$\l$perturbed" ground state space $G_{\e , \ell}$ (which is itself
stable to perturbation) uniquely as a known anyonic system or in
mathematical parlance a \underline{modular} \underline{functor}.
The functor is the Drinfeld double of the even $-$ label $-$
sector of the $SU(2)$-Chern-Simons unitary topological modular
functor at level $\ell , D E \ell$.  Even labels corresponds in
physical terms to the integer spin representations so the
even-label-sector derives from the group $SO(3)$.

The Hamiltonian $H_{\c,\ell}$ defines a \underline{quantum}
\underline{loop} \underline{gas} which can be compared (see Figure 3.12) with the
classical analog. The statistical mechanics of classical loop
gases [Ni] identifies a known critical regime and from this we
infer that for $\ell = 1$ and $2$, $G_{\c , \ell}$ is unstable and
the collapse to $G_{\e , \ell} \cong DE \ell$ is truly by
perturbation, for $\ell \geq 3, \,\, G_{\c , \ell}$ is stable and
in this case finding $G_{\e , \ell} \cong DE \ell$ requires $\e
> \e_\ell > 0$, or some other device (see section 3),
hence the initial use of quotes $\l \quad$".  Figure 0.1 is a
hypothetical phase diagram. The stability of $G_{\c, \ell}$ at
$\ell = 3$ is probably very slight $-$ see footnote 6 in section
3 and the corresponding discussion.

The reader should not be alarmed that a $\l$doubled" Chern-Simons
theory arises.  The doubled structure makes it a gauge theory and,
as we will explain, the double, being achiral, is more likely to
have a robust physical realization. The modular functor $DE \ell$
has $\ld = \left(\lfloor \f{\ell+1}{2}\rfloor\right)^2\,\,
\l$labels" or , physically, $\ld$ super selection sectors for
quasiparticle excitations (including the empty particle.)
Physically this means that a local bit of material, a two
dimensional disk with a fixed boundary condition, which is in its
\underline{unique} ground state $G_{\e , \ell}$ can have $\ld$
types of point-like anyonic excitations (presumably with
exponentially decaying tails) which can only be created in pairs.
$\ld$ is the number of order integers pair $(x,y)$ with $0 \leq x,
y\leq \ell$, and $x, y=$ even. By mathematically deleting small
neighborhoods of such excitations a ground state vector is
approximately achieved in the highly degenerate ground state space
$G_{\e , \ell}$ associated to a punctured disk with boundary
conditions. It is known [FLW2 ] and [FKLW] that for $\ell \geq
3,\ell \neq 0 \mod 4$, there is a universal, inherently
fault-tolerant, model for quantum computation based on the ability
to create, braid, fuse, and finally distinguish these excitations
types.  Thus $H_{\c, \ell}$ could be of technological importance:
a physical system, a $\l$quantum loop gas", in this (perturbed)
universality class could be the substrate of a universal fault
tolerant quantum computer.

Any unit vector $\Psi \i G_{\c , \ell}$ is a superposition of
classical $\pm -$spin states $|\Psi \rangle$ which are
distinguished by the eigenvalues $\pm 1$ of a commuting family of
Pauli matrices $\sigma_z^v$ equal to $\left|\begin{smallmatrix} 1&0 \\ 0&-1 \end{smallmatrix}\right|$
at vertex $v$. Sampling $\Psi = \SS a_i |\Psi_i
\rangle$ by observing $\{\sigma_z^v\}$, we $\l$observe" a
classical $|\Psi_i\rangle$ with probability $|a_i |^2$.  The
domain wall $\Pa_i$ separating the $+$ - spin regions from ${-}$ -
spin regions of $\Psi$ may be thought of as a random, self dual,
loop gas [Ni]. This random state is self dual because there is a
symmetry between $\l$up" and $\l$down". It is a Gibbs state with
parameters $k=0$, self dual, and $n= \left( 2\cos \f{\pi}{\ell
+2}\right)^2$, where the weight of a configuration $\Pa$ is
$w(\Pa) = e^{-k ({\tn{total length }} \Pa)} n^{\# {\tn{ components
}}\Pa}$.  It is known that for $0 <n \leq 2$ and $k= 0$ the loop
gas is critical, sitting at a $2^{\tn{nd}}$ order phase transition
as $k$ crosses from negative to positive. This information
together with sections $3$ or $4$ support a phase diagram like the
one shown in Figure 0.1 with parameters $d:=2\cos \f{\pi}{\ell+2} =
\sqrt{n}$ and $\e$. The parameter $\e$ scales a local perturbation
term $\e V$.  We will argue that the simplest choice for $V$, $V =
\left(\us{\tn{site }i}{\SS} \sigma_x^i \right)$, is a likely
candidate.  The diagram is labelled $\l$hypothetical" since there
is no proof of its accuracy.

The challenge to solid state physics  is to find or engineer a
two dimensional quantum medium in the universality class, $DE3$
below.
\begin{figure}[hbtp]
 \begin{center}
 \includegraphics[width=11cm, height=11cm]{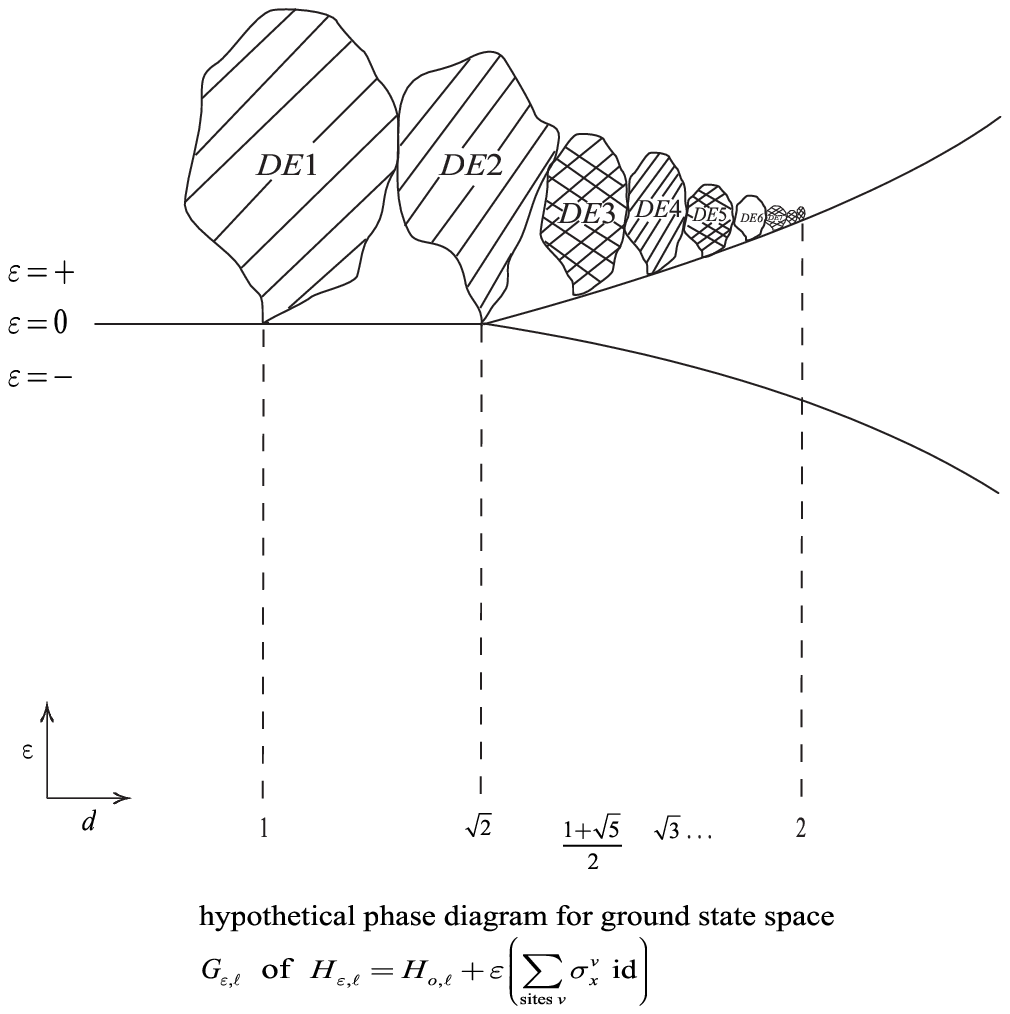}
  \caption{Shaded regions are the topological
phases $DE1, DE2, DE3, DE4, \ldots$.  Doubly shaded regions are the
computationally universal phases $DE3, DE5, DE7, \ldots$. We have
no way of predicting if the topological phases are actually in
contact with each other as drawn. Solid lines are phase boundaries
between inequivalent systems.}
 \end{center}
 \end{figure}

\begin{figure}
\begin{center}
\begin{tabular} {l|c|c|c|c|c}
       {\scriptsize{Theory}}
      & {\scriptsize{\shortstack{dim  \\ on $T^2$}}}
      & {\scriptsize{\shortstack{number of \\ constant  color \\ particles = labels \\ and their\\ braid reps.}}}
      & {\scriptsize{\shortstack{number of \\ additional \\ color reversing \\ particles \\ and the total \\ braid reps.}}}
      & {\scriptsize{\shortstack{specific\\ heat}}}
      & {\scriptsize{\shortstack{nonsingular \\ unitary  topological\\ modular functor? \\(UTMF)}}} \\\hline

  DE1 & 1 & 1,T & 1,T & 2 & {\shortstack{yes, \\ but trivially}} \\
  DE2 & 4 & 4,T & 1,N & 5 & {\shortstack{no,  rank \\$(S-\tn{matrix}) =2$}} \\
  DE3 & 4 & 4,U & 4,U & 8 & yes  \\
  DE4 & 9 & 9,N & 4,N & 13 & yes  \\
  DE5 & 9 & 9,U & 9,U & 18 & yes  \\
  DE6 & 16 & 16,? & 9,? & 25 & no  \\
  DE$\ell$ & $\lceil \f{\ell +1}{2}\rceil^2$& {\shortstack{$\lceil \f{\ell +1}{2}\rceil^2$,\\ $U$ for $\ell \geq 5$, \\ $\ell \neq 2 \mod 4$}}
                                            & {\shortstack{$\lceil \f{\ell}{2}\rceil^2$, \\ $U$ for $\ell \geq 5$, \\ $\ell \neq 2 \mod 4$}}
                                            & $\lceil \f{(\ell +1)^2}{2}\rceil$
                                            & {\shortstack{yes if \\ $\ell\equiv 0,1,3 \mod 4$ \\ no if \\
                                             $\ell \equiv 2 \mod 4$}}

\end{tabular}\\
\caption{For sufficiently many particles we have recorded if the
(generalized) braid group representations are: T (trivial),
A(abelian), N(nonabelian), U(computationally universal), we have
called the total number of elementary particles, including those
that reverse the $(|+\rangle , |-\rangle)$ coloring, $\l$specific
heat" as it counts local degrees of freedom above the ground
state. The color constant elementary particles are the irreducible
representations of the corresponding linear category (see $\S
\,2$). Coloring-reversing particles are explained at the end of
$\S \,2$.}
\end{center}
 \end{figure}

The presumptive approach $-$ nearly universal in the literature
$-$ to building a quantum computer is quite different from our
topological/anyonic starting point. It is based on manipulating
and protecting strictly local $-$ as opposed to global or
topological $-$ degrees of freedom.  It may be called the
$\l$qubit approach" since often a union of $2-$level systems (with
state space $\us{i}{\ot} \C^2_i$) is proposed. Actually, the
number of levels $-$ or even their finiteness $-$ is not the
essential feature, it is that each tensor factor of the state
space $-$ call it a qunit $-$ is physically localized in space (or
momentum space). The environment will -- despite the best efforts
of the experimentalist $-$ interacted directly with these $\l$raw"
qunits. It has long been recognized ([S], [U]) that the raw qunits
must encrypt fewer $\l$logical qubits."  The demon in this
approach is that very low initial (or raw) error rate - perhaps
one error per $10^{-5}$ operations - and large ratios of raw to
logical qunits $\sim 10^3$ seem to be required [Pr], to have a
stable computational scheme.  This problem pervades all approaches
based on local or $\l$qubit" systems: liquid NMR, solid state NMR,
electron spin, quantum dot, optical cavity, ion trap, etc$\ldots$.

Kitaev's seminal paper [K1] on anyonic computation, amplified in
numerous private conversations, provides the foundation for the
approach described here.  Anyons are a $(2+1)-$dimensional
phenomena: when sites containing identical particles in a
$2-$dimensional system are exchanged (without collision) there
are, up to deformation, two basic exchanges; a clockwise and a
counter-clockwise half turn - or $\l$braid" if the motion is
considered as generating world lines in $2+1 -$dimensional
space-time.  The two are inverse to each other but of infinite
order rather than order $=2$.  So whereas only the permutation
needs to be recorded for exchanges in $R^3$, in $R^2$
$\l$statistics" becomes a representation $\rho$ of a braid group
$B$ into the unitary group of a Hilbert space $h$ encoding the
internal degrees of freedom of the particle system:
\[
\rho:B\la U(h).
\]
Since a representation into unitary transformations, $\l$gate set"
in (quantum) computer science language, is the heart of quantum
computation it is not really a surprise that any kind of particle
system with a sufficiently general (it certainly must be
nonabelian) image $\rho(B)$ can be used to build a universal model
for quantum computation.  This has been shown in [FLW1], [FLW2],
and [FKLW].

What are the advantages and disadvantages of anyonic verses qubit
computation?  The most glaring disadvantage of anyons is that no
one is absolutely sure that nonabelian anyons exist in any
physical system.  Two dimensional electron liquids exhibiting the
fractional quantum Hall effect FQHE are the most widely studied
candidates for anyonic systems.  The Laughlin state at filling
fraction $\nu =1/3$ has observed excitations charges of $(1/3) e$
and these are convincingly linked by the mathematical model with a
statistical factor of $\omega = \pm e^{2\pi i/3}$ for the exchange
of such pairs. Quasiparticle excitations with nonabelian
statistics is one of the most exciting predictions of Chern-Simons
theory as a model for the FQHE.  With a few low level (e.g. $\ell
= 1, 2$ or $4$, when $G = SU(2)$) exceptions nonabelian anyonic
systems are capable, under braiding, of realizing universal
quantum computation [FLW2]. The essential point is that the
$\l$Jones representation" of the braid group (on sufficiently many
strands) associated to the Lie group $SU(2)$ has a dense image at
least in $SU(h)\subset U(h)$, $h$ an irreducible summand of the
representation. At $\nu= 5/2$ according to [RR] the Hall fluid is
modelled by a $U(1)$ theory coupled to CS$2$ [the Chern-Simons
theory of $SU(2)$ at the $4^{\tn{th}}$ root of unity (level $\ell
=2$)]; the latter is a theory with a nonabelian $\l$Clifford
group" representation.  This model was selected from conformal
field theories to match expected ground state degeneracies and
central charge, and is further supported by numerical evidence on
the overlap of trial wave functions. Though very interesting, this
representation is still discrete and is not universal in the sense
of [FLW1]. However at $\nu=8/5$,  with perhaps weaker numerical
support [RR], it is thought that the Hall fluid model contains by
$CS3$ (level$=3$, $5^{\tn{th}}$ root of unity). Here braiding and
fusing the excitation would yield universal quantum computation
[FLW1].

So let us, for the sake of discussion, accept that FQHE systems
have computationally universal anyons, we are still a long way
from building a quantum computer. FQHE systems are very delicate:
\begin{enumerate}
    \item The required crystals have been grown successfully only in
a few laboratories.
    \item The temperatures at which the finer plateaus are stable  are
order milK.
    \item The chiral asymmetry intrinsic (For CS2
and CS3 the central charge is $\f{3}{2}$ and $\f{9}{5}$
respectively.) to the effect requires an enormous transverse
magnetic field, order 10 - 15 Tesla to reduce magnetic length to
where conduction plateaus are observed.
\end{enumerate}
At feasible magnet lengths\footnote{In semiconductors with
dialectic constant $\e \approx 10$ and $|B|\approx 10$ Tesla the
characteristic length $\ell =\sqrt{\e \hbar /e \beta}\approx 150
\AA$ compared to about $4\AA$ separation between the ions in a
crystalline solid.}, the Coulomb interaction between electrons is
at least three orders of magnitude weaker than in solids.
Corresponding to the weakness of these interactions the spectral
gap protecting topological phases is necessarily quite small.
Perhaps for this reason, even the most basic experiments to prove
existence of $\l$nonabelions" have not been carried out, and the
use of these systems for computations appears unrealistic.

For applications such as breaking the cryptographic scheme RSA, it
can be estimated that several thousand anyons must be formed,
braided at will (perhaps implementing tens of thousands of half
twists), and finally fused. This appears to be a nearly impossible
task in a FQHE system.

The main point of this paper is that computationally universal
anyons may be available in more convenient systems. $H_{\c ,
\ell}$ is a local model for a paramagnetic system of Ising spins
with short range antiferromagnetic properties.  Written out in
products of Pauli matrices $H_{\c, \ell}$ is seventh order (on the
standard triangular lattice) and thus looks complicated compared
to, say, the Heisenberg magnet.  But geometrically it is quite
simple and its ground states are known exactly. A $2-$dimensional
material in the universality class $DE \ell$ proposed as the
ground state space for $H_{\e , \ell}, \ell \neq 1,2,$ or $\equiv
0 \mod 4 $, will have excitations - $\l$quasiparticles" - capable
of universal fault tolerant quantum computation within a model
that allows creation, braiding, fusion and measurement of
quasiparticle type.

A topological feature is not too easy to detect; by definition,
topological properties cannot be altered or measured by purely
local operators but instead require something akin to an
Aharonov-Bohm holonomy experiment.  So perhaps the universality
class of $H_{\e , \ell}$ already exists in surface layer physics
but is waiting to be discovered. Or perhaps with $H_{\c , \ell}$
in mind something in its (perturbed) universality class can be
engineered. If this is possible there would be no reason to expect
the system to be particularly delicate.  The characteristic
energies for magnets are often several hundred Kelvin [NS].
Furthermore the modular functor $DE \ell$ (this includes the
information of the various braid group representations, $6_j -$
symbols, $S$ and fusion matrices) which arises is amphichiral, the
central charge $c=0$, so there is no reason that time symmetry
must be broken and no apparent need for a strong transverse
magnetic field. These two features are in marked contrast to the
delicate FQHE systems.

Subsequent to the initial draft of this paper a different local
Hamiltonian $H'_{\c, \ell}$ was found which bears the same
relation as $H_{\c, \ell}$ to the topological modular functors
$DE\ell$, but has potential advantages:
\begin{enumerate}
  \item it is expressible as $4^{\tn{th}}$ rather than $7^{\tn{th}}$ order
interactions and \\
  \item its classical analog is the much studied self dual Potts model
    for $q=\left(2 \cos \f{\pi}{\ell+2}\right)^4$.
\end{enumerate}
We have added a section $1'$ following section 1 to explain this
alternative microscopic model.

We make no proposal here for a specific implementation of $H_{\c ,
\ell}$ or for how to trap and braid its excitations but we hope
that models in the spirit of [NS] for the high $T_c$ cuprates may
soon be proposed. In this regard, we note that relatively simple -
but still non classical-braiding statics have been proposed [SF]
in conjunction with the phenomena of spin-charge separation [A]
for high $T_c$ cuprate super conductors above their $T_c$. Certain
$-1$ phases are predicted to occur when braiding the electron
fragments $\l$visions" and $\l$chargeons" around each other and
around ground state defects called $\l$holons". Also contained in
this paper is the suggestion that topological charges might in
passing though a phase transition become classical observables,
e.g. magnetic vortices.  Similarly other phase transitions might
link higher $(\ell =3)$ to lower $(\ell =1)$ topological phases
and might be useful in measuring quasiparticle types. Whether even
the simplest topological theory is realized in any known
superconductor is open, but [SF] is cited as precedent for anyonic
models for solid state magnetic systems with high characteristic
energies.   So while the FQHE motivates this paper, we hope we
have steered toward its mathematical beauty and away from it
experimental difficulties.

What are the generic advantages of anyonic computation?  First
information is stored in topological properties $\l$large scale
entanglement" of the system that cannot be altered (or read) by
local interaction. This affords a kind of physical stability
against error rather than the kind of combinatorial error
correction scheme envisioned in the qubit models - $\l$hardware"
rather than $\l$software" error correction. Second, at least in
the simplest analysis\footnote{Kivelson and Sandih[KS] find that
Landau level-mixing in FQHE can thicken the tails to polynumerical
decay, but this is not a fundamental effect.}, one expects
excitation of a stable system to be well localized with
exponentially decaying tails.  Thus physical braiding should
approximate mathematical braiding, $\rho : B \la U(h)$ up to a
$\l$tunnelling" error of the form $e^{-c L}$, where $c$ is a
positive constant, and $L$ is a microscopic length scale
describing how well separated the excitations are kept during the
braiding process.  This error scaling is highly desirable and
seems to have no analog in qubit models.  While tunnelling treats
virtual errors, errors which borrow energy briefly from the
vacuum, actual errors would be expected scale like $e^{-c_1
T/T_\c}$ where $T_\c$ is a character energy for the system and $T$
the operating temperature.  This is essentially the error analysis
Kitaev made for his anyonic system, the toric code [K1].

This paper draws on three sources of inspiration: 1) Kitaev's
paper [K1] on anyonic computation, 2) the FQHE, and 3) rigidity in
the classification of von Neumann algebras subfactors.  Rigidity
implies that certain monoidal tensor categories have very few
ideals. But when interpreted physically, $\l$ideal" means
$\l$definable by local conditions", so we find that a certain
locality assumption (Ansatz 3.4) strongly limits the physics. This
provides an algebraic approach to the perturbation theory of
$H_{\c , \ell}$ - and perhaps yields greater insight than would be
possible by analytic methods. We find that for $H_{\c , \ell}$ the
polylog extensively degenerate space of $0-$modes $G_{\c , \ell}$
possess in addition to its linear structure, an important
$\l$multiplicative" structure $-$ the structure of a monoidal
tensor category - which we argue, should be preserved under a
perturbation. The rigidity of type II$_1$ factor pairs, an aspect
of which is stated as Thm 2.1, provides a unique candidate for the
(still finitely degenerate) $\l$perturbed" ground state space
$G_{\e , \ell}$ of $H_{\e, \ell}$.  The space $G_{\e , \ell}$ is a
braid group representation space with the representation induced
by an adiabatic motion of quasiparticle excitations.

Throughout, the excitations on a surface $Y$ are assumed to be
localized near points so excited states of $H_{\e , \ell}(Y)$
become ground states of $H_{\e , \ell}(Y^-)$ but now on a
punctured surface $Y^{-}$ with $\l$boundary conditions" or more
exactly $\l$labels," (see section 2.) We treat excited states
indirectly as ground states on the more complicated surface
$Y^{-}$.

The existence of a stable phase $G_{\e , \ell} \cong DE \ell$ will
be argued by analogy with the FQHE where topological phases are
found to be stable, from algebraic uniqueness, and via
$\l$consistency checks". But these arguments constitute neither a
mathematical proof nor a numerical verification.  The latter may
be exactly as far off as a working quantum computer. It was
precisely the problem of studying quantum mechanical Hamiltonians
in the thermodynamic limit, e.g. questions of spectral gap, that
lead Feynmann [Fe] to dream of the quantum computer in the first
place. It is curiously self-referential that we may need a quantum
computer to $\l$prove" numerically that a given physical system
works like one.

We turn now to the definition of $H_{\c , \ell}$ and $H'_{\c ,
\ell}$; returning later to amplify on the relations to quantum
computing, $\C^\ast -$ algebras, Chern-Simons theory, and
topology. (The connection between Chern-Simons Theory and
complexity classes is discussed in [F1].)

In addition to Alexei Kitaev, I would like to thank Christian
Borgs, Jennifer Chayes,  Chetan Nayak, Oded Schramm, Kevin Walker,
and Zhenghan Wang for stimulating conversations on the proposed
model.

\section{The model}

The model describes a system of spin $= \f{1}{2}$ particles
located at the vertices $v$ of a triangulated surface $Y$.  The
Hilbert space is $\mathcal{H}={\us{v=1}{\overset{n}{\ot} }}
\C^2_v$ where $\C^2_v$ is the local degree of freedom $\{\m +
\rangle, \m -\rangle\}$ at the vertex $v$. The basic Hamiltonian
$H_{\c , \ell}$ is written out below as a sum of local projections
and thus is positive semidefinite.  The ground state space (energy
$=0$ vectors) $G_{\c , \ell}$ of $H_{\c , \ell}$ can be completely
understood (this is unusual since the projector above do not
commute) and identified (as $n \la \iy$) with what we call the
even Temperley-Lieb surface $\l$algebra" ${\tn{ETL}}^{s}_{d}$
where $d= 2\cos \f{\pi}{\ell +2}$.

Ultimately our focus will be on the ground states on a multiply
punctured disk $-$ the puncture corresponding to anyonic
excitations (see section 5).  Two issues arise: (1) non-trivial
topology and (2) boundary conditions.  The boundary conditions are
quite tricky so it is best to work first with closed surfaces of
arbitrary genus (even though these are not our chief interest) to
understand the influence to topology alone $\l$liberated" from
boundary conditions.

$Y$ will denote a compact oriented surface throughout.  In
combinatorial contexts, $Y$ will be given a triangulation $\tr$
with dual cellulation $\mathcal{C}$. Initially, we consider the
case where $Y$ is closed, boundary $Y= \partial Y = \emptyset$. We
will speak in terms of the dual cellulation by $2-$cells or
$\l$plaques" $c$. For example, if $Y$ is a torus it may be
cellulated with regular hexagons. This is a perfectly good example
to keep in mind but higher genus surfaces are also interesting,
while the sphere is less so. Soon we will consider surfaces with
boundary.

Distributing $\ot$ over $\bigoplus$, one writes $\mathcal{H}=$
span $\{$classical spin configurations on plaques$\}=:$ span
$\{s_i\}$.  Let $c$ be a plaquet, $s$ a classical spin
configuration and $\overline{s}^c$ that configuration with
\underline{reversed} spin $(+ \la -$ and $- \la +)$  at $c$. For
$1< i,j\leq 2^n$ define $h_{ij}(c) = 1$ if $(1) s_j =
\overline{s}_i^{c}$ and $(2)s_i$ assigns the same spin $\pm$ to
$c$ and all its immediate neighbors, and $h_{ij}(c) = 0$
otherwise. Define $g_{ij} (c) = 1$ if $(1) s_j = \overline{s}^c_i$
and (2) the domain wall $\Pa_{s_i}$ between $+$ and $-$ plaques,
in the spin configuration $s_i$, meets $c$ in a single connected
topological arc, and $g_{ij} (c) = 0$ otherwise. We define:

\begin{equation}
\begin{aligned}
H_{\c, \ell} = & \sum_{\begin{subarray}{1}
            \tn{plaques $c$, pairs}\\
            \tn{of spin states } s_i, s_j
            \end{subarray}} g_{ij} (c)\left( \left(|s_i\rangle - |s_j\rangle \right) \left(\langle s_i| - \langle s_j|\right) \right)+\\
    \kappa   &  \sum_{\begin{subarray}{1}
            \tn{plaques $c$, pairs}\\
            \tn{of spin states } s_i, s_j
            \end{subarray}}
            h_{ij}(c)\left( \left(|s_i\rangle -  \f{1}{d}| s_j\rangle \right)\left( \langle s_i|  - \f{1}{d}\langle
            s_j|\right) \right)
\end{aligned}
\end{equation}

The constant $\kappa$ is positive and may, in this paper, be set
as $\kappa=1$.  To help digest the notation each of the two sums
has $n2^{2n}$ terms most of which are zero.  It is easy to see
that $g_{ij} = g_{ji}$.  If the domain wall $\Pa$ meets $c$ in a
topological arc reversing the spin of $c$ isotopes the domain wall
across $c$ to the complementary arc $=\overline{\partial
c\smallsetminus \Pa}$.  Contrariwise if $h_{ij} = 1$ then $h_{ji}
= 0$. The parameter ${d}$ could be any positive real number but we
will be concerned mainly with $d = 2\cos \f{\pi}{\ell+2}, \ell =
1, 2,3, \cdots$.  The cases $\ell =2$, $d = \sqrt{2}$ and
$\ell=3$, $d=\f{1+\sqrt{5}}{2}$, the golden ratio, are of
particular interest. Finally, each term in the definition of
$H_{\c , \ell}$ should be read, according to the usual ket-bra
notation, as orthogonal projection onto the indicated vector:
$|s_i\rangle - |s_j\rangle$ or $|s_i\rangle - 1/d\,| s_j\rangle$.
These vectors (whose projectors occur nontrivially in the sums)
are certainly not orthogonal to each other (using the inner
product $|+\rangle$ hermitian orthonormal to $| -\rangle$ in
$\C^2$, extended to define the tensor product Hermitian structure
on $\mathcal{H}$) so those individual projectors do not commute.
It is therefore surprising at first that we can completely
describe the (space of) zero modes $G_{\c , \ell}$ of this
positive semidefinite form, $H_{\c , \ell}$. However once the
description is given the surprise will evaporate for it will be
clear how $H_{\c , \ell}$ was $\l$engineered" precisely to yield
this result.  Identifying $G_{\c , \ell}$ is the next goal.

Associate to the closed oriented surface $Y$ an infinite
dimensional vector space ETL$_d (Y)$, the even Temperley-Lieb
space of $Y$.  It is the $\C -$span of $\l$isotopy classes" of
closed bounding $1-$ manifolds $\Pa$ modulo a relation called
$d-$isotopy. The $\l$bounding" condition means that $\Pa$ is a
domain wall separating $Y$ into two regions which could be
labelled $\l | +\rangle$" and a $\l |-\rangle$". Neither $\Pa$ nor
the regions are presumed to be connected. We do not orient $\Pa$,
so we do not distinguish here between states which differ by
globally interchanging $|+\rangle$ and $|-\rangle$. The term $\l
1-$manifold" means $\Pa$ does not branch or terminate at any
point.  Isotopy, of course, means gradual deformation.  The $d
-$isotopy relation: $\Pa -d (\Pa \setminus \Pa_\c)$, when imposed,
says that if a component $\Pa_\c$ of $\Pa$ bonds a disk in $Y$
then $\Pa = d(\Pa \setminus \Pa_\c )$, $d$ times the value on the
submanifold with $\Pa_\c$ deleted. We often work with the dual
$ETL^\ast_d (Y)$, which are the functions $f$ on bounding isotopy
classes satisfying $f(\Pa) = d (f(\Pa \setminus \Pa_\c))$. Let
$\Pa^!$ be a $\Pa$ as above enhanced by one of the two choices for
$\l$signing" the complementary regions. Define $ELT^{!}_d (Y)$ to
be $\C-$span $\{\Pa^! \}$, so that $ELT^{! \ast}_d (Y)$ are the
functions from $\{ \Pa^! \}$ obeying the $d-$isotopy relation.



Both the definition of $H_{\c, \ell}$ and $ETL^\ast_d$ can easily
be extended to $Y$ a compact surface with boundary $= \partial Y$,
given a fixed boundary condition, the points where $\Pa$ meets
$\partial Y$ (transversely). So if $\tr$ is a triangulation of
$(Y,
\partial Y)$ with dual cellulation $\mathcal{C}$ and if the spin
configuration $| + \rangle$ or $| - \rangle$ is fixed at every
vertex ($=$ dual cell) on $\partial Y$ then formula (1.1) defines
a Hamiltonian operator on the configurations with that boundary
condition provided, in both terms, we restrict the sum to plaques
$c$ which do \underline{not} meet $\partial Y$.  This prevents
$\l$fluctuations" from altering the boundary conditions. Define
$H_{\c, \ell} (Y, \partial Y)$ in this way.  Similarly if a
$2-$coloring (or $+$, $- \,\l$signing") of $Y$ is fixed along
$\partial Y$ we may consider a relative $\Pa^!$ as a extension of
this signing to a division of $Y$ into $+$ and $-$ signed regions
(which are presumed to lie in $Y$ as subsurfaces).  Now relative
to the boundary condition (the signing) $ETL^{! \ast}_d (Y,
\partial Y)$ is defined as functions from $\{ \Pa^!\}$ to $\C$
which obey the $d-$ isotopy relation; $ETL^\ast_d (Y, \partial Y)$
is the set of such function invariant under $^-$, the global $|+
\rangle \longleftrightarrow |- \rangle$ interchange.

If $\tr$ is a triangulation on a surface $Y$, with or without
boundary then we have the combinatorial versions of ETL$_d^! (Y)$
and ETL$_d (Y)$, ETL$^{\tr !}_d(Y)$ and ETL$^\tr_d (Y)$ (resp.)
define using only $(|+\rangle , |- \rangle )2-$colorings in which
each dual $2-$cell (plaquet) is $+$ or $-$.

There are natural maps of $\C-$vector spaces:
\begin{equation}
  \tn{ETL}^{\tr !}_d (Y)\rightarrow \tn{ETL}^!_d (Y) \tn{ and } \tn{ETL}^{\tr}_d (Y)\rightarrow \tn{ETL}_d (Y).
\end{equation}
These maps are of course never onto (only the simpler $d-$isotopy
classes are realized).  Also for certain triangulations $\tr$ the
kernel can also be non-zero (due to $\l$stuck" configurations).
However, it is easy to see that as $\tr$ is subdivided.
$ETL^{\tr}_d (Y)$ approximates $ETL_d (Y)$ in the sense that the
direct limit $\underrightarrow{\lim} \,ETL^\tr_d (Y) \cong  ETL_d
(Y)$, similarly $\underrightarrow{\lim}\, ETL^{\tr !}_d (Y) \cong
ETL^!_d (Y)$.

Let $\tr$ be  a fixed triangulation of $Y$ (with fixed boundary
condition $-$ a $^- \,\, -$ projective $(|+ \rangle, |-\rangle )
\, 2-$coloring $-$ if $\partial Y \neq \emptyset $), set $G_{\c
,\ell}(Y, \tr)=$ zero modes (ground state space) of the positive
semidefinate $H_{\c ,\ell}$ defined above (1.1). Clearly $H_{\c
,\ell}$ is $^{-}-$invariant and so $G_{\c ,\ell}$ is
$^{-}-$invariant. Note that $^{-}$ is not always fixed point free:
on $Y = T^2$, the configuration which is $|+\rangle$ on an
essential annulus $A\subset T^2$ and $|-\rangle$ on $T^2
\smallsetminus A$ is a $^- \,\, -$ fixed point. Let $G_{\c
,\ell}^{+} (Y, \tr )$ denote the $+1-$eigenspace of $^{-}$.

\begin{pro}
For $Y$ a closed surface or a surface with fixed boundary
conditions, there are natural isomorphisms $G_{\c , \ell} (Y , \tr
)\cong \tn{ETL}^{\tr !}_d(Y)\tn{ and } G_{\c , \ell}^{+} (Y , \tr
)\cong \tn{ETL}^\tr_d(Y)$.
\end{pro}

\noindent{\bf Proof:}  From line (1.1), $\Psi \i G_\c (Y, \tr)$
iff $\Psi = \us{i}{\SS} (f| s_i\rangle) |s_i \rangle$ for some
linear functional obeying the $d-$isoptopy relation, thus $G_{\c ,
\ell} (Y, \tr) \cong \tn{ETL}^{\tr !}_d (Y)$. The involution
$^{-}$ acts compatibility on both sides so $\tn{ETL}^\tr_d(Y)$ may
be identified as the $+1$ eigenspace of $^{-}$ on the r.h.s. \qed

When we come (section 3) to imposing the mathematical structure of
a modular functor (or TQFT) on the ground state spaces $G^{+}_{\c,
\ell} (Y)$ for various surfaces $Y$ we will need to impose a base
point $\ast$ on each boundary component $C \subset
\partial Y$. This is directly analogous to the framing of Wilson loop in [Wi], in fact
the base point moving in time defines the first direction of a
normal frame to the Wilson loop in the $2+1$ dimensional
space-time picture.  As in the previous application, the base
point is introduced for mathematical rather than physical reasons.
It allows the state vectors in each conformal block to be
identified precisely and not merely up to a (block-dependent)
phase ambiguity.  Concretely in our model the base point prevents
domain walls from spinning around a puncture.  Note that if (a
superposition of) domain walls $\Pa$ represent an eigenspace for
Dehn twist around the puncture with eigenvalue $ \lambda \neq 1$
and if twisting is not prevented then the relation $| \Pa \rangle
=\lambda |\Pa\rangle$ will occur, killing the state $|\Pa \rangle$
which is certainly not desired.  I thank Nayak for pointing out
that although choosing base points breaks symmetry, none of the
physics depends on which base points are chosen.  The Hamiltonian
has a $\underset{k}{{\underbrace{U(1) \times \cdots \times
U(1)}}}-$gauge \break symmetry where $k = \#$ boundary components
of $Y$.


\renewcommand{\thesubsection}{$\bf 1^\prime$}
\subsection{An alternative microscopic model.}
\renewcommand{\thesubsection}{\thesection.\arabic{subsection}}
In this subsection
we present an alternative Hamiltonian, $H'_{\c, \ell}$, on a
cellulated surface $(Y, \mathcal{C})$.  We do not restrict to
triangulation since the square lattice actually yields the
simplest form.  It has the same relation, in the infrared, to
topological theories as does $H_{\c,\ell}$.  In this model the
degrees of freedom are on bonds and the loops lie in a
$\l$midlattice" separating the $|+\rangle$ clusters from the $|-
\rangle$ dual clusters (isolated vertices and isolated dual
vertices count as clusters). There are perhaps three advantages:
\begin{enumerate}
  \item On a square lattice, all terms in the Hamiltonian have order $4$
        (as compared to seven in the previous model).  It is simple enough that we expand it as a product of Pauli
        matrices. \\
  \item The corresponding classical statistical mechanical model is the Potts model in cluster expansion $(FK)$ form
        with $q= (2 \cos\f{\pi}{\ell +2} )^4$.; and \\
  \item The loops in this model are $\l$fully packed" so no isotopy is possible, only $d-$isotopy.  In particular the total
        length of the loops separating $|+\rangle$ from $|- \rangle$ is configuration independent.  Here is $H'_{\c, \ell}$ ; the
        notation is explained below.
\end{enumerate}

\begin{equation}
\begin{split}
H'_{\c, \ell} = & \sum_{\includegraphics[width=.25cm,height=.25cm]{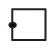}}
\left(|3\rangle
-\f{1}{d}\,|4\rangle \right) \left( \langle 3 | - \f{1}{d} \langle
4 |\right)+  \\
&\kappa\,\,\sum_{\includegraphics[width=.25cm,height=.25cm]{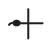}}
\left(|\widehat{1} \rangle - \f{1}{d} |\widehat{0} \rangle\right)
\left( \langle \widehat{1} | - \f{1}{d} \langle\widehat{0}
|\right)
\end{split}
\end{equation}

$\kappa$ is a positive constant which for symmetry we suppose to
be $\kappa =1$.  Again $d= 2\cos \f{\pi}{\ell +2}$.  On each bond
there is a spin $=\f{1}{2}$ degree of freedom $= \C^2 =$ span $\{
|+\rangle, |-\rangle \}$. The first summation is over all plaques
($2-$ cells) with a market edge (So, if the surface is a $10
\times 10$ torus cellulated with $100$ squares, the first summand
contains $400$ terms.)  Each term \break in the  first sum is orthogonal
projection  onto the vector  \break  $\left(|- \rangle \otimes | + \rangle
\otimes |+ \rangle \otimes \dots \otimes |+ \rangle - \f{1}{d}
|+\rangle \otimes |+\rangle \otimes |+\rangle \otimes \dots
\otimes |+ \rangle\right)$  where the tensor factors begin with the
bond containing the dot and proceed counterclockwise around the
plaquet.  (Of course this projector is understood to be tensored
with the identity over all remaining bonds.)  Perhaps this is
confusing, but we have used the notation $|3\rangle$ for the first
and $|4\rangle$ for the second basis vector in this combination
because in the \underline{square} lattice case, those numbers
count the $+$ signs: a more elaborate notation would be $| n_i -
1\rangle - \f{1}{d}| n_i \rangle$.  The second term is the
$\l$double dual" of the first where one duality swaps cellulation
with  dual cellulation (homology with cohomology) and the other
duality swaps $| +\rangle$ and $|- \rangle$.  Thus the second
summation is over vertices  with a marked incoming bond; the
vector $| \widehat{1}\rangle$ denotes $| + \rangle \otimes |
-\rangle \otimes |- \rangle \otimes \dots \otimes |- \rangle$ and
$| \widehat{0} \rangle$ denotes $|- \rangle \otimes |- \rangle
\otimes |- \rangle \otimes \dots \otimes |- \rangle$, again
reading counterclockwise from the dot. The $^\wedge$ reminds us
that we are reading around a site not a plaquet.  In the case of
the square lattice the two types of terms may be expressed as a
$4^\tn{th}$ degree polynomial in Pauli matrices: $ \s_z =
\begin{smallmatrix}
 _{| + \rangle } \\
 _{|- \rangle }\end{smallmatrix}\overset{\begin{smallmatrix}
 _{| + \rangle } & _{|- \rangle }\end{smallmatrix}}
{\left| {\begin{smallmatrix}
  {1} & {\,\,\, 0} \\
  {0} & {-1}
\end{smallmatrix}}\right|}
$
 and
$\s_x = \left|\begin{smallmatrix}
  {0} & {1} \\
  {1} & {0}
\end{smallmatrix}\right|$.

$H'_{\c, \ell}$ has two types of terms:
\begin{align*}
= & \left[ \f{1}{16} (I - \s^0_z ) - \f{1}{8d} \s^0_x + \f{1}{16
d^2} (I + \s^0_z )\right] \otimes  \left[ (I + \s^1_z ) \otimes
(I + \s^2_z )\otimes (I + \s^3_z)\right], \tn{ and} \\
& \left[ \f{1}{16} (I - \s^0_z ) - \f{1}{8d} \s^0_x + \f{1}{16
d^2} (I + \s^0_z )\right] \otimes  \left[ (I - \s^1_z ) \otimes
(I - \s^2_z )\otimes (I - \s^3_z)\right]\\
= & \f{1}{16}\left[ \f{d^2 +1}{d^2} I + \f{1-d^2}{d^2} \s^0_z
-\f{2}{d^2} \s^0_x )\right] \otimes  \left[ (I + \s^1_z ) \otimes
(I + \s^2_z )\otimes (I + \s^3_z)\right], \tn{ and}\\
& \f{1}{16}\left[ \f{d^2 +1}{d^2} I + \f{d^2- 1}{d^2} \s^0_z
-\f{2}{d^2} \s^0_x  \right] \otimes  \left[ (I - \s^1_z ) \otimes
(I - \s^2_z )\otimes (I - \s^3_z)\right]
\end{align*}

The proper context for understanding $H'_{\c, \ell}$ is Baxter's
$\l$mid lattice" [B].  If $ \mathcal{C}, \mathcal{C}^\ast$ are
cellulation and dual cellulation, let $c'=c \cap c^\ast$ be the
general intersection of a plaquet and a dual plaquet.  Put a
center $\ast$ in $c'$ and a center point $\ast_1 , \dots \ast_n$
in each of its boundary $1-$ cells. Subdivide $c'$ by the cone of
$\us{i=1}{\overset{n}{\bigcup}} \ast_i$ to $\ast$ and let $c''$
denote the general plaquet of this subdivision.  The collection
$\{c''\}$ are precisely the plaquets of the $\l$mid lattice".  As
an example, for the square lattice of unit size the dual lattice
is shifted by $(1/2, 1/2)$ and the resulting mid lattice is
spanned by vectors $\{ (0, 1/4), (1/4, 0)\}$.

A classical configuration $s: \{\tn{bonds of } \mathcal{C}\} \la
\{ |+\rangle, |- \rangle\}$ is encoded as the union of bonds on
which $s$ is $|+ \rangle$, the components of which are called
\underline{clusters} and the union of the duals of bonds on which
$s$ is $|-\rangle$, whose components are called dual clusters.
There is a well defined $1-$ manifold (multi$-$loop) $\Pa_s$ in
the mid lattice which separates clusters from dual clusters.

The Hamiltonian $H'_{\ell, \c}$ builds in dynamics which
fluctuates broken $(|3\rangle)$ and complete $(|4 \rangle)$ boxes
and broken $|\widehat{1}\rangle$ and complete
$(|\widehat{0}\rangle)$ dual boxes with a prescribed weight factor
$=d$. The vector $|4\rangle$ encodes a small face$-$centered loop,
$O_f$ in the mid lattice while $|\widehat{0}\rangle$ encodes a
small vertex$-$centered loop, $O_v$. The first term projectors, by
annihilating $|3\rangle + d |4 \rangle$, enforce a relation .  If
$\Psi = \sum_i a_i \Psi_i \in G'_{\c, \ell}$, the zero modes of
$H'_{\c, \ell}$, and $i$ is written out as (index on boundary
plaquet, distant indices $\overset{\rightharpoonup}{\kappa}$)
then,
\begin{equation}
 d\,a_{|3\rangle,\overset{\rightharpoonup}{\kappa}} = a_{|4\rangle,\overset{\rightharpoonup}{\kappa}}, \tn{ for all }
\overset{\rightharpoonup}{\kappa}.
\end{equation}
Examining this relation on mid lattice multiloops $\Pa$ (and
suppressing $\overset{\rightharpoonup}{\kappa}$) we see that
$\Pa_{|4\rangle}$ differs from $\Pa_{|3 \rangle}$ in that an $O_f$
has been added to the isotopy class of $\Pa_{|3\rangle}$ by
$\l$pinching off" a small bend in $\Pa_{|3\rangle}$.
Correspondingly $\Pa_{|4\rangle}$ has it's coefficient
$a_{|4\rangle}$ equal to $d$ times the coefficient $a_{|3\rangle}$
of $\Pa_{|3\rangle}$. Similarly for the double dual: up to isotopy
$\Pa_{|\widehat{0}\rangle} = \Pa_{|\widehat{1} \rangle} \cup O_v$
and for a zero mode the coefficients much satisfy:
\begin{equation}
  d \, a_{|\widehat{1} \rangle, \widehat{\kappa}} =a_{|\widehat{0} \rangle, \widehat{\kappa}}
\end{equation}
analogous to line (1.2) and Proposition 1.1 we have:

\begin{pro}
There are natural maps: $ETL^{\mathcal{C}!}_d (Y) \la ETL^!_d (Y)$
and $ETL^\mathcal{C}_d (Y) \la ETL_d (Y)$.  In the (direct) limit
they become isomophisms.  There are natural isomorphisms: $G'_{\c,
\ell} (Y, \mathcal{C}) \cong ETL^{\mathcal{C}!}_d (Y)$ and  $G^{'
+}_{\c, d} \cong ETL^{\mathcal{C}}_d (Y)$. \qed
\end{pro}

Proposition 1.2 replaces the triangulation $\tr$ with the
cellulation $\mathcal{C}$, so $ETL^{\mathcal{C}!}_d$ means formal
configurations, $\SS a_s s$, where $s$ assigns $|+\rangle$ or $|-
\rangle$ to the bonds of $\mathcal{C}$ and $a_s$ obeys (1.4) and
(1.5). $ETL^\mathcal{C}_d$ are formal configurations which are also
invariant under the global swap, $(^- ), |+ \rangle
\longleftrightarrow |- \rangle$. In the limit this relation
expresses $d-$isotopy of the mid lattice domain wall $\Pa$.  Note,
however that the first two maps mentioned in the proposition 1.2
are not necessarily injective. The situation is summed up by the
following example. On a $2\times 2$ square torus the two possible
staircase diagonals i.e. $|+\rangle$ on one positively sloping
diagonal and $|- \rangle$ on the complement, to not fluctuate (are
not in the same ergodic component) whereas already in the $3
\times 3$ torus there is enough room that any two staircases of
slopes $=1$ are connected by fluctuations.

\vskip.2in \epsfxsize=6in \centerline{\epsfbox{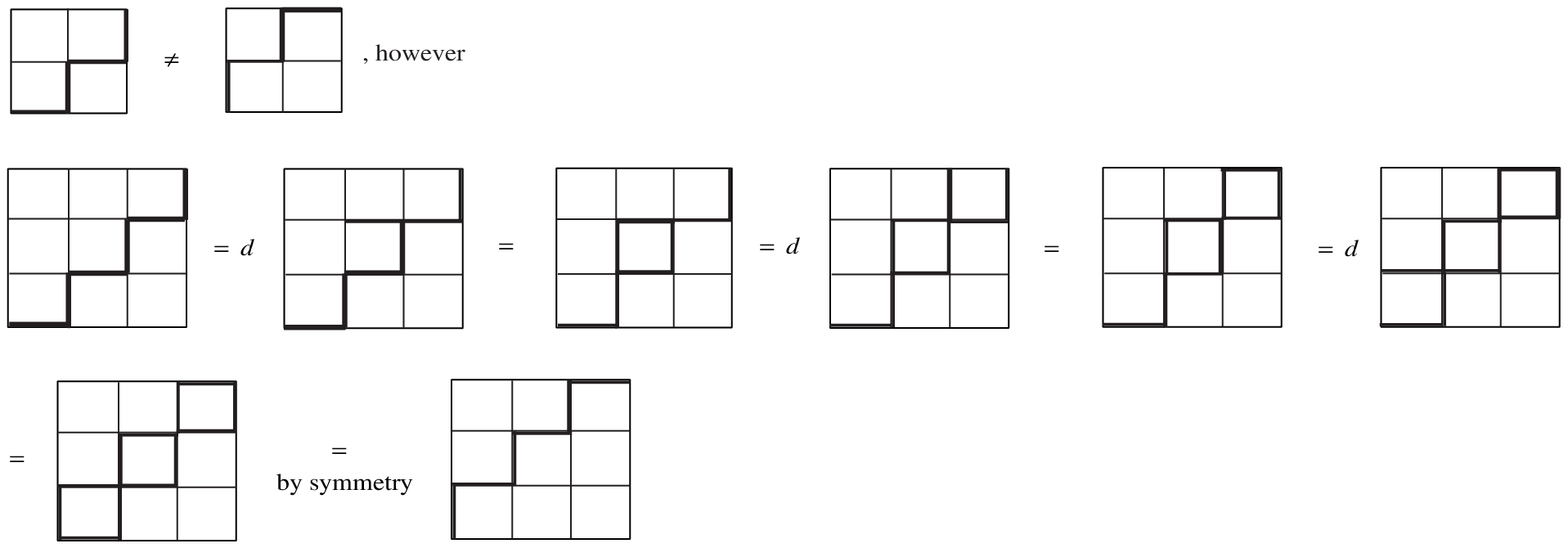 }}
{\centerline{Figure 1.3}} \vskip.2in

\begin{rem}\tn{
Because of their importance in solid state physics, we observe
that a certain \underline{ring} \underline{exchange} Hamiltonian
$H''_\c$ is the parent of all $H'_{\c, \ell}$ in that the
zero$-$modes $G''_\c$ contain the zero modes $G'_{\c, \ell}$, for
all $\ell$.  Each $G'_{\c, \ell}$ arises from a distinct linear
constraint on $G''_\c$.}
\end{rem}
\begin{align*}
 H''_\c&= \sum_{\includegraphics[width=.25cm,height=.25cm]{drawingb.eps}} \left(|3 \rangle - |3'\rangle\right)(\langle 3| - \langle 3'|) + \kappa \sum_{\includegraphics[width=.25cm,height=.25cm]{drawingc.eps}}
          \left(|\widehat{1} \rangle - |\widehat{1}'\rangle\right)(\langle \widehat{1}| - \langle \widehat{1}'|)\\
 |3'\rangle & \tn { is like } |3\rangle \tn{ except cycled one step: }\\
 |3'\rangle & =|+\rangle \otimes |-\rangle \otimes |+\rangle \otimes \dots, \otimes |+ \rangle, \tn{ similarly}\\
 |\widehat{1}'\rangle & =|-\rangle \otimes |+\rangle \otimes |-\rangle\otimes\dots\otimes |-\rangle .
 \end{align*}
Note that $(|3\rangle -|3'\rangle ) \in$ span
$\left(\left(|3\rangle - \f{1}{d} |4\rangle \right), |3'\rangle
-\f{1}{d} |4\rangle\right)$, etc..., so $G'_{\c , \ell} \subset
G''_\c$.  The zero modes $G''_\c$ can be identified with the
(combinatorial) isotopy classes of domain walls between $|+
\rangle$ and $|-\rangle$ regions.

Measuring spins (by a family of $\sigma_z$'s) converts a ground
state vector $\Psi \in G_{\c, \ell}$ or $G'_{\c, \ell}$ into a
classical probabilistic state $=$ meas.$(\Psi)$ which turns out to
be a Gibbs state.  The statistical physics of meas.$(\Psi)$ plays
an important role in section 3.  First, however, we use section 2
to lay down the algebraic framework.

\numberwithin{equation}{section}
\section{Things Temperley-Lieb }

The generic Temperley-Lieb algebra is a tensor algebra over the
complex numbers adjoined an indeterminate $d$.  Often $d$ is
written in terms of another indeterminate $A$ as $d= -A^2 -
A^{-2}$. In degree $n$ the algebra $\tn{TL}_n$ has generators $1,
e_1 ,\dots , e_{n-1}$ and the relations $e_i^2 = e_i , e_i e_j =
e_j e_i$ if $| i-j| \geq 2$ and $e_{i} e_{i \pm 1}e_{i} =
\f{1}{d^2}e_{i}$. Pictorially, after V. Jones and L. Kauffman, we
may think of the generators as pictures of arcs disjointly
imbedded in rectangles (multiplied by the coefficient $1/d$) and
multiplication as vertical stacking. For example for $n=4$, we
have:


\vskip.2in \epsfxsize=3in \centerline{\epsfbox{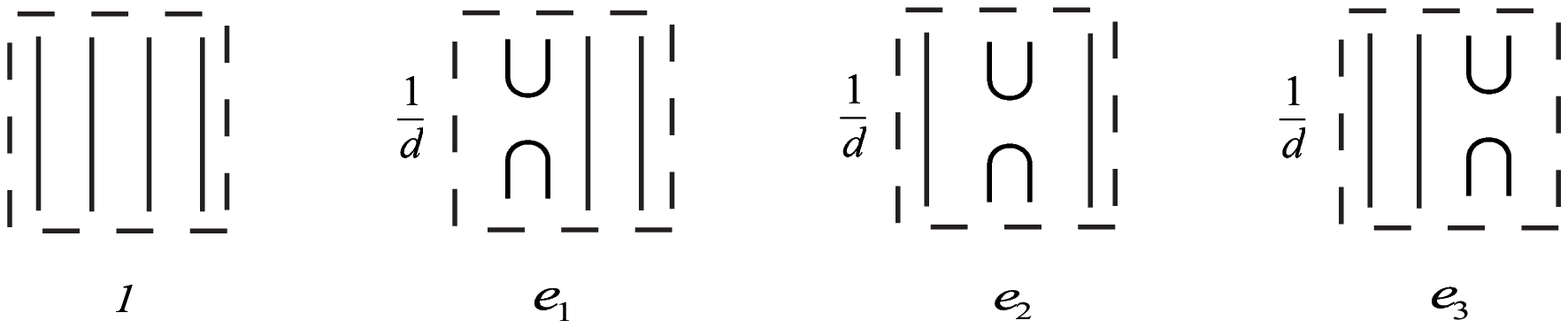 }}
{\centerline{Figure 2.1}} \vskip.2in

There is a convention that any closed circle ( and these may arise
when the pictures are stacked)  should be regarded as a factor of
$d$. All closed circles in a picture should be deleted and the
resulting picture should then be formally multiplied by $d^{( \#
\tn{circles})}$.  The reader can now easily verify the relation by
stacking pictures.  Kauffman proved the algebra of such pictures
has no other relations [K]. A tensor structure between grades
$\tn{TL}_n \ot \tn{TL}_m \la \tn{TL}_{n+m}$ is created by
horizontal stacking.  An inclusion $\tn{TL}_n \la \tn{TL}_{n+k}$
is obtained by adding $k$ vertical strands on the right.  The
union of grades is the (generic) Temperley-Lieb algebra, $\tn{TL}
= \us{n=1}{\overset{\infty}{U}} \tn{TL}_n$. The structure of this
algebra is completely worked out in [J]: Each grade $\tn{TL}_n$
has  $\dim (\tn{TL}_n) = \f{1}{n+1} \, \binom{2n} {n}$, the
$n^{\tn{th}}$ Catalan number, and is a direct sum of matrix
algebras that fit together via a rather simple Brattelli diagram.
Also of interest are specializations where the indeterminate $d$
is set to a fixed nonzero real number. Here the structure differs
from the generic case only when $d$ assumes a $\l$special value"
$d=2\cos \f{\pi}{\ell +2}$, $\ell$ a positive integer, and has
been worked out by Goodman and Wenzl [GW].

There is an involution - on $\tn{TL}$ which acts by reflecting the
rectangle in a horizontal line and conjugating coefficients
$(\overline{d} = d)$ making TL a $\ast$ - algebra. Using this, the
$\l$Markov trace pairing $<a,b> :=$ trace $(a\overline{b})$ may be
defined. $\l$Trace," on pictures, means closing a rectangular
diagram by a family of arcs sweeping from top to bottom and then
evaluating each circle as a factor of $d$ times $1 \in \C$. Extend
this definition to a Hermitian pairing on $\tn{TL}$

\vskip.2in \epsfxsize=2in \centerline{\epsfbox{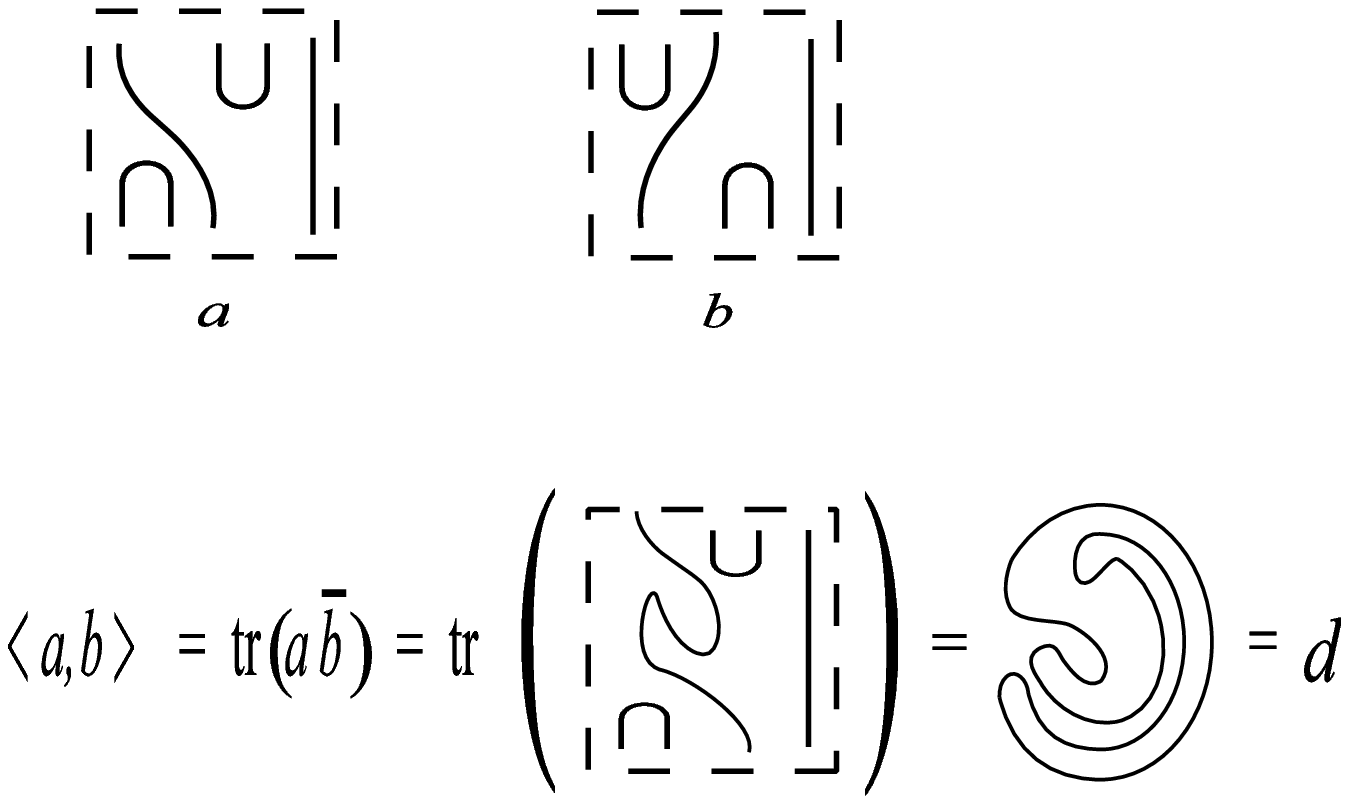 }}
{\centerline{Figure 2.2}} \vskip.2in

\begin{thm} ([J]) The trace pairing $\langle\, , \,\rangle:\tn{TL}\ot \tn{TL} \la \C[d]$,
when $d$ is specialized, to a positive real number, becomes a
positive definite Hermitian pairing $\langle\, , \, \rangle_d
:\tn{TL}_d \ot \tn{TL}_d \la \C$ exactly for $d\geq 2$.  For $d =
\l$special" = $2 \cos \f{\pi}{\ell + 2}$,  $\ell$ a positive
integer $\langle\, , \,\rangle_d$ is positive semidefinate.  For
other values of $d \in \mathbb{R}\smallsetminus  0$, $\langle\, ,
\,\rangle_d$ has mixed signs. \qed
\end{thm}

For $d = 2\cos \f{\pi}{\ell + 2}$ define the radical $R_d \subset
\tn{TL}_d$ by $<R_d , \tn{TL}_d >_d \equiv 0$. The radical $R_d$
has first non-trivial intersection with the $(\ell +
1)^{\tn{th}}$ grade where it is $1-$dimensional: $R_d \cap
\tn{TL}_{d , \ell} = 0$ and $R_d \cap \tn{TL}_{d, \ell + 1} =
\tn{span}_\C (p_{\ell + 1})$.  The elements $p_{\ell +1}$
belonging to $ \tn{TL}_{{-}, \ell+1}$ (the $\ell +1$ grade of the
generic algebra) are called the Jones-Wenzl [W] projectors and a
simple recursive formula for these is known.

In this paper we will be particularly concerned with $p_3$ and
$p_4$ (for $d = \sqrt{2}$ and $\f{ 1+ \sqrt{5}}{2}$ respectively)
which can be computed (from the formula on page 18 [KL]).

\vskip.2in \epsfxsize=4.5in \centerline{\epsfbox{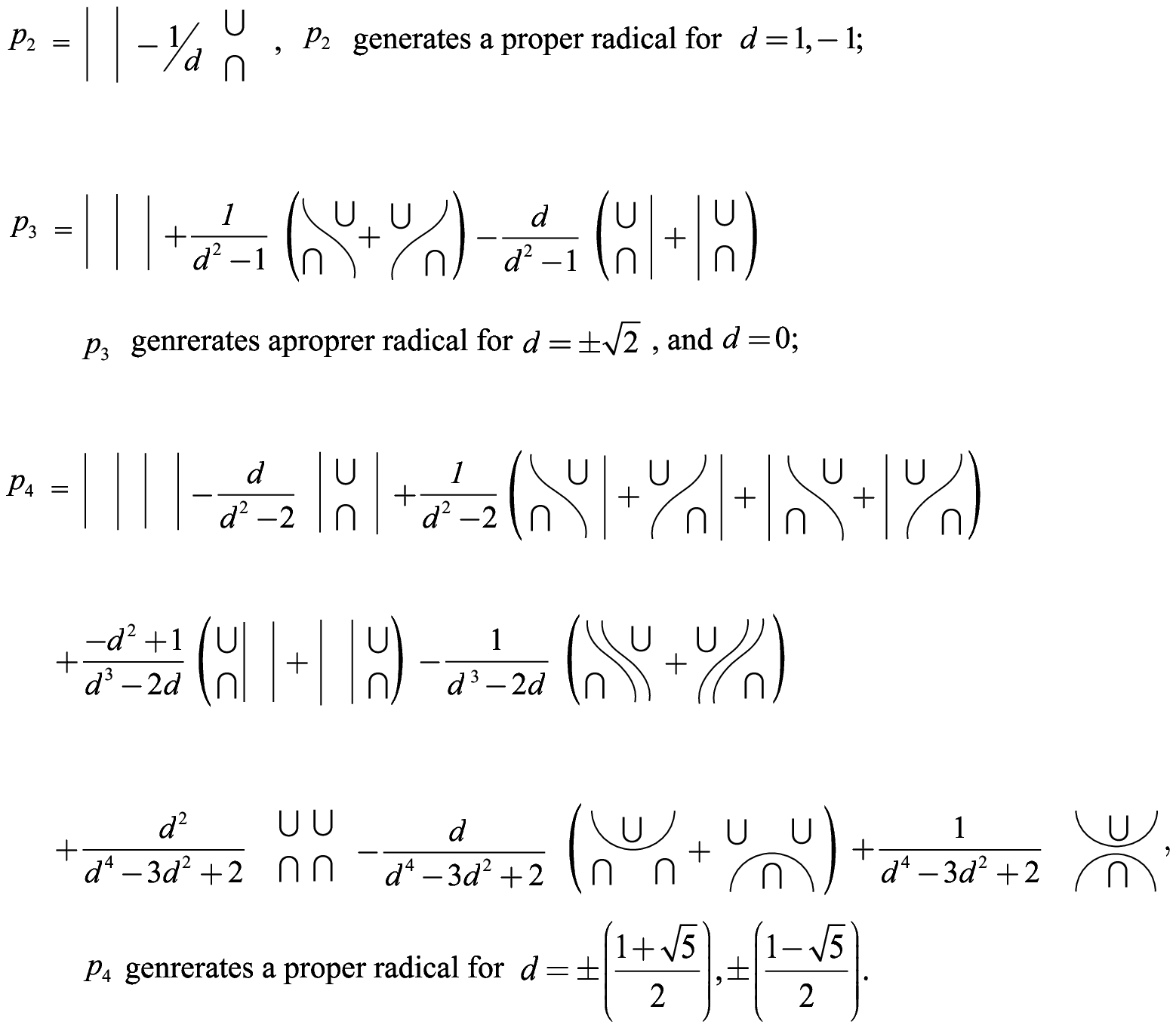 }}
{\centerline{Figure 2.3}} \vskip.2in


It is known that when $d$ is special, the ideal $J(p_{\ell + 1})$
generated by $p_{\ell + 1}$ in $\tn{TL}_d$ (the specialized
$\tn{TL}$ algebra) is $R_d$.  The notion of ideal closure in
different algebraic contexts is essential to all that follows so
we will be explicit here; $J (p_{\ell +1})$ is the smallest subset
of $\tn{TL}_d$ containing $p_{\ell + 1}$ so that if $c_1 , c_2
\i\, \C; a, b$ belong to the subset and $x,y \,\i \, \tn{TL}_d$
then: $c_1 a + c_2 b, ax, xa, a \otimes y$, and $y \otimes a$
\underline{all} belong to the subset. $J(p_{\ell +1})$ is linear
subspace of $\tn{TL}_d$ and is a two sided ideal under
$\centerdot$ and $\otimes$.  So $J$ is, by definition, closed
under formal linear combination and all types of picture stacking:
top, bottom, right, and left.

For $d$ special the algebra $\tn{TL}_d$ contains many other ideals
besides $R_d$ (e.g. the ideal generated by diagrams with at least
two $\l$horizontal" arcs) but we find that when we move to the
\underline{category}, $R_d$ becomes unique (see Appendix). This
motivates the definition of the Temperley-Lieb category
$\tn{TL}_d^c$.

The generic Temperley-Lieb category $\tn{TL}^c$ is a strict
monoidal  tensor category over $\C(A)$ with objects $N_\c
=\{0,1,2, \ldots\}$ thought of as that number of marked points in
the interior of a horizontal interval. The indeterminate $A$
determines $d$, above, by the formula $d=-A^2 -A^{-2}$.  The
morphisms Hom$(m,n)$ is a $\C(A)$ vector space spanned by all
pairing of the $n + m$ points that can be realized by disjointly
imbedded arcs in a rectangle for which the $m$ points are on the
top and the $n$ points on the bottom edge.  The only difference
from the algebra is that we do \underline{not} demand that a
nontrivial morphism have $m=n$. Again composition $(^{\centerdot}
)$ is vertical stacking and $\otimes$ is horizontal stacking. The
involution, the specialization of $d$ and the notions of
$\l$ideal" are defined using exactly the same words as before. Now
the Markov trace $ <a,b> = \tn{tr}(a\overline{b})$ becomes a
Hermitian pairing Hom$(m, n) \times \tn{Hom}(m,n) \la \C$. Theorem
2.1 continues to hold with $\tn{TL}^c$ and $\tn{TL}_d^c$ replacing
$\tn{TL}$ and $\tn{TL}_d$ respectively and for $d$ special the
radical $R_d$ is still the ideal closure of $p_{\ell + 1}$.  But
in the categorical setting there is a new result conjecture by the
author and proved by Goodman and Wenzl (see Thm 3.3. in the
appendix to this paper), which when combined with Theorem 2.1
yields.

\begin{thm}\tn{{\bf (Goodman, Wenzl)}
For $d= $ special value $ = 2\cos \f{\pi}{\ell +2}$, $\tn{TL}_d^c$
has a unique non-zero, proper ideal $= R_d = J(p_{\ell + 1})$ and
on the quotient TL$^c_d /R_d$ the pairing $\langle, \,\,\rangle_d$
becomes positive definite.  If $d\neq$ special value but is of the
form $d= \a + \overline{\a}$, $\a$ a root of unity, $\a \neq \pm
1$ or $\pm \,\,i$, then TL$^c_d$ has a unique non-zero proper
ideal, the pairing $\langle, \,\,\rangle_d$ descends to the
quotient but has mixed sign.  For other values of $d \in
\mathbb{R}\smallsetminus 0$, TL$^c_d$ has no non-zero proper
ideal. } \qed
\end{thm}

We can continue to make the algebraic structure more flexible,
more suited to both topology and physics, while retaining the key
notion of $\l$ideal" and uniqueness property set out in the
proceeding theorem.  One step in this direction is Jones theory of
$\l$planar algebras" [J2].  These are generalized categories with
an operad structure replacing the notion of morphism.  The
TL-planar algebra, $\tn{TL}^p$ or $\tn{TL}_d^p$, if $d$ is
specialized, begins with a Hilbert space $h_{2k}$ associated to an
even number $2k$ of points marked on a circle: $h_{2k} \cong$
span(imbeddable arc pairings in a disk $D$ with $2k$ marked points
on $ \partial D$). To a disk with $j-$ internal punctures $D^-$
and a relatively imbedded $1-$ manifold $\Pa \subset D^{-}$, where
$\Pa$ has $2k_i$ endpoints on the $k^{\tn{th}}$ interval boundary
component and $2k$ endpoints on the outer boundary component,
Jones associated (in an obvious way\footnote{Let closed loops in
$D$ be assigned the multiplicative factor $d$}!) a homomorphism
$\us{i=1}{\overset{j}{\ot}} h_{2k_i} \la h_{2k}$. In the planar
algebra context the distinction between times $(\centerdot)$ and
tensor $(\ot)$ has been lost because there is no up, down, right,
left.  Instead we have $\l$subpictures" of $\l$pictures", i.e.
restrictions of imbedded $1-$manifold on a surface to a
subsurface.

\begin{defi}
A \underline{picture} $\Pa$ in $Y$ is an imbedded $1-$submanifold
(multi curve), proper  if $\partial Y \neq \emptyset$. A formal
picture is a linear combination of pictures with identical
boundary if $\partial \Pa \neq \emptyset$.
\end{defi}

In Jones' theory there is no action by Dehn twist because surfaces
are considered up to homeomorphism.

We take a further step, and allow surfaces with genus $>0$, here
Dehn twist becomes crucially important. Consider an oriented
compact surface $Y$, and the possible imbedded $1-$manifolds
($\l$multi-curves") $\Pa$ in $Y$. Picking a special value $d$ for
closed circles which bound a disk ($\l$trivial circles") defines
$d-$isotopy.  In section 1, we have defined ${\tn{ETL}}_d (Y)$ to
be the $\C -$vector space of $d-$isotopy classes of closed
null-bounding $1-$manifolds modulo $d-$isotopy, on a surface $Y$.

\begin{defi}
Suppose $a = \Sigma a_i \Pa_i$ is a formal picture in a disk
$\delta \subset$ interior $(Y)$ with fixed endpoints $\partial
\Pa_i \subset \partial \delta$.  The ideal $J(a)$ or $\langle a
\rangle \subset ETL_d (Y)$ generated by $a$ are the $d-$isotopy of
 formal pictures of the form $a x, \,x = \Sigma x_j \chi_j$,
$\chi_i$ a picture in $Y\setminus \delta$ with $
\partial \chi_j =\partial \Pa_i$, for all $i$ and $j$, $ax= \underset{i, j}{\Sigma} a_i x_j(\Pa_i \cup \chi_j )$.
Dually, $\langle a \rangle^\ast \subset ETL^\ast_d (Y)$ are the
functions annihilating $\langle a \rangle$. Concretely, $y \in
\langle a \rangle^\ast$ iff $y (ax) =0$ for all $x$ as above. The
definition of ideal is the same in $TL_d (Y)$ and similar in the
combinatorial settings: $ ETL^\tr_d (Y)$ and $ETL^\mathcal{C}_d
(Y)$.
\end{defi}




One finds that the quotient $ETL_d /J(p_{\ell +1}) =: QE\ell$ has
(or better $\l$recovers") the structure of a TQFT, or more
precisely, a $2+1-$dimensional unitary topological modular functor
(UTMF). For this, we must extend the definition of $QE\ell$  to
the case of a surface with labelled boundary
$(Y,\overset{\rightharpoonup} {t})$. The essential feature is that
$QE\ell$  may be calculated by $\l$gluing rules" applied to these
smaller pieces. When we wish to emphasize the modular (UTMF)
structure on $QE\ell$  we use the notation $DE\ell = QE\ell$  to
recall the doubled $SO(3)$ or $\l$even" theory discussed in the
introduction. Up the global $|+\rangle\leftrightarrow |-\rangle$
involution $^-$ on configurations, $DE \ell$ will be our model for
the perturbed ground state space $G_{\e , \ell}$, $ DE\ell \cong
G^+_{\e, \ell}$.

A UTMF is a very natural way to model the topological properties
of a two dimensional particle system without low lying modes in
the bulk. Knowing that the ground state has the structure of a
particular modular functor $(DE \ell)$, tells us all the
topological information about, excitation types, braiding rules
(nonabelian Barry phase), $6_j-$symbol, $S-$matrix, and fusion
rules. It is this structure that we have been seeking.

The statement $DE\ell = QE\ell$ is a purely topological one and it
is possible to piece it together from the topological literature
using [BHMV], [Prz] and [KL].  An exposition [FNWW] of the easiest
modular functors is in progress and will explicate this
isomorphism and contain a proof of theorem 2.5 below.

But let us take a step back and explain this structure (UTMF) in a
context where the gluing rules are obvious. Then we will summarize
the axioms and finally explain the labels, pairing, and
cutting/gluing operations  in $DE \ell$ in terms of functions on
pictures.

Let $M(Y)$ be the vector space spanned $1-$submanifolds ($=$
pictures) $\Pa$ of $Y$ with \underline{no} equivalence relation.
Suppose $Y$ is cut into two pieces by a circle $\a \subset Y$, $Y
= Y_1 \cup_\a Y_2$.  The uncountable set $X$ of all finite subsets
of $\a$ will be the $\l$labels" or $\l$superselection sectors" of
this theory. Neglecting the measure zero event that $\Pa$ and $\a$
are not transverse, we can formally write:

\begin{equation}
M(Y) = \us{x \e\, X}{\bigoplus} M(Y_1 , x) \ot M(Y_2 , x)
\end{equation}
where $M(Y_i , x)$ is the vector space of $1-$manifold in $Y_i$
meeting $\a = \partial Y_i$, in the finite point set $x$. Equation
(2.6) is the essential feature of a TMF as used by Witten [Wi] and
formalized by Segal [S], Atiyah [A], Walker[W], and Turaev [T].
Many enormous $\l$classical spaces" have this kind of formal
structure but it requires beautiful algebraic $\l$accidents" to
find finite dimensional $\l$quantizations" of these.



Bounding or $\l$even" pictures span another (huge) vector space
$EM(Y)$.  Let us set $d= 2\cos \f{\pi}{\ell+2}$ and constrain the
functions, in $M^\ast(Y)$ and $EM^\ast(Y)$, first by the $d-$
isotopy relation and then annihilation by the ideal generated by
Jones-Wenzl relation $p_{\ell +1}$.  This yields the following
quotients and inclusions in lines (2.2) and (2.3)
\begin{equation}
\begin{split}
DK &(A= ie^{\pi i/2\ell +4})\longleftarrow TL_d(Y)\longleftarrow M(Y), \\
DK^\ast &\left( A= i e^{\pi i/ 2\ell + 4 }\right)= \langle p_{\ell +1}
\rangle^\ast \hookrightarrow TL^\ast_d (Y) \hookrightarrow
M^\ast(Y)
\end{split}
\end{equation}

\begin{equation}
\begin{split}
DE\ell & \longleftarrow ETL_d(Y) \longleftarrow EM(Y) \\
 DE\ell^\ast &= \langle p_{\ell +1}
\rangle^\ast \hookrightarrow ETL^\ast_d (Y)\hookrightarrow
EM^\ast(Y)
\end{split}
\end{equation}

\begin{thm}
The annihilating subspace $\langle p_{\ell +1} \rangle^\ast$  of
$M^\ast(Y)$, we wrote it $DK^\ast$, is in fact, the Drinfeld double [Dr] of
the unitary topological modular functor (UTMF) derived from the
Kauffman bracket at $A= i e^{\pi i/ 2\ell + 4 }$.  This is true
even at odd levels, $\ell =$ odd, where the undoubled Kauffman
bracket MF is flawed by having a singular $S-$matrix. $DE \ell$ is
a UTMF for $\ell \neq 2 \mod 4$ and in these cases is a
\underline{trivial} double: $V^\ast \otimes V$.
\end{thm}

\begin{rem}\tn{
For the corresponding even space, $DE\ell$ the same MF arises for
$A, i A, -A$, and $-iA; A= i e^{\pi i/ 2\ell + 4 }$ so the
notation agrees with the introduction.}
\end{rem}

\begin{rem}\tn{
The Kauffman bracket TMF (or TQFT), constructed in [BHMV], is not
identical to the TMF derived from $SU(2)$.  In physics there is
the loop group $L(SU(2))$ approach and in representation theory
there is the quantum group $(Usl_{2 , q})$ approach and these lead
to the same representation categories.  Globalization of these
representation categories (this view point is explained in [Ku])
yields the same MF.  The pictures underlying the Kauffman bracket
are unoriented arcs.  The Rumer-Teller-Weyl theorem  shows these
\underline{almost} correspond to Rep $q(SU(2))$. However an
important minus sign, the Frobenius-Schur indicator, corresponding
to the quaternionic (not real) structure of the fundamental
representation is missing. This minus sign propagates into the $S$
matrix making the $K$ and the $SU(2)$ (TMFs) distinct. A different
microscopic model which allowed arbitrary $1-$manifolds (not just
bounding $1-$manifolds) could, depending on the local details,
lead to $\mathcal{D}K\ell$ or $DSU(2) \ell$ so solid state
physicists looking for anyons will need to be aware of this
distinction in detail [FNWW].  The present models $H_{\c, \ell}$
and $H'_{\c, \ell}$ address only \underline{bounding}
$1-$manifolds which correspond to (endomorphism of) the even
symmetric powers of the fundamental representation which are all
real. Thus $K \ell$ restricted to even labels $EK \ell \cong SO(3)
\ell$, the $SU(2)$ theory at even labels.  The same holds, of
course, for the doubles of these TMFs.}
\end{rem}

\begin{add} \tn{In [FLW2] it is shown that the braid representation of the $\l$Fibanocci
category"\footnote{Greg Kupperberg's term for the even label
sub-theory of $(SU(2), 3)$ also called the $SO(3)-$theory at level
$3$.} (F) is universal for quantum computation. $DE 3(A= e^{\pi
i/10})$ has $4$ labels $0, 0; 0, 2; 2,0;$ and $2, 2$ and is
isomorphic to $F \otimes F^\ast$ implying that $DE 3$ is also
universal.}
\end{add}
\begin{add}\tn{
If $Y$ has a fixed triangulation $\tr$ then combinatorial versions
of the six vector spaces connected by maps (2.2) and (2.3) in
Theorem 2.5 are defined.  Provided $\tr$ is sufficiently fine, no
information is lost; the left most combinatorial spaces are
actually isomorphic to $DK\ell$ and ${D} E \ell$ respectively. The
proof is the same as in the main topological theorem of [F1].
Furthermore an estimate on the required fineness of $\tr$  (it is
linear in $\ell$) can be extracted from that proof. Also if $Y$
has boundary and labels $\overrightarrow{t}$(see the discussion of
labels which follows immediately) are specified, then the left
most combinatorial spaces are again defined and these map onto the
TMFs with the given boundary labels $\overrightarrow{t}$.}
\end{add}

To appreciate the last statement we set out Walker's axioms [Wa]
for a UTMF. Fortunately these can be abbreviate due to two
simplification: 1) the theories are unitary and 2) both are
quantum doubles (i.e. the endomorphisms, of another more primitive
UTMF), so the central charge $c=0$ $( c (V \otimes V^\ast ) = c(V)
+ c(V^\ast) = c(V) - c(V) =0.$ Thus no $\l$extended structures" or
projective representations need be mentioned. For a concrete
appreciation of these examples, see figures 2.4 and 2.5 where the
particle types $(=labels)$, fusion algebra, and braiding, and
$S-$matrices in the cases $DE3$ are given.

A labelled surface $Y$ is a compact oriented surface $Y$ possibly
with boundary, each boundary component has a base point marked and
a label $t \e \mathcal{L}$ from of finite label set $\mathcal{L}$
with involution $\widehat{\,\,}$ containing a distinguished
trivial element $0$, fixed by $\widehat{\,\,}$. For Kauffman
$SU(2)$, and $SO(3)$ theories the labels are self dual
$a=\widehat{a}$ but we include the hats in the formulas anyway. A
UTMF will be a functor $V$ from the category of label surfaces,
and isotopy classes of diffeomorphisms (preserving labels and base
points) to the category of finite dimensional Hilbert spaces over
$\C$ and unitary maps.

\noindent\underline{Axiom 1 (disjoint union):}  $V(Y_1 \amalg Y_2
, t_1 \amalg t_2)= V(Y_1 , t_1 , )\ot (Y_2, t_2)$, the equality is
compatible with the mapping class groupoids:
\[
V(f_1 \amalg f_2)=V(f_1) \ot V(f_2).
\]
\underline{Axiom 2 (gluing):} If $Y_g$ is obtained by gluing $Y$
along dually labeled $(x, \widehat{x})$  boundary components then:
\[
V(Y_g, t) =\bigoplus_{\begin{subarray}{1}
            (x,\widehat{x})\,\e \tn{ labels on the}\\
            \tn{paired circles}
            \end{subarray}} V(Y, t, x, \widehat{x})
\]
The identification is also compatible with mapping class
groupoids - and is associative (independent of order of gluings).

\noindent\underline{Axiom 3 (duality):} $V(Y, t) =V(-Y, t)^\ast$,
where $-$ is orientation reverse on $Y$ and $\widehat{\,\,}$ on
labels, and $\ast$ denotes the space of complex linear
functionals.  The Hermitian structures on $V$ give vertical maps
and the diagram below must commute:

\begin{align*}
\begin{CD}
  V(Y) &\longleftrightarrow & V(-Y)^\ast \\
  \updownarrow & & \updownarrow \\
  \overline{V(Y)^\ast}&\longleftrightarrow &\overline{V(-Y)}
\end{CD}
\end{align*}

All these identifications are compatible with the mapping class
groupoids.  The Hilbert space pairings are compatible with maps:
\[
\langle x,y\rangle =\langle V(f) x, V(f)y\rangle, \tn{ where }  x,y \, \e V(Y, t),
\]
and disjoint union: $\langle\a_1 \ot \a_2, \bt_1 \ot \bt_2 \rangle
>=\langle\a_1 , \bt_1\rangle\langle\a_2 , \bt_2\rangle$. Writing \break
$\a , \bt \e V(Y_g, t)$ as $\a =\us{x}{\bigoplus} \a_x$ and $\bt =
\us{x}{\bigoplus} \bt_x$ according to axiom 2, then $\langle\a,
\bt\rangle = \us{x}{\SS} S_x \langle\a_x , \bt_x\rangle$. Where
$S_x = \us{x_i \e \, \overset{\rightharpoonup}{x} =(x_1, \ldots,
x_n) }{\prod} S_{0, x_i}$.  The symbols $S_{0, x_i}$ are the
values of a fixed function $\mathcal{L} \la \C \smallsetminus
\{0\}$ which is a part of the definition of $V$. Experts will
recognize $S_{0,  x_i}$ as the $(0, i)$ - entry of the $S-$ matrix
of $V$: this is the map that describes exchange of meridian and
longitude of a torus in the natural $\l$label" bases.

\noindent\underline{Axiom 4 (Empty surface):}  $V(\emptyset) \cong
\C$

\noindent\underline{Axiom 5 (Disk):} Let $D$ be a disk,
$V(D,a)\cong
    \begin{cases}
        \C,  & a=0 \\
        0, & a\neq 0
    \end{cases}$

\noindent\underline{Axiom 6 (Annulus):} Let $A$ denote an annulus.

Then
\begin{equation*}
    V(A, a, b)\cong
    \begin{cases}
        \C, \,\,  a=\widehat{b} \\
        0, \,\, a\neq \widehat{b} .
    \end{cases}
\end{equation*}

To complete these axioms to a theory incorporating $3-$manifolds
Walker adds axioms $7-10$.  We will not need these here except to
note that a $3-$manifold $X$ determines a vector $Z(X)$ belonging
to $ V(\partial X)$.  If $X= Y \times I$, $V(X)=$ id $\in$
Hom$\left(V(Y \times + 1), V(Y \times -1)\right) = V(Y) \ot
V(Y)^\ast = :\mathcal{D}\left( V(Y) \right)$.

In the $SU(2)$ theories it has been known since [Wi] that if $X$
contains a labeled link $\l$Wilson loop" (or suitable labeled
$\l$trivalent graph") then this pair also defines an element of
$\mathcal{D}V(Y)$. The simple idea is to regard the $1-$manifold
$\Pa \subset Y=Y \times 0$ as a link labeled by $\l 1$", the
$2-$dimensional representation of sl$(2, q)$ inside $X=Y \times
[-1, +1]$.  This defines a map $M(Y)\la {D} \ell(Y)$. If $\Pa$ is
null bounding (in $Z_2 -$homology) on $Y$ then there is a
subsurface $Y_0 \subset Y$ with $\partial (Y_\c ) =\Pa$. Let $G$
be a generic spine (trivalent graph) for $Y_\c$. Derived from
Witten's theory and it bracket variation are combinatorial
recoupling rules ($6_j$ symbols) which are exposited in detail by
Kauffman and Lins in [KL]. We have adopted their notations (which
caused us to rename Walker's trivial label $\l 1$" by $\l 0$")
except in the choice of $A$, $A^4 = q = e^{2\pi i/\ell +2}$. To
make $d$ positive we choose $A= i e ^{\pi i/2 \ell +4}$, that is
$i$ times the primitive $4(\ell +2)^{\tn{th}}$ root chosen in
[KL]. For $\ell =$ even our $A$ is still a primitive $4(\ell
+2)^{\tn{th}}$ root of unity, for $\ell = $ odd, it is a primitive
$2(\ell +2)^{\tn{th}}$ root of unity but still defines a
nonsingular TUMF on the even labels.

Applying recoupling, $\Pa$ yields a formal labelling of $G$ in
which only even labels - odd dimensional representations $-$
appear. This means that the set of possible morphism $Z(X, G)$ is
isomorphic to the endomorphism algebra of the sum of even labelled
blocks. Restricted to even levels, the $4$ choices for $A$
differing by powers of $i$ give the same Kauffman bracket UTMF and
this agrees with the $SO(3)$ UTMF, which we call $DE\ell$.

The recoupling relations on labelled trivalent graphs $G \subset Y
\times 0$, i.e. the $6_j$ symbols, are consequences of projector
the relation $p_{\ell+1}$ applied to formal $1-$manifolds (and
conversely $p_{\ell+1}$ follows from $6_j$). What is less direct
[Prz] is that on a surface $Y, \, p_{\ell +1}$ alone generates the
same relation as including $Y\times 0 \subset Y \times [-1, +1]$
and then employing both $p_{\ell+1}$ and the Kauffman bracket
relation $
{\includegraphics[width=.30cm,height=.30cm]{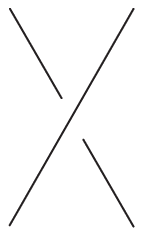}}\,\, A
)( +A^{-1}{\cup\atop\cap}$.


Abstractly we know the label sets for $\mathcal{D}K \ell$ and $DE
\ell$, but we need to interpret these labels in $\mathcal{D}K \ell
:= TL_d/ \langle p_{\ell +1} \rangle$ and $DE\ell := ETL_d /
\langle p_{\ell+1}\rangle$ resp. and in this context of recover
the gluing formula. From a physical point of view it would be
surprising if we could \underline{not} localize because we expect
the Hamiltonian $H_{\e, \ell}$ to define a stable topological
phase for which the superselection sectors of excitations define
the label set. But such reasoning is in the end circular; it is
better to have a mathematical proof that the candidate ground
state space $G_{\e , \ell}$ has the structure of a UTMF and view
this as evidence for or a $\l$consistency check" on  the physical
stability of $G_{\e, \ell}$.

We now explain the $\l$labels" for the theories $DK \ell$ and $DE
\ell$ in terms of $\l$pictures". A conceptual point is that the
label has a kind of symplectic character: $\l$half" the label's
information  is a non negative integer $\leq \ell$ which counts
$\l$essential" strands of $\Pa$ passing inward from a component
$C\subset \partial Y$.  Think of this as $\l$position"
information. (Any $\l$excess" strands correspond to a descendent
field or gapless boundary excitation.)  The other half of the
information ($\l$momentum") is expressed as a symmetry condition
on $\Pa$ in the bulk $Y$. The formal picture $\Pa$ must lie in the
image of certain minimal idempotents $-$ certain eigenspaces or
projector images as constructed below.


Abstractly, the label set $\mathcal{L}$ for $DE\ell = {Q}E \ell:=
ETL_d / \langle p_{\ell +1}\rangle$ may be written as:
\[
\mathcal{L} = \{(0,0); (0,2);(2,0); \ldots;
\left(2\left\lceil\f{\ell + 1}{2}\right\rceil, 2\left\lceil\f{\ell
+1}{2}\right\rceil\right)\}
\]
The $\l$position" part of the doubled label $t= (a, b)$ for $\Pa
\in QE\ell$ on a boundary components $C \subset \partial Y$ is
$|a-b|$.  This quantity is the smallest number $\#$ of domain wall
$(\gamma)$ intersections with $C'$, $\# = |a -b|$, as $C'$ varies
over all imbedded loops parallel to $C$ (i.e. cobounding an
annulus with $C$) and the domain wall $\Pa$ also varies over all
$\langle p_{\ell +1}\rangle -$ equivalent pictures. The
$\l$momentum" part of the label is an eigenvalue.


Let us do this more carefully.  We follow [BHMV] to define what
Walker calls an $\l$annulus category" $\wedge^A_\ell$.  $A = S^1
\times I$, obj$(\wedge^A_\ell) =\{$set of even number of points
on  $S^1\}$ then an element of morph $(\wedge^A_\ell)$ are all
formal combinations of pictures in $A$, which beginning on the
object in $S^1 \times -1$ and end on the object in $S^1 \times
+1$, and which obey the relations: $d-$isotopy and $p_{\ell + 1}$.
(Recall $\langle p_{\ell +1}\rangle=$ negligible morphisms of
$TL_d$. Also see appendix.) Suppose that $Y$ is a surface with
connected boundary $\partial Y=C$, then there is a gluing action
of $\wedge^A_\ell$ on $DE\ell (Y): f \in \tn{ morph
}\left(\wedge^A_\ell\right)$, $g \in DE \ell(Y)$, $f \c g \in DE
\ell(Y); \break f\c g(x_i \otimes z_j) = \chi_i \omega_j$, where
the coefficient of $f$ on the picture $x_i$ is $\chi_i$, $f(x_i) =
\chi_i$ and $g (z_j) = \omega_j$. For this action to be defined we
must pick an identification $Y \cong Y \underset{C}{\cup} A$. Also
$C$ has a fixed parameterization and an orientation; these tell us
which end of $A$, $S^1 \times -1$ or $S^1 \times +1$ to glue to
$C$. Technically, this means one of $\wedge^{A \tn{ opp }}_\ell$
or $\wedge^A_\ell$ is acting according to orientation. Since we
will not make calculations, we will not be careful in choosing
orientations and in distinguishing categories and their opposites.
If $Y$ has $k$ boundary components the $k-$fold product
$\underset{k}{\LARGE{\textsf{X}}} \wedge^A_\ell$ acts on $DE
\ell(Y)$. The reader should note that in the cases where there is
a miss match between boundary conditions on $\partial Y$ and
$\underset{k}{\coprod} A$ there is by \underline{definition} no
action defined.  This would not make sense if we were dealing with
algebras and is precise by the extra flexibility that make linear
categories, sometimes called algebroids, a useful generalization
of algebras.

So $\underset{k}{\LARGE{\textsf{X}}} \wedge^A_\ell=:\mathcal{C}$
has an action or $-$ to use the usual terminology when actions are
linear $-$ a \underline{representation} on $DE \ell(Y) =:V$.

To clarify, for each $\partial -$condition on
$\underset{k}{\coprod} A$ there are two corresponding finite
dimensional vector spaces $DE \ell (Y_{\tn{into $A \,\,
\partial-$condition}})=:V_{\tn{in}}$ and \break $DE \ell(Y_{\tn{out $A \,\,
\partial-$condition}})=:V_{\tn{out}}$.  A morphism
$_\tn{in}\gamma_\tn{out} \in$ morph$(\mathcal{C})$ induces (by
gluing) a linear map $: V_\tn{in} \la V_\tn{out}$. The
construction is so natural that all the required diagrams commute,
and gluing, indeed, defines a representation.  (There are some
technical points but these are well understood and will cause us
no trouble.  As annular collars are added to $Y$ various
$\l$association" must be chosen so the action is $\l$weak" not
$\l$strict".  Also it is sometimes convenient to forget the
parameterization of $C$ and only remember the base point $\ast =1
\in S^1$, and further to replace the uncountable object set $-$
finite subsets of $S^1-$ with the countable object set consisting
of one exemplar object for each finite cardinal $0,1,2,\ldots$.
This processes is called \underline{skeletonizing} the category.).

The positivity properties of the pairing $\langle \, ,\rangle$,
figure 2.2 and Thm 2.1, and its extension implies that its finite
dimensional representations decompose uniquely into a direct sum
of irreducibles.  The arguments for this are nearly word for word
what is said to prove that a finite dimensional $\mathbb{C}^\ast
-$ algebras is isomorphic to a direct sum of matrix algebras.  The
algebroid context changes little.

So let us decompose $V$ as a representation of $\mathcal{C}$ into
a direct sum of its irreducibles.  We record multiplicities by
tensor product with a vector space $W_i$ on which no action of
$\mathcal{C}$ exist:
\begin{equation}
V\cong \bigoplus_{\tn{irreps. }\mathcal{C}} W_i \bigotimes V_i.
\end{equation}

The index $i$ is a multi index $i =(i_1 , \ldots , i_k )$ and
counts the $\l$admissible labellings" of $\partial Y = C_1 U \dots
U C_k$.  In the theory $DE \ell$, the possible components of $i$
are the
\begin{equation}
\{\tn{irreps of }\wedge^A_\ell\} =: \mathcal{L},
\end{equation}
the label set of $DE\ell$. The involution $^\wedge$ is induced by
orientation reversal on $A$ and conjugates representations: it
happens to be trivial in these theories.

A final categorical comment:  the form of r.h.s. 2.4 suggest it
represents a $2-$vector: a linear combination (in this case of
irreps.) with vector space, rather than scalar coefficients.  This
is, in fact, the correct categorical setting for $MF(Y)$ when $Y$
has boundary.

So far this discussion of the label set has been rather abstract
but it is possible to make explicit calculations by considering
the annular categories as operators on $TL^c_d$, the
Temperley-Lieb, or $\l$rectangle" categories.  A label $a \in
\tn{irrep} \left(\wedge^A_\ell \right)$ is generated some
idempotent $\overline{a} =\,  _k\overline{a}_k =$ morph $(k, k)
\subset \wedge^A_\ell$, which is a linear combination of annular
pictures. Quoting a result which will appear in [FNWW] with full
details, we describe (Figure 2.4) the idempotent $\overline{a}$
for the four labels of $DE3, a= (0,0), (0,2), (2, 0)$ and $(2,2)$.
In the language of rational conformal field theory these labels
are the 4 \underline{primary} \underline{fields}. They are given
below as formal pictures in annuli.

Previously, we only considered ideals to be generated by formal
pictures in a disk.  But now we can let $\overline{a}$, by
stacking formal pictures in annuli, generate an $\l$annular
ideal", $J$. Elements $b\in J$ may have more than $k$ boundary
points on $S^1 \times \pm 1$; such $b$ are the
\underline{descendent} \underline{fields}.


\vskip.2in\epsfxsize=3.5in \centerline{\epsfbox{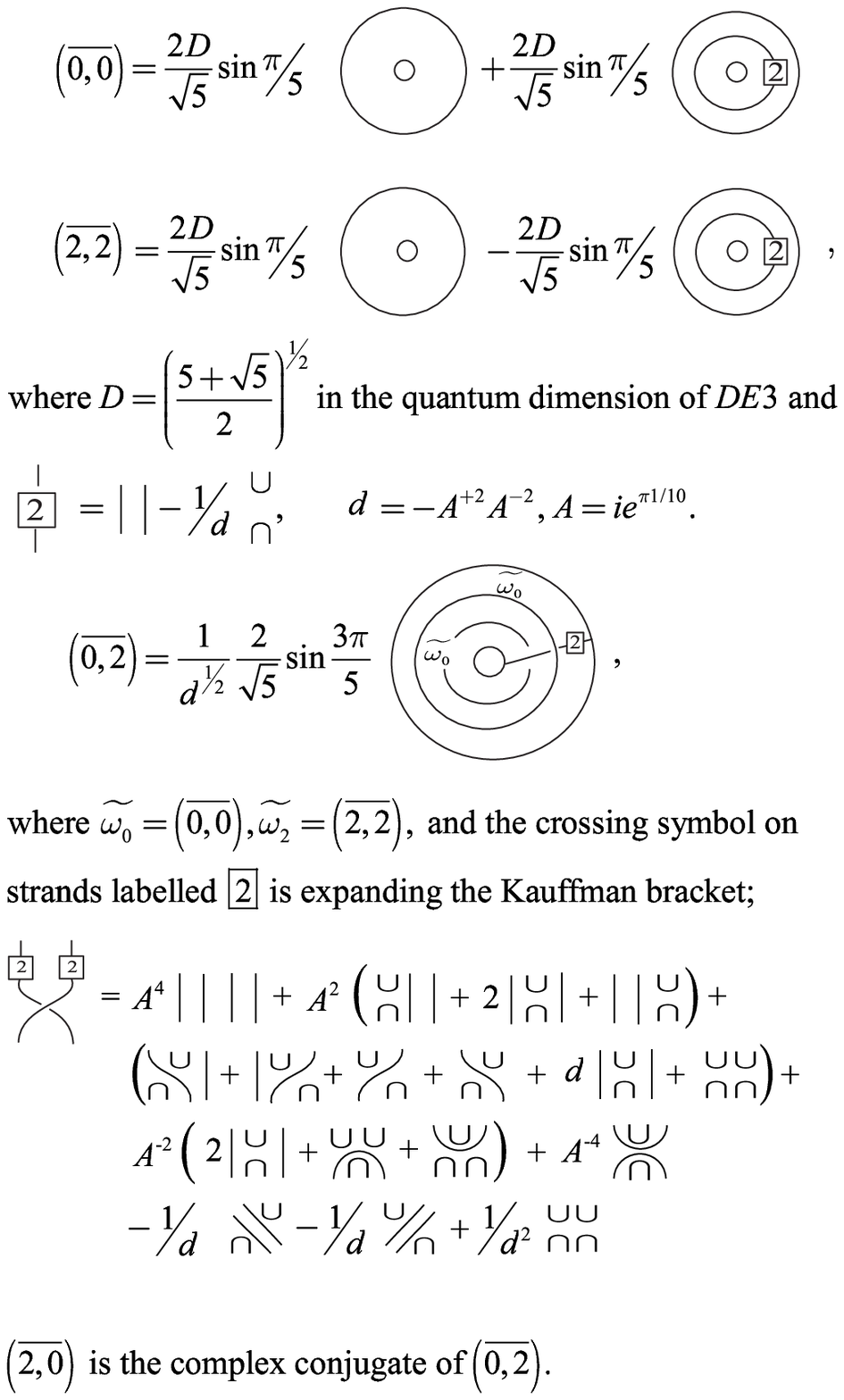 }}
\centerline{Figure 2.4} \vskip.2in


The idempotent $(\overline{0, 2})$ has among its many terms five
\underline{principal} terms, pictures containing only arcs going
between inner and outer boundaries of $A$ and no arcs which are
boundary parallel.  The other terms enforce orthogonality of
$(\overline{0, 2})$, to the descendents of $(0,0)$ and $(2,2)$.
The principal terms written out below.
\begin{equation}
(\overline{0, 2}) ::  I + e^{3 \pi i/5} F + e^{6 \pi i /5} F^2+
e^{9 \pi i/5} F^3  + e^{12 \pi i/5} F^4 + \tn{ lower terms } F
\end{equation}
fractional Dehn Twist: $F=
${\includegraphics[width=.50cm,height=.45cm]{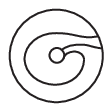}},
 $I=$ {\includegraphics[width=.50cm,height=.45cm]{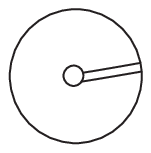}}. $(\overline{0,
2})$ is a $e^{2\pi i/5}$ eigenvector of $F$.  The powers of $F$
are obtained by radial stacking of annuli.

In the case $|a_i - b_i | =2k_i$, $k_i \neq 0$, consider the
commuting actions of $F=e^{\pi i/k_i}$ twists of $C_i$ on
${Q}E_\ell$. The resulting eigenvalues turn out to be distinct
within the $\l$position" $= |a -b|$ summand of $V(Y)$.  These
eigenvalues add the $\l$momentum" information which determines the
labels: the minimal projections to eigenspaces.

For $k_i =0$, Dehn twist acts by the identity, so here
prescription must be different. Suppose $s$ is a configuration on
$Y$ which has constant spin $\l$monochromatic" either $| +\rangle$
or $|-\rangle$, near $C_i$ and let [s] be its image in $QE\ell$.
Define an action on [s] by adding an annular ring of the opposite
spin in interior $(Y)$ immediately parallel to $C_i$ and define
$\bt_n = \us{x=0}{\overset{\lfloor\f{\ell +2}{2}\rfloor }{\SS}}
S_{2n, 2x} R^x$ where $R^x$ consists of $x$ parallel annular rings
of the opposite spin stacked up parallel to $C_i$ but in interior
$(Y)$, and where $S_{y, x}$ is the $S$-matrix $= \f{2}{\sqrt{\ell
+2}}\sin \f{ \pi xy}{\ell +2}$  of the undoubled theory. For $n
=0, \ldots, \lfloor\f{\ell +2}{2}\rfloor$, the maps $\bt_n$ are
commuting projectors (up to a scalar), which also commute with
twists around other boundary components.  For $k_i =0
\,\,\l$position" is refined to a label by applying the idempotent
$\beta_p$. The trivial label is $\beta_0$.  The image of $\beta_p$
turns out to be the $-(A^{2p +2} + A^{-2p -2})$ eigenspace of the
actions of $R$ on the $|a-b|=0\,\,\l$position" summand of $V(Y)$.
Thus when $\partial Y \neq \emptyset$, $DE\ell
(Y,\overrightarrow{t})$ is an orthogonal summand of $QE\ell (Y)$
determined by specifying a minimal even number $n_i$, of arcs
reaching each boundary component $C_i$ (but not parallel to it),
$n_i \e\{ 0,\ldots , \ell\}$, and an eigenvalue $\lambda_i$ at
$C_i \subset
\partial Y$. Note that gluing different eigenspaces images
automatically implies a trivial result as required by the gluing
axioms.  This is immediate from the commutivity of the minimal
idempotents.  For surfaces with boundary, we may write
$DE\ell(Y):= \underset{\tn{admissible labelings}
\overset{\rightharpoonup}{t}}{\bigoplus}(Y,
\overset{\rightharpoonup}{t})$, then $DE\ell(Y) = QE\ell(Y)$ in
this case as well.

For the first computationally universal case, level $\ell = 3$,
the $S-$matrix, $F-$matrix ($= 6_j-$symbol), the action of Dehn
twist, and Verlinda (fusion) relations are listed below.  The only
interesting $F-$matrix in the undoubled theory occurs when all
four \underline{external} label $=2$, so $F$ reduces to a
$2-$tensor.
\vskip.2in\epsfxsize=3.4in \centerline{\epsfbox{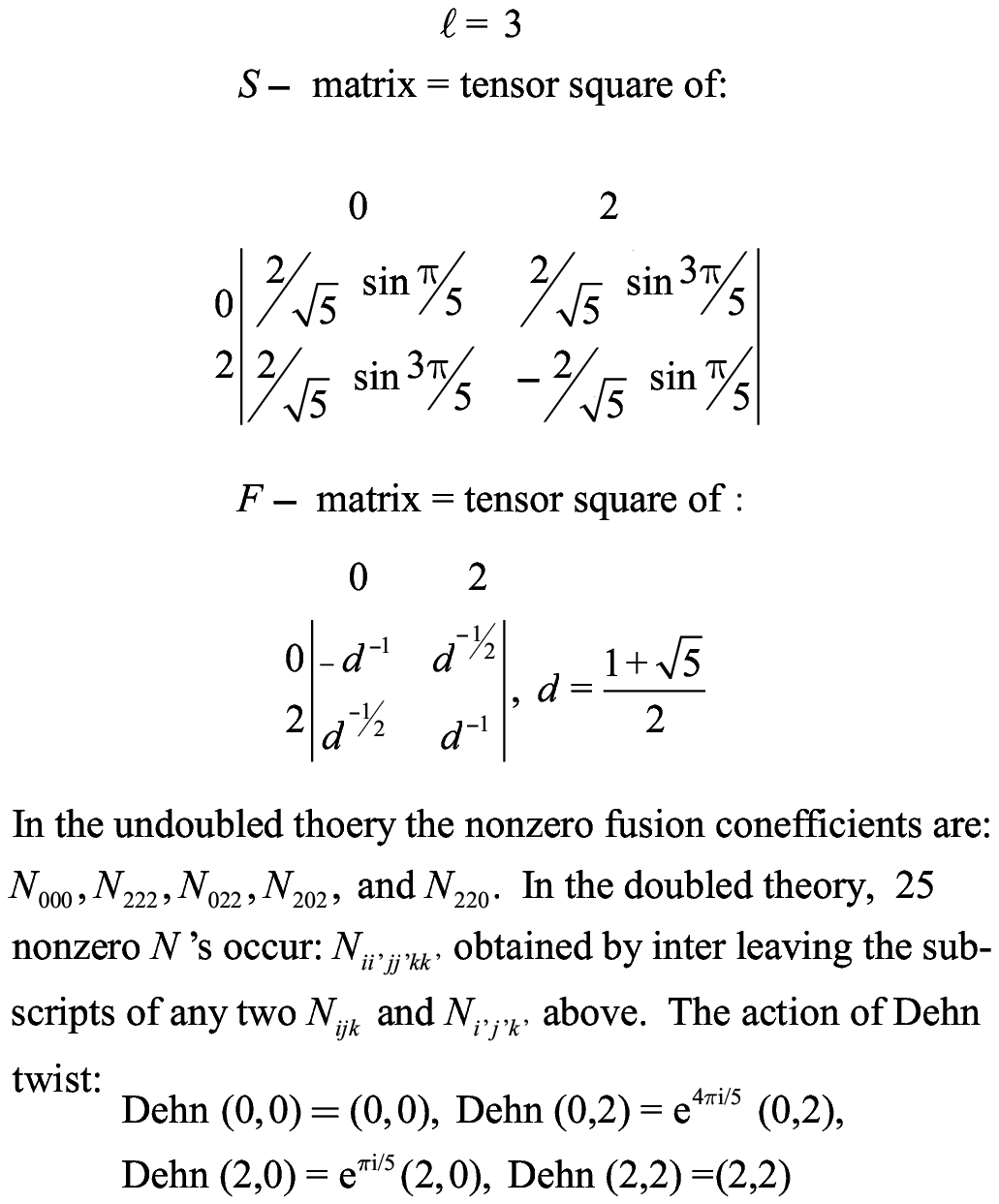 }}
\centerline{Figure 2.5} \vskip.2in

We have come to a point where we can study the difference between
span $\{2-\tn{colorings }(|+\rangle , |-\rangle)\}= G_{\c, \ell}$
and the $^- \, -$ invariant combinations $G^+_{\c, \ell}$. We may
begin with the enhancement of $ETL_\ell (Y)$ to $ETL^!_\ell (Y)$,
manifold $2-$coloring modulo $d-$isotopy (for $d= 2\cos
\pi/\ell+2$).  The enhancement leads the $\l$color reversal
particles" of figure 0.2 which do not fit exactly into the
UTMF$-$TQFT formalism, (but perhaps a $Z_2 -$graded version?) as
they do not raise the ground state degeneracy on the torus. They
should, however, arise physically and contribute to specific heat.
We will return to these shortly.

First, we show that there is only one lifting to the enhancement
of the projector relation $p_{\ell + 1}:$ for $\ell$ odd let
$p_{\ell +1}^{\tn{black}}$ and $p_{\ell +1}^{\tn{white}}$ denote
the extension to $2-$colorings of $p_{\ell +1}$ by the relation
applied to a $2-$coloring in a neighborhood of a transverse arc
which crosses $\ell+1$ strands of $\Pa$ from black to black and
white to white respectively. The reader may take black
$=|+\rangle$ and white $=|-\rangle$.  If $\ell$ is even, noting
that all the projectors $p_i$ have left-right symmetry there is
only one way to lift $p_{\ell +1}$ to $ETL^!_\ell$.

\begin{pro}
$J \left(p_{\ell +1}^{\tn{black}}\right) = J \left(p_{\ell +1
}^{\tn{white}}\right)$.
\end{pro}

\noindent{\bf Proof:} We may use under crossings to indicate
formal combinations of ${\tn{TL}}-$diagrams which are consistent
with the Kauffman relation:

\vskip.2in \epsfxsize=1.8in \centerline{\epsfbox{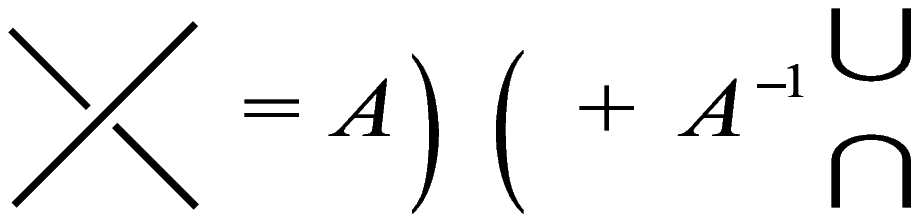 }}
{\centerline{Figure 2.7}} \vskip.2in


Now consider (for $\ell =3$) the following sequence.

\vskip.2in \epsfxsize=4.5in \centerline{\epsfbox{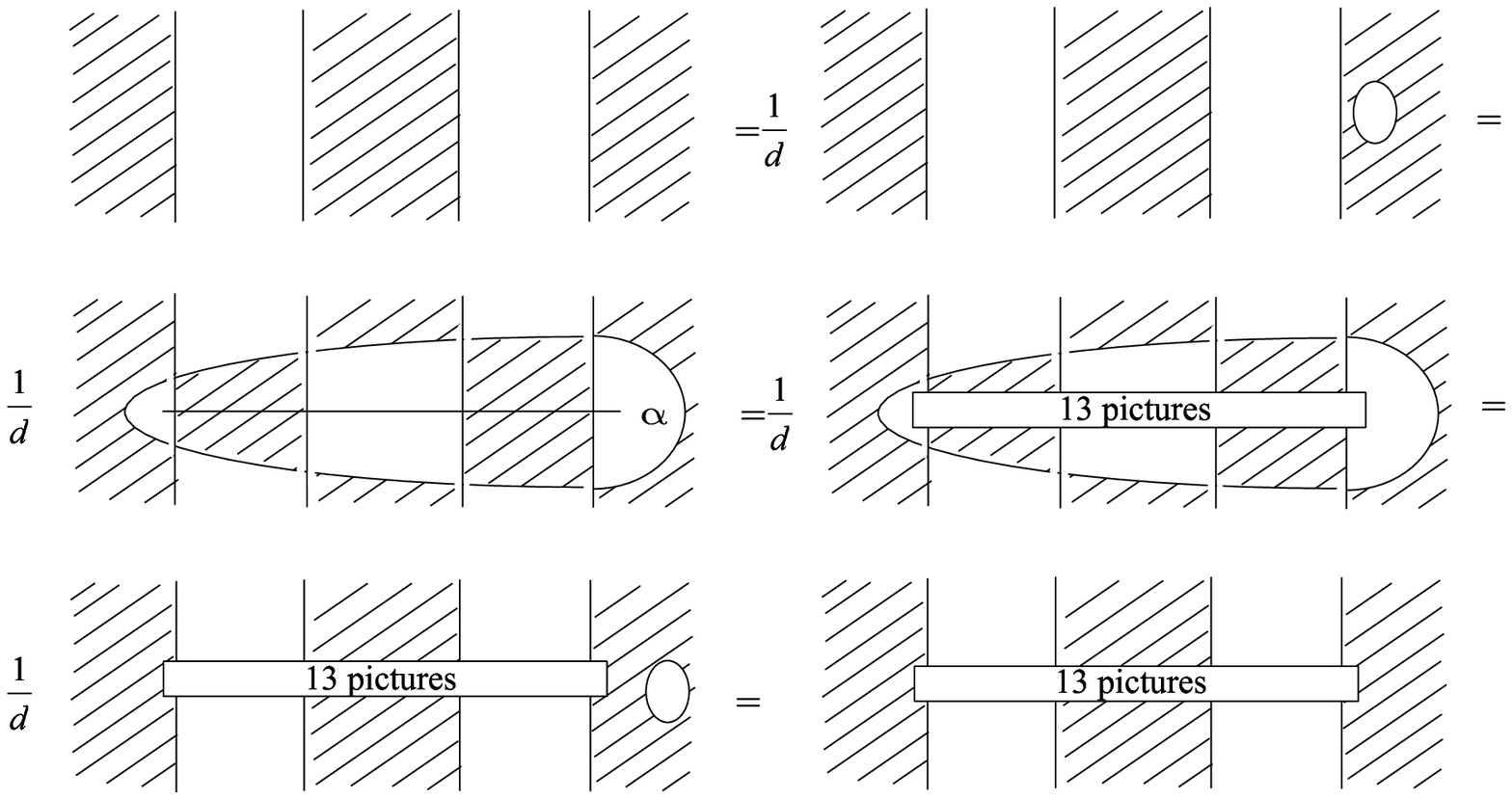 }}
{\centerline{Figure 2.8}} \vskip.2in

These 6 steps effect $p_{\ell
+1}^{\tn{black}}$ across the arc $\a$ with an application of
$p_{\ell +1}^{\tn{white}}$.  $\square$

For a closed surface $Y$ let $QE^!_\ell (Y)$ be the enhancement
$ETL^!_\ell (Y)/\left<p_{\ell +1}^{\tn{black}}\right> \break
=ETL^!_\ell (Y)/\left<p_{\ell +1}^{\tn{white}}\right> $ and
$QE^!_\ell (Y) \la QE_\ell(Y)$ the forgetful map. We do not know
if this map is always an isomorphism. However for the case of most
interest $\ell =3$, Proposition 2.11, below shows that $QE^!_3
(T^2)\cong QE_3 (T^2)$, $T^2$ the $2-$torus. Since $\dim
\left(V(T^2)\right) = |\mathcal{L}|$ the cardinality of the label
set, this implies, that largest quotient of $QE^!_3$ having the
structure of a UTMF is isomorphic to $QE_3$.

\begin{pro}
$QE^!_3 (T^2) \cong Q_3 (T^2 )$.
\end{pro}

\noindent{\bf Proof:} Let $B$ be the black and $W$ the white
coloring of $T^2$ while $M$ and $M^2$ are the one and two
meridional ring colorings respectively:

\vskip.2in \epsfxsize=4in \centerline{\epsfbox{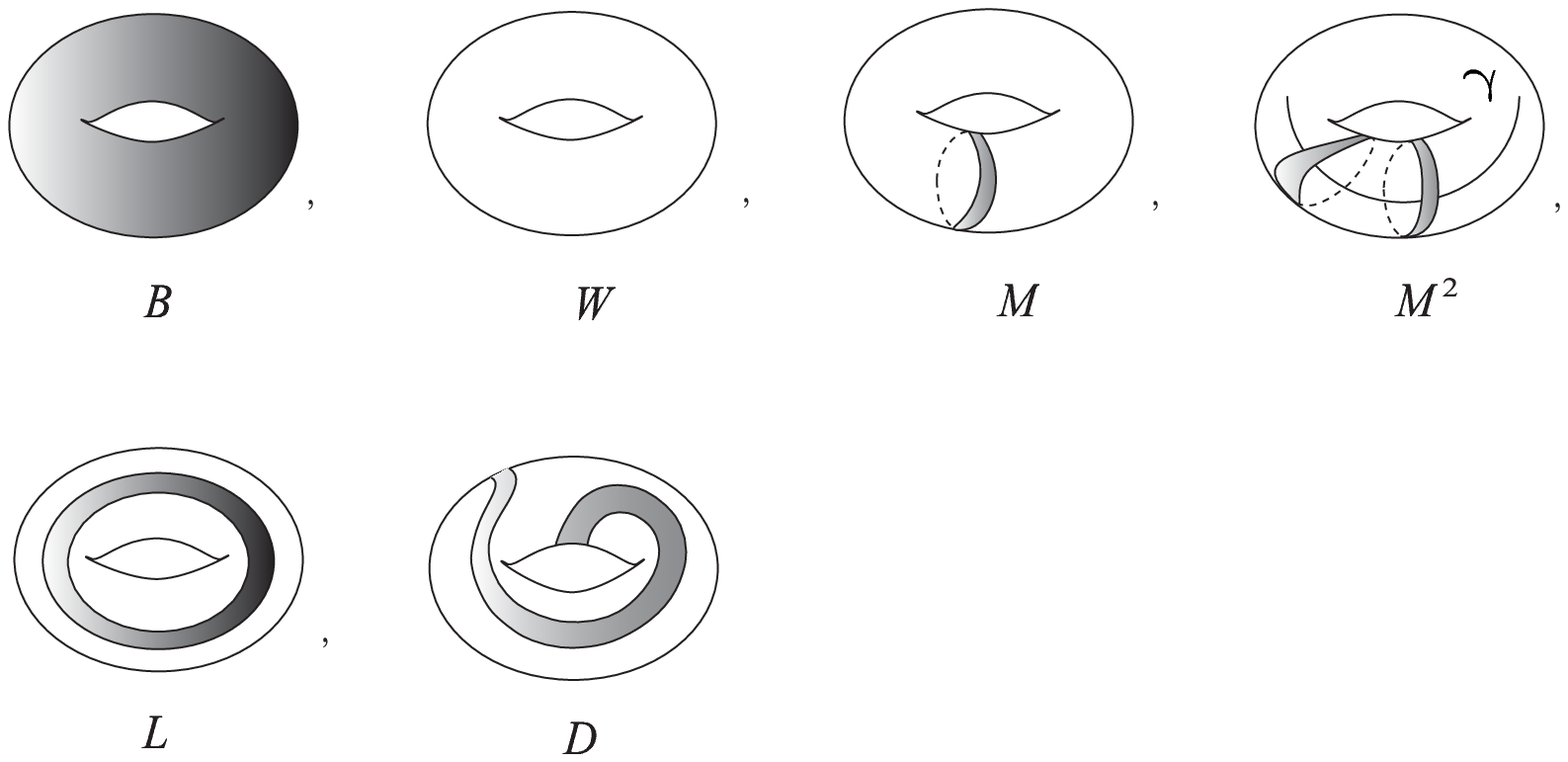 }}
{\centerline{Figure 2.9}} \vskip.2in


Applying $p_4$ across the arc $\Pa$ (and a short calculation using
Figure 2.3) yields: $M^2 = 3M -W$. Since $M$ and $M^2$ are
black-white symmetric the third term must be symmetric as well,
hence $W=B$.

It follows quickly that $QE^!_3(T^2)= \C -{\tn{span }} (W, M, L,
D)$ where $L$ is a longitudinal and $D$ a diagonal ring.
Forgetting the $2-$coloring and retaining only the domain wall we
get a basis for $QE_3 (T^2)\cong DE3 (T^2)$. \qed

It is possible to brake the color symmetry $^-$ by adjusting the
Hamiltonian to fix the color $=|- \rangle$ at some plaquet on each
component of $Y$.  This adjustment creates a new ground state $G$
canonically isomorphic to the former $G^+$, so we drop the $^+$
from the notation. However this does not obviate the need to study
the enhancement. The point is that localized \underline{color-
reversing} excitations remain and are expected physically. These,
when realized on an annulus algebra, have opposite coloring on
$S^1 \times -1$ and $S^1 \times +1$, and so \underline{cannot} be
glued into a ground state on $T^2$.

Let us see how this works in the simplest example, the level $=1$
theory $\ell =1$, $ d=1$, $A = ie^{2\pi i/12}$.  When we make no
$\l$evenness" restriction this theory, $D1$, is also called $Z_2-$
gauge theory [SF] and [K1].  It has four labels: $0=
\f{1}{2}(\emptyset -R)$, $m=\f{1}{2} (\emptyset -R)$, $e= \f{1}{2}
(I+T)$, and $em = \f{1}{2} (I-T)$ where the pictures in these
combinations are:

\vskip.2in \epsfxsize=3.5in \centerline{\epsfbox{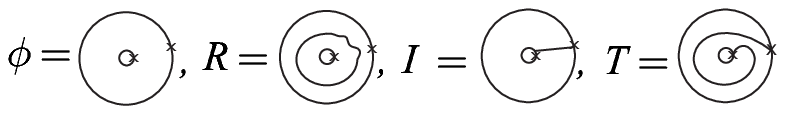 }}
{
\vskip.2in

Notice that the labels are orthogonal under stacking annuli.  One
may check that braiding $m$ around $e$ or $em$ introduces a phase
factor $= -1$, as does braiding $e$ around $m$ or $em$. The even
theory $DE1(A=\,ie^{2\pi i/12} )$, has only one particle $(0,0)-$
which is the trivial particle, and has dimension $=1$ on $T^2$,
and so is quite trivial.  In quantum systems with other
microscopics (e.g. [K1]) can easily realize $D(1)$ but in our set
up the pictures do \underline{not} arise directly but
\underline{indirectly} as a domain wall so $I$ and $T$ make no
sense.  However $R$ does make sense as a domain wall between
$|+\rangle$ and $|-\rangle$ boundary conditions at opposite ends
of $A$.  In fact, we may define an elementary excitation of
$DE(1)$ at a plaquet $c$ by using the local projector at $\C^2_c
\,\, \left( |+\rangle_c + |-\rangle_c \right) \otimes
\left(\langle + |_c + \langle -||_c \right)$, instead of the
ground state projector $\left( |+\rangle_c - |-\rangle_c  \right)
\otimes \left(\langle + |_c - \langle -|_c \right)$.  Thus the $\l
m$" particle can arise as an excitation of $DE1$ even though it
does not contribute to ground state degeneracy. This is the
prototypical color-reversing particle.

Regarding the other information in figure 0.2, the coloring
preserving or $\l$label" excitations (the irreps. of
$\wedge^A_\ell$) are counted by the (even, even) lattice points in
$[0, \ell +1] \times [0, \ell +1]$. The color reversing labels are
(odd, odd) $\subset [0, \ell +1][0, \ell +1]$.  The $S-$matrix of
the (undoubled) $SU(2)$ or Kauffman theory when restricted to even
labels is singular (precisely) for $\ell \equiv 2$ mod $4$, for
example at $\ell =2$

\begin{equation*}
   S=
  \begin{vmatrix}
    \f{1}{2} & \f{\sqrt{2}}{2} &\f{1}{2}\\
    \f{\sqrt{2}}{2} & 0 & \f{-\sqrt{2}}{2}\\
     \f{1}{2} & \f{-\sqrt{2}}{2} &\f{1}{2}\\
  \end{vmatrix}
\end{equation*}
\begin{equation*}
S_{\tn{even}}::
  \begin{vmatrix}
    \f{1}{2} & \f{1}{2}\\
    \f{1}{2} & \f{1}{2}
  \end{vmatrix}
\end{equation*}
  When $S$ is nonsingular, $\ell \neq 1, 2, 4$ and
  the number of braid stands $\geq 5$ it is known [FLW2] that the braid representations are dense
  in the corresponding  special unitary groups.


\section{Perturbation and deformation of $H_{\c, \ell}$}

As remarked near the end of Section 0, excited states, i.e. anyons,
will be studied as ground states on a punctured surface with
labelled boundary. In the large separation limit, the braiding of
anyons can be formulated as an adiabatic evolution of the
\underline{ground} state space on a labelled surface $Y$ with
boundary. So in the present section we confine the discussion to
ground states. Although boundary is assumed to be present and
labelled we will nevertheless consider only perturbations acting
in the bulk so the role of the boundary is peripheral in this
section.

The passage from $G_{\c, \ell}$, the ground state space of $H_{\c,
\ell}$, to the deformed ground state space $G_{\e, \ell}$ of
$H_{\e, \ell}$ does not result from the breaking of a symmetry, in
fact $G_{\c, \ell}$ has no obvious symmetry.  Rather it is the
creation of new $\l$symmetry": topological order.  If a
perturbation $V$ is breaking an existing symmetry then only the
original ground state and the effective action of $V$ at the
lowest nontrivial order  is relevant.  But in the present case, to
understand the effect of a perturbation $\e V$, one should first
describe all low lying (gapless) excitations above $G_{\c ,\ell}$
and then see how $V$ can act effectively on $G_{\c ,\ell}$ through
virtual excitations. For example in the toric codes [K1] the
ground state space may be rotated in an interesting way if a
virtual pair of $e$ (electric) particles appear, tunnel around an
essential loop (of combinatorial length $=L$), and then
annihilate. In the case of toric codes there is an energy gap to
creation of $(e, e)$ pairs so the above process has exponentially
small amplitude in the refinement scale $L\sim e^{- L/L_\c}$. In
contrast, for level $\ell \leq2$ we expect the ground state space
$G_{\c, \ell}$ to be gapless\footnote{For $\ell=3$ the gap may be
extremely small as explained latter in this section.}(in the
thermodynamic limit) and processes which act through virtual
excitations will be important in the perturbation theory because
excitations are cheap. However it seems hopeless to analytically
describe these gapless excitations so we skip this step and resort
to an ansatz (3.4) stated below.  It asserts that $G_{\e, \ell}$
is modelled as the common null space of local projectors acting on
$G_{\c, \ell}$. We argue for this via an analogy to FQHE,
uniqueness considerations and $\l$consistency checks".

From section 2, the reader knows that we wish to identify $G_{\e,
\ell}$ with $G_{\c, \ell}/ \langle p_{\ell+1}\rangle$ (for
suitable values of $\e$), and this is what the ansatz implies. The
fact that $G_{\c ,\ell}/\langle p_{\ell+1}\rangle \cong DE_\ell$
(see $\S$ 2 and [FNWW]) has the structure of an anyonic system
(mathematically a UTMF) is the first consistency check. There will
be one more presented in section 4.  Let us prepare to state
ansatz 3.4 carefully.

\begin{defi} \tn{
An operator $\mathcal{O}$ on a tensor space $\mathcal{H}=\us{v \in
V}{\ot} \C_v^2$ is $k-$ \underline{local} if it is a sum of
operators $\mathcal{O}_i$ each acting on a bounded $(\leq k)$
number of tensor factors and id on remaining factors. We say
$\mathcal{O}$ is \underline{strongly local} if the index set
$\{V\}$ are vertices of a triangulation $\tr$ and $\mathcal{O}
=\us{i}{\SS} \mathcal{O}_i$, where each $\mathcal{O}_i$ is
$k-$local with the $k$ active vertices spanning a connected
subgraph $G_i$ of $\tr$. All $G_i$ are assumed isomorphic and with
fixed isomorphisms $G_i \la G_j$ inducing isomorphisms
$\mathcal{O}_i \cong \mathcal{O}_j$.  In the latter case, we call
$(\tr, \mathcal{O})$ a \underline{quantum} \underline{medium}.}
\end{defi}

\begin{No}\tn{  In the special case that a family of strongly
local operators $\{\mathcal{O}_i\}$ are projection onto
$1-$dimensional subspaces, the system $\{\mathcal{O}_i\}$ is
equivalent to what topologist call a combinatorial
\underline{skein} \underline{relation} ([L] [KL]), though in the
topological context equivalence of isotopic pictures is implicitly
assumed.  An example of a (14 term) skein relation is $p_{\ell
+1}=0$ (see fig. 2.3), applied to $\Pa$, the dual$-1-$cell domain
wall between $|+\rangle$ and $|-\rangle$. A skein relation is a
local linear relation between degrees of freedom, domain walls in
our case. The intersection of all the null spaces
$\us{i}{\bigcap}{\tn{null}}(\mathcal{O}_i
)={\tn{null}}(\mathcal{O})$ is the subspace perpendicular to the
equivalence classes in $\mathcal{H}$ defined by the combinatorial
skein relation.}
\end{No}

\begin{defi} \tn{
The \underline{joint} \underline{ground} \underline{space} (jgs)
of $\{\mathcal{O}_i\}$ is $\us{i}{\bigcap} E_{\c , i}$ where
$E_{\c , i}$ is the eigenspace corresponding to the lowest
eigenvalue $\ld_\c$ of $\mathcal{O}_i$.}
\end{defi}

The jgs $\{\mathcal{O}_i\}$ is not necessarily the lowest
eigenspace of $\mathcal{O}=\underset{i}{\SS}\mathcal{O}_i$ because
jgs $\{\mathcal{O}_i\}$ can easily be $\{0\}$. In this case the
Hamiltonian is $\l$frustrated". It may happen that $\mathcal{O}$
has long wave length excitations at the bottom of its spectrum
which do not show up in the spectrum of $\mathcal{O}_i$. However
if $\mathcal{O}$ defines a stable physical phase it is an
optimistic but not unrealistic assumption that jgs
$\{\mathcal{O}_i\} =$ ground state space $(\mathcal{O})$. For
example, this occurs in the $\l$ice model" or $\l$perfect matching
problem" on the honeycomb lattice [CCK] and in the fractional
quantum Hall effect (FQHE).

The FQHE begins with a $\l$raw" state space $\mathcal{H}$, the
lowest eigenspace for an individual electron confined to a
$2-$dimensional disk $D$ and subjected to a transverse magnetic
field $B$.   This $\mathcal{H}$ is called the lowest
\underline{Landau} \underline{level}. Each level can hold a number
of spin $+$ and $-$ electrons $\approx {\tn{ area
}}D/({\tn{magnetic  length}})^2$ and the fraction of that number
actually residing at the level is called the filling fraction
$\nu$.  In a spherical model, the Coulomb interaction $H$ between
pairs of electrons $(e_i , e_j)$ can be written [RR] as a sum of
projectors onto various $\l$joint angular momentum subspaces"
$p_{ij} ((2k + 1) N_\phi)$. The null space, ${\tn{null}}(H_q)$,
for
\begin{equation}
H_q=\sum_{k=0}^{\f{q-3}{2}} \sum_{i<j} p_{ij} \left((2k +1) N_\phi
\right)
\end{equation}
is nontrivial and, of course, is the joint ground space jgs of the
individual projectors in the sum. In fact, null $(H_q)$ is
\underline{Laughlin's} $\l$odd denominator" state space at $\nu =
\f{1}{q}$.

{\bf Ansatz 3.4: }  For well chosen $\e$, the perturbed ground
state space $G_\e$ will be stable and can be written as $G_\e=$
jgs $(|s_i\rangle\langle s_i|) \cap G_\c$ for some strongly local
family of projectors $\{|s_i\rangle\langle s_i|\}$ acting on
$\mathcal{H}$. Equivalently if $s_i^\c$ is the orthogonal
projection of $s_i$ into $G_\c$ and $|s_i^\c\rangle\langle
s_i^\c|: G_\c \la G_\c$ is the corresponding projector then $G_\e
\cong$ jgs $(|s_i^\c\rangle\langle s_i^\c|)$.

In topological terms the ansatz asserts that the reduction
\linebreak$G_\c \la G_\e$  occurs by imposing a skein relation.
The ansatz is essentially a strong locality assumption.

As discussed above, the Laughlin $\nu = \f{1}{q}$ states, $q$ odd,
follow this pattern with the Coulomb interaction between electrons
playing the role of the perturbation on the disjoint union of
single electron systems. Since the Landau level has no low lying
excitation the analogy is closest with $G_{\c, \ell}$, $\ell \geq
3$. Theorem 2.5 gives us the following:

\begin{con}
\tn{Suppose $H_\c$ is subjected to a sufficiently small
perturbation $\ell \leq 2$ or an appropriate deformation, $\ell
\geq 3$, which partially lifts the log extensive degeneracy of the
ground state $G_{\c, \ell}$ to yield a strictly less degenerate
ground state $G_{\e, \ell}$. If we assume the stability of $G_{\e,
\ell}$ we expect $G_{\e, \ell}$ to be modelled as $G_{\e, \ell}
\cong G_{\c, \ell} /\langle p_{\ell +1} \rangle = $ the modular
functor $DE \ell$. }
\end{con}

For the projector $p_{\ell+1}$ to arise as an effective action of
$\e V$ on $G_{\c, \ell} = \C < d-$ isotopy classes of domain walls
$>$, for $\e >0$, various sets of $\ell+1$ walls must have a
polynomially large probability of simultaneously visiting the
support $U_i$ of some local combinatorial $\mathcal{O}_i =
p_{\ell+1}^i$ enforcing $p_{\ell+1}$.  The walls must visit a site
$U_i$ or $p^i_{\ell +1}$ cannot enforce orthogonality to the
relation vector $p_{\ell +1}$ (as depicted in figure 2.3 for $\ell
\leq 3$.)

The notion of a combinatorial instance of $p_{\ell +1}$ was
developed in [F2].  It amounts to a discretization of the smooth
domain wall diagrams (Figure 2.3) by choosing specific
superpositions of local plaquet spin configurations (with the spin
state of the plaques at the boundary of the configuration
constant) to represent the smooth relation $p_{\ell +1}$.
Evidently there are many distinct combinatorial patterns which are
instances of a fixed $p_{\ell +1}$.  The simplest of these amount
to geometric rules for simplifying $\ell +1$ domain wall when
these run parallel for (roughly) $\ell +1$ plaques.  As discussed
in [F2], imposition of such a combinatorial relation in the
presence of mild assumptions on the triangulation $\tr$, produces
a result isomorphic to the smooth quotient.  It is sufficient that
the triangulation must have injectivity radius $>>
\ell +1$ and bounded valence. So $\tr$ should subdivided (to
approach a thermodynamic limit) as shown in Figure 3.1.

\noindent{\bf Heuristic:} There is a curious pattern observed in
Figure 2.3 (and further computer calculation of Walker (private
communication), for $\ell +1 =$ even and $d=2 \cos
\f{\pi}{\ell+2}$, the sum of the coefficients in $p_{\ell +1}$, in
the geometric basis $\{g\}$ is zero, and for $\ell +1 =$ odd, the
sum is small. The geometric pictures $\{ g\}$ may be filtered by
an integer weight $n$ which counts the fewest sign changes on
plaquets $-$ in topological terms, the fewest domain wall
reconnections on $\l$surgeries" $-$ required to transform the
straight (identity) picture to $g$.  So, referring to Figure 2.3,
the first term for $p_2$ has weight $=0$, the second term has
weight $=1$.  The terms for $p_3$ have weights $0,2,2,1,1$
respectively.  Notice that sign (coefficient $(g)$)
$=(-1)^{\tn{weight } (g)}$.

This suggests that $V=\underset{c, \tn{ plaquet}}{\Sigma} \s^c_x$
is a reasonable choice for $V$ to obtain figure 0.1.  The Pauli
matrix $\s_z$ has $+1-$ eigenvector $|+\rangle + |-\rangle$ and
$-1-$ eigenvector $|+\rangle - |-\rangle$, and thus assigns a
lower energy  to combinations of geometric pictures $g$ which have
a $(-1)$ phase shift associated to domain wall surgery.  The
perturbation \break $V=\underset{c}{\Sigma} ( |+\rangle +  |-
\rangle )(\langle +| + \langle -|)=\f{1}{2} \underset{c}{\Sigma}
(id + \s^c_x )$ contains terms which annihilate antisymmetric
combinations of approaching domain walls of $\Pa$:
$\left(|)(\rangle-|{\smile \atop \frown} \rangle\right)$.

Nonzero entries coupling all the terms of $p_{\ell +1}$ occur
first at order $\ell$, i.e. in $V^\ell$, so one may expect that
this is the order at which an effective action arises on $G_{\c
,\ell}$.

It is now time to treat the \underline{statistical physics} of a
general ground state vector $\Psi \in G_{\c, \ell}$. A perturbed
Hamiltonian $H_{\e, \ell}$ will \underline{not} have a ground
state modelled by $G_{\c, \ell}/\langle p_{\ell+1}\rangle$ if the
domain walls of $\Psi$ have an effective tension. The simplest
place to see this is on a closed surface $Y$ with the
triangulation $\tr$ determining a metric. When the domain walls
$\gamma \subset Y$ are pulled tight under tension they will stand
a bounded distance apart and have exponentially small amplitude
for simultaneously entering a small locality $U_i$ so
$p_{\ell+1}^i$ will be unable to act.

Measuring any $\Psi\i G_{\c, \ell}$ via the community family
$\{\s^c_z \}$ projects $\Psi$ into the geometric basis.  This
results in a probabilistic spin configuration which is Gibbs with
probabilities proportional to $n^{\# {\tn{ loops}}}$ where $n= d^2
= (2 \cos \pi/\ell + \nolinebreak 2)^2$.

Let us write $\Psi \i \mathcal{H},  | \Psi |=1$, in the classical
basis of spin configurations, $\Psi = \SS {a_k} |\Psi_k \rangle$.
We say, consistent with measurement of any observable which is
diagonal in the $|\Psi_k \rangle$ basis, that $|\Psi_k \rangle$
has \underline{probability} $|a_k|^2$. Thus $\Psi_k$ becomes a
\underline{random} classical component of the random
configuration, meas.$(\Psi)$. Just as one asks about the typical
Brownian path, we ask what a typical $\Psi_k$ looks like. There
will be a competition between energy and entropy. Since $d>1$, the
Hamiltonian $H_{\c , \ell}\, \l$likes" trivial circles and will
place a high weight on configurations with most of the surface
area of $Y$ devoted to a $\l$foam" of small circles. However
entropy favors configurations with longer, fractal loops which
exhibit more variations.  From the critical behavior of loop gases
we know that for $d\leq \sqrt{2}$ entropy dominates and $\Psi\i
G_{\c , \ell}$ is a critical Gibbs state with typical loops
fractal.  For $d> \sqrt{2}$ energy dominates and the Gibbs state
is stable: To free up dual lattice bonds to build this foam the
topologically essential part $\Pa^+$ of $\Pa$ will be pulled tight
by an effective $\l$string tension".

Recall from the introduction that the Gibbs weight on a loop gas
state $\gamma$ is proportional $w(\gamma) = e^{-k ({\tn{total
length }}\gamma)} n^{\# {\tn{ components }} \gamma}$, our is the
self dual, $k=0$ case.

Let us be explicit. In any ground state vector $\Psi \i G_{\c,
\ell}, \Psi= \SS a_i \Psi_i$, we have seen that the coefficients
$a_1$ and $a_2$ of $d$ isotopic configurations $\Psi_1$ and
$\Psi_2$ satisfy $a_1/a_2 =d^{\#_1}/d^{\#_2}$ where $\#_i$ is the
number of trivial domain wall components, $\l$ trivial loops", of
$\Psi_i$. The Pauli matrices $\sigma_z^v = \left|{{1}{\quad 0}
\atop{0}{\,\,\,-1}}\right|$ applied at vertex $=v$ form a
commuting family of observables so we may $\l$observe" in the
geometric basis $\{\Psi_i\}$ of classical spin configurations to
obtain meas.$(\Psi)$ and we see that the ratio of probabilities of
observing $\Psi_1$ verses $\Psi_2$ is:
\begin{equation}
\f{p(\Psi_1)}{p(\Psi_2)} = (d^{\#_1 -\#_2})^2 = (d^2)^{\#_1 -\#_2}
\end{equation}

Thus observing $\Psi$ in the classical basis yields a Gibbs states
\break meas.$(\Psi):: e^{-\beta E(\Psi_i)} |\Psi_i \rangle$ for $E
(\Psi_i) = -\#_i$ and $\beta = 2 \log d$.

Such probabilistic states are called $\l$loop gases" and have been
extensively studied [Ni], e.g. in the context of the
$\mathbf{O}(n)-$ model. It is believed that, in the self dual case
$k=0$, $d \leq \sqrt{2}$, $\ell \leq 2$, there is no string
tension and that the Gibbs state is critical: typical loops are
1/polynomial in size and correlations decay polynomially.
Furthermore the familiar $\l$space = imaginary time ansatz" (see
lines 3.7-13) suggest that this regime, $\ell \leq 2$, should have
$G_{\c, \ell}$ gapless. For $\ell \geq 2$, the loop gas is beyond
the critical range. For these values correlations of the loop gas
decay exponentially and it is believed the loops are in $\l$bubble
phase" where any long loop forced by topology would be pulled
tight by an effective string tension. The corresponding $H_{\c,
\ell} \,\,\ell >2$ should be gapped, above its (polylog)
extensively degenerate ground state space $G_{\c, \ell}$ (compare
with line 3.8). It is in this case, specifically for $\ell=3$,
that we still may hope the ansatz describes $G_{\e, \ell}$ for
some family of deformations $H_{\e, \ell}, \,\, \e_1
>\e >\e_\c >0$ as suggested by the phase diagram, Figure (0.1).  This would imply $G_{\e, \ell} = G_{\c,
\ell}/<p_{\ell+1}> \cong G_{\c, \ell}/R_\ell = DE_\ell$.

The gap above $G_{\c, 3}$ and therefore $\e_\c$ might be quite
small.  A loop gas with $k=0$ (defined on a mid-lattice see Ch. 12
[B]) is closely related to the $FK$ representation of the
self-dual Potts model at $q=n^2 =d^4$. For $\ell=3$, $d=\f{1+
\sqrt{5}}{2}$ and $q \cong 5.6$.  Although the self-dual lattice
Potts model in $2-$dimensions have second order transitions
(critical) for $q \leq 4$ and first order transitions (finite
correlation length) for $q >4$, exact calculations show that for
$q=5.6$ the correlation length $\zeta$ though finite is several
hundred lattice spacing [BJ]\footnote{I thank Steve Kivelson for
pointing out the existence and relevance of these calculations.}.

Our $\l$alternative" Hamiltonian $H'_{\c,\ell}$ resulted from an
effort to sharpen the relation between meas.$(\Psi)$ and the (FK)
Potts model. I would like to thank Oded Schramm for helpful
conversations on this relation.  Recall that $H'_{\c, \ell}$ is a
Hamiltonian on Hilbert space of spin $=1/2$ particles on the bonds
of a surface triangulation, cellulation, or lattice. We will work
locally and so ignore contributions of the global Euler
characteristic $\chi (Y)$ to the formulas below.  Also, we write
$\l=$" to mean that the equation holds up to a fixed extensive
constant, like the total number of bonds.  Recall from the
introduction that we consider the union of $|+\rangle$ bonds and
disjoint from this the union of $|-\rangle-$ dual bonds. Our
$\l$loop gas" is on the mid lattice separating the $|+\rangle -$
from the $|- \rangle -$ clusters.  Let $E$ be the number of
$|+\rangle -$ edges and $E^\ast$ the number of $|-\rangle -$
edges, $C(C^\ast ))$ the number of clusters (dual clusters) with
the convention that isolated vertices (dual vertices) count as
clusters (dual clusters).  Let $L$ be the number of loops in the
loop gas.  The Potts model parameter $q$ (number of colors), it
turn out, should be set as $q=d^4 = (2 \cos \pi/\ell +2)^4$.
Finally, $0\leq p \leq 1$, denotes a probability.

We have two basic equations.  Every loop (in the plane) is the
outermost boundary of either a cluster or dual cluster so:
\begin{equation}
  L= C + C^\ast
\end{equation}
Also there is an Euler relation:
\begin{equation}
  C^\ast \,\l=" C+E
\end{equation}
Now we can re-express the loop gas Gibbs weight $\omega$ in terms
of clusters in the FK Potts model:
\begin{equation}
\begin{split}
\omega (\tn{spin conf.})\,&\l=" (d^2 )^L = (d^2 )^{C+C^\ast }=(d^4
)^C (d^2 )^E 1^{E^\ast}\,  \l=" \\ &(d^4 )^C \left(\f{d^2}{d^2
+1}\right)^E \left(\f{1}{d^2 +1}\right)^{E^\ast} = q^C p^E
(1-p)^{E^\ast}
\end{split}
\end{equation}

Because of the $^-$ - symmetry between $|+\rangle$ and $|-\rangle$
we expect $p$ to be the self dual point for this value of $q$, and
this we check this below:
\[
\omega\, \l=" q^C p^E (1-p)^{-E} = q^{C^\ast} p^{-E} (1-p)^E\,
\l=" \omega^\ast ,
\]
using (3.4) we get $p^{2E} (1-p)^{-2E} =q^E$, so
\begin{equation}
 p=\f{\sqrt{q}}{1+ \sqrt{q}}
\end{equation}
we have proved.

\begin{thm}
Observing with the family $\{\s^b_z \tn{ for all bonds } b\}$ maps
any ground state vector $\Psi \in G'_{\c, \ell}$ into a Gibbs
state meas.$(\Psi)$ of the self dual Potts model for $q=(2
\cos\f{\pi}{\ell +2} )^4$. \qed
\end{thm}

This justifies thinking of $H'_{\c , \ell}$ as a $\l$quantum
Potts" model and contemplating a diagram of relations:

\begin{equation*}
\begin{aligned}
\stackrel{\textstyle\tn{conf. field
theory,}}{\textstyle\tn{$sl_2$, level $\ell$}}& \xrightarrow
{\text{  infra-red limit  }}(SU, \ell) \tn{UTMF}
\xrightarrow {\text{   double   }} DE \ell \\
 {\scriptstyle{\tn{ correlation functions }}} & {\Big\downarrow} \hspace{6cm}  \Big\uparrow {\scriptstyle{\tn{ zero modes }}}\\
\tn{Potts } q =& \,(2 \cos \f{\pi}{\ell +2} )^2    \\
&\hspace{.7cm}\tn{ Potts  } q=\big(2 \cos \f{\pi}{\ell +2} \big)^4
\xleftarrow{\text{ observe }}  \stackrel{\textstyle\tn{quantum
Potts }}{ d =(2 \cos \f{\pi}{\ell +2} )}
\end{aligned}
\end{equation*}
\centerline{Figure 3.1 }

Let us return to the heuristic connecting the spectral gap above
$G_{\c, \ell}$ and the statistical properties of a ground state
vector $\Psi \i G_{\c, \ell}$.  Let $A$ and $B$ be two strongly
local operators on a quantum medium $(\tr , H)$. For example
measuring $A$ might be $\s^c_z$ at plaquet $c$. Assume, first, that there is a unique up to a phase ground state
vector $\Omega$ and a spectral gap $=\delta$ above $\langle\Omega
| H | \Omega\rangle =0$. Then if we evolve in
\underline{imaginary} time:
\begin{equation}
\l{\tn{imaginary time correlation}}" = \langle \Omega e^{t H} A e^{-t H} B
\Omega\rangle =   \langle \Omega A e^{-t H} B \Omega\rangle.
\end{equation}
Writing $|B\Omega\rangle= \langle\Omega B \Omega \rangle \Omega + \Psi_B$ and
$\langle\Omega A|^* = \langle\Omega A\Omega \rangle \Omega + \Psi_A$ note that
non$-\Omega$ summands $\Psi_B$ and $\Psi_A$ decay under imaginary
time evolution at a rate $\geq e^{-\delta t}$.  Thus applying
$e^{-tH}$ to either the bra or the ket in (7) gives:
\begin{equation}
\l{\tn{imaginary  time correlation"}} \la \langle\langle\Omega A \Omega \rangle \Omega
  \langle\Omega  B \Omega \rangle \Omega \rangle  = \langle \Omega A \Omega \rangle\langle \Omega B \Omega \rangle
\end{equation}
 where the convergence ($\la$) is exponential. Denote the spatial
 translations of $A$ (by $A_\ell$), the
 analogous operator to $A$ acting near a site at distance $\ell$ from the support the
 original of $A$. The simplest expectation is
 that spatial correlations should
 behave in the same way as imaginary time correlations,
\begin{equation}
  \l{\tn{space correlation"}} = \langle \Omega A_\ell B \Omega \rangle \xrightarrow {\text{ exponentially }}
   \langle \Omega  A \Omega \rangle\langle \Omega B \Omega \rangle
\end{equation}
This statement is precise in a Lorentz invariant context but is
expected also to hold in greater generality provided that there is
some linkage between temporal and spatial scales.

A code space $G \subset \mathcal{H}$ is an important
generalization of a $1-$dimensional subspace (see [G] for examples
and discussion in other notations).

\begin{defi}
We say $G \subset \mathcal{H}$  is \underline{$k-$code}, with $k$
measuring the strength of the encryption, if for any strongly
$k-$local operator $\mathcal{O}_k$ the composition:
\begin{equation}
G\xrightarrow{\tn{ inc }} \mathcal{H}
\xrightarrow{\mathcal{O}_k}\mathcal{H} \xrightarrow{\Pi_G} G
\end{equation}
must be multiplication by some scalar (perhaps zero). $\Pi_G$ is
orthogonal projection onto $G$.
\end{defi}

The importance of a code space is that it resists local
perturbations. In [G] Gottesman states, in different language his,
Thm 3:
    \begin{thm}\tn{
     Suppose $\mathcal{L} \subset \mathcal{H}$ is a subspace of a
     Hilbert space and $\mathcal{E}$ is a linear space, called $\l$errors", of
     the operators HOM $(\mathcal{H},\mathcal{H})$ so that the
     composition below, for any $E \i \mathcal{E}$ is always multiplication
     by a scalar $c (E)$: $\mathcal{L}\,\xrightarrow{\tn{ inc }} \mathcal{H}\,
     \xrightarrow{E} \mathcal{H}\,\xrightarrow{\Pi_\mathcal{L}}
     \mathcal{L}$.  Then $\mathcal{L}$ constitutes a $\l$code space" protected from
     errors in $\mathcal{E}$.  This means that there is a physical operator
     (composition of measurements and unitary transformations)
     which corrects errors coming from $E \i \mathcal{E}$. ($\Pi_\mathcal{L}$ is orthogonal
     projection onto $\mathcal{L}$.)} \qed
      \end{thm}

We have already encountered a code space. The space $DE \ell \cong
G_{\c , \ell}/ \langle p_{\ell+1}\rangle \cong \subset
\mathcal{H}$ has the code property (and similarly for $G'_{\c ,
\ell}/ \langle p_{\ell +1}\rangle$). From the theorem and the disk
axiom (section 2) we see immediately that the dual space $G_{\c ,
\ell} \cap \langle p_{\ell +1} \rangle^\ast \subset \mathcal{H}$
is a code space for operators (errors) supported in any fixed disk
$D \subset Y$. This is true long before the refinement limit, we
need only a modest level of refinement before the combinatorial
quotient \underline{exactly} assumes the structures of a unitary
topological modular functor (UTMF)[F2]. Then axiom $4$ (section 2)
says that $V(D , 0) \cong \C$ and $V(D , a) \cong 0$ whenever a
label $a\neq 0$. Thus any operator supported on $D$ must act as a
scalar.

Observe that even when a ground state space $G$ is degenerate, if
it is nevertheless a code space and also has a spectral gap of $\d
>0$ between it and the first excited state, the argument for the decay of spatial correlations
\begin{equation}
\langle\Omega_\c A_\ell B \Omega_\c\rangle
\end{equation}
remains valid for any $\Omega_\c \e G$:
\begin{equation}
|\Omega_\c B \rangle =\langle\Omega_\c B \Omega_\c \rangle
\Omega_\c + 0 \Omega_1 + 0 \Omega_2 +\cdots+ \Psi_B , \Psi_B \perp
G,
\end{equation}
\begin{equation}
\langle\Omega_\c A|^\ast = \langle \Omega_\c A \Omega_\c \rangle
\Omega_\c + 0 \Omega_1 + 0 \Omega_2 +\cdots+ \Psi_A , \Psi_A \perp
G,
\end{equation}
$\{\Omega_\c , \Omega_1 , \Omega_2 \ldots\}$ an orthonormal basis
for $G$. Applying $e^{-\tn{tH}}$ to say ket (3.12) and then
pairing with (3.13) observe that $\langle\Omega_\c
A\Omega_i\rangle = 0, i>0$, so:
\begin{equation}
\l\tn {imaginary time correlation"}\xrightarrow{\tn{ exponentially
}} \langle\Omega_\c A \Omega_\c \rangle\langle\Omega_\c B
\Omega_\c \rangle
\end{equation}
Thus the usual heuristic: gap $\longleftrightarrow$ finite
correlation length, gapless $\longleftrightarrow$ polynomial
decay, is not less valid for code spaces than simple,
non-degenerate ground states. Hence, the expectation is that code
spaces $G$ are protected by a spectral gap and meas.$(\Psi )$,
$\Psi \in G$ has finite (or even zero) correlation length.





Curiously, it is the \underline{square} of the Beraha numbers,
$n^2=d^4 = q=\left(2\cos \f{\pi}{\ell+2}\right)^4 , \,\,
\ell\geq1$, which enter as the weight of a circle in the loop gas
Gibbs state.  This means that for $\ell \geq 3$ these systems are
outside the critical range in the thermodynamic limit. But recall
[BJ] that, for $\ell =3$ the resulting theoretical stability is
extremely weak.

We would like to propose the possibility that the transition $n_c$
from critical to bubble phase on a surface $Y$ might be sensitive
to roughening, i.e. an increase of Hausdorff dimension. Roughening
appears to increase the entropy of long domain walls giving them
more dimensions to fluctuate in, so one might expect the
energy/entropy balance point to increase to $n_c >2$. There are
two arguments for $n_c =2$. One is an explicit study of the
spectral gap of the corresponding transfer matrix in the $6-$
vertex model. This needs a geometric product structure and so will
not apply to a typical rough surface.  The second (ch 12 [B]) is
topological and seemingly does apply.  It used the Eluer relation
to create a translation between the Potts model, a loop model, and
an ice-type model.  Formally, when the Potts parameter $q$ crosses
$4$, $\theta =\log (z)$ ($z$ is the parameter in the ice type
model) passes from imaginary to real $q^{1/2} =e^{z}+ e^{-z}$.
This shows a singularity in the \underline{coordinates} but not
necessarily the \underline{model} itself at $q=4$, $n=\sqrt{q}
=2$, so it seems that a role for surface roughening has not been
excluded.

If $n_c$ can be promoted to $\f{3 +\sqrt{5}}{2} \approx 2.62$ by
surface roughening, then a $G_{\e, 3} \cong G_{\c , 3}/\langle p_4
\rangle \cong DE3$ might be available as the ground state space of
an honest perturbation of $H_{\c , 3}$. Alternatively, the
stability may at $n=2.62$ be so slight as to be physically
irrelevant.  In either case, roughened or not, the quantum medium
must certainly have topological dimension $=2$ to admit anyons but
the Hausdorff dimension of $Y$ might approach $3$.

To be susceptible to the imposition of a topological symmetry (to
force the system to be perpendicular to the ideal $\langle p_{\ell
+1}\rangle$) we need $G_{\c , \ell}$ to be unstable (or in the
$\ell =3$ cases, nearly so). What will be the properties of a
$G_{\e , \ell}\cong G_{\c , \ell}\cap \langle p_{\ell
+1}\rangle^\ast$ if in fact such a ground state space is achieved?
In $\S$ 2 the global properties of $DE3$ as a UTMF were discussed.
These have implications for the local properties of a unit vector
$\Psi \i G_{\e , \ell} \cong G_{\c , \ell}\cap\langle p_{\ell
+1}\rangle^\ast$ and these contrast with a vector $\Phi \i G_{\c ,
\ell}$.

First, because $ G_{\e , \ell}$ has the structure of a UTMF no
information about the state $\Psi$ can be determined from
observables acting on a disk $D$ imbedded in $Y$, $D\subset Y$.
 Usual
correlations such as $\sigma_z^v \otimes \sigma_z^{v'}$ must be
zero (or at least exponentially decaying) or else measurements in
a large disk $D \subset Y$ and $\l$extrapolation" would reveal
information about the state $\Psi$ on $Y$. One might think with
rapid decay of correlations, that observing $\Psi$ in the
classical basis we would see a $\l$bubble phase" with a few tight
global walls $\gamma_\c \subset \gamma$ forced by topology. But
this is \underline{impossible} such global lines could be locally
detected on a disk $D \subset Y$ and would reveal information on
$\Psi$ which is forbidden. In other words, a local operator could
split the ground state whereas no local operator should effect
more than an exponentially small splitting of $G_{\e , \ell}$.

How is the paradox resolved? The domain walls $\l$loops" in a
typical (according to $L^2 -$norm) component $\Psi_i$ of $\Psi$
will be very long probably space filling but at the same time not
locally correlated.  This behavior is seen already in the typical
classical (i.e. observed) states of any toric code word [K1], so
the phenomena is not a surprise.

Next we show that on $(Y, \tr )$ the ground state space $G_{\c,
\ell}$ of $H_{\c, \ell}$ is polylog extensively degenerate.
Understanding this scaling is an important ingredient in the
perturbation theory. Fix a closed surface $Y$ of genus $(Y) =g$
and use the number of vertices $v(\tr)$ as a measure of the
combinatorial complexity of the triangulation $\tr$.  Assume, for
studying the $v(\tr_i):=v_i \la \infty$ asymptotics of $G_\c$,
that the triangles of the triangulations $\tr_i$ have bounded
similarity type.  This means that triangle shapes should not be
arbitrarily distorted. In this regard $\l$barycentric subdivision"
is $\l$bad" but more regular subdivisions are $\l$good".


\vskip.2in \epsfxsize=3.5in \centerline{\epsfbox{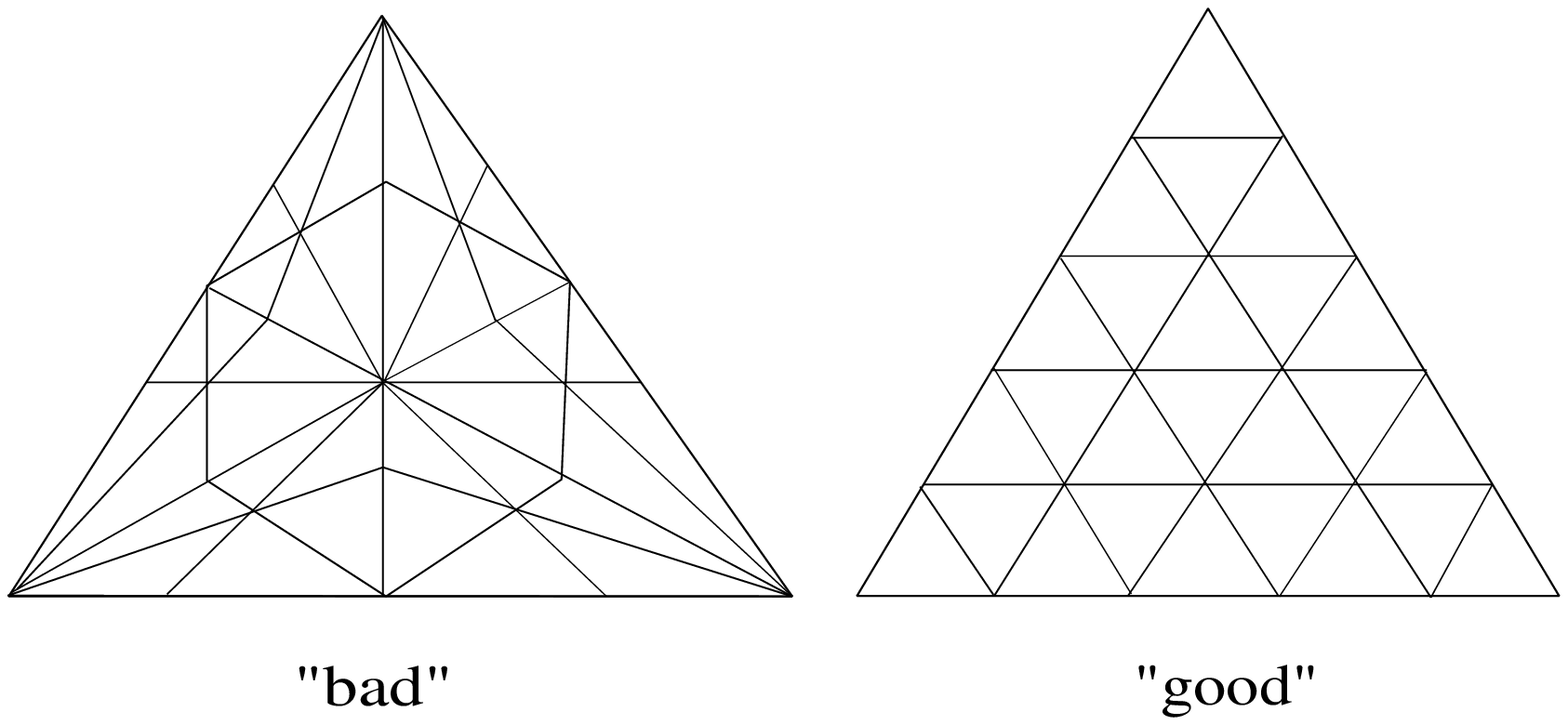 }}
{\centerline{Figure 3.2}} \vskip.2in

\begin{pro}\tn{
If genus $(Y) =g>1$  then $\dim(G_{\c, \ell} (Y, \tr))$ and
$\dim(G'_{\c, \ell} (Y, \tr))$ are $\mathcal{O}
\left(v(\tr)^{3g-3}\right)$. If genus $(Y)=1$, the dimensions are
$\mathcal{O} (v  (\tr ))$.}
\end{pro}

{\bf Proof}  A \underline{pant} is a $3-$punctured $2-$sphere. Fix
a hyperbolic metric on $Y$. In a hyperbolic metric, $d-$isotopy
classes have unique geodesic representatives.  Using the
Fenchel-Nielsen  coordinates [Th] on a geodesic pants
decomposition of $(Y, \tr)$, we find, for each pant, 3
multiplicity parameters and $3$ twist parameters each assuming
$\mathcal{O}(v^{1/2})$ values, a definite fraction of which are
mutually consistent. Since the Euler characteristic $\chi$
(pant)$=-1$, there are $2g -2 = |\chi (Y)|$ pants in the
decomposition.  The consistent parameter settings define geodesic
patterns, $\mathcal{O}\left(v^{3g-3}\right)$ geodesic
$1-$manifolds compatible with $\tr$.  These $1-$manifolds are
unique up to isotopy and have no trivial circles and so are also
unique up to $d-$isotopy.  But $G_{\c, \ell}$ is defined as the
perpendicular to the $d-$isotopy relation on $\mathcal{H}$.\qed

\begin{thm}
 Assume $\tr$ of $Y$ is sufficiently fine.
 If $\{\mathcal{O}_i\}$ is a family of strongly local Hermitian
 operators on $\mathcal{H}$ with jgs $X \subset \mathcal{H}$
 then there are three possibilities for $X \cap G_{\c, \ell}$:
 \begin{itemize}
 \item [(1)] $X \cap G_{\c, \ell} =  \{ 0\}$,
 \item [(2)] $X \cap G_{\c, \ell} = G_{\c, \ell}$, and
 \item [(3)] $X \cap G_{\c, \ell} \cong G_{\c, \ell}\cap \langle p_{\ell +1}\rangle^\ast = DE \ell (Y)$.
 \end{itemize}
 The choice  $\mathcal{O}_i = p_{\ell +1}^i$
 realizes possibility (3).
\end{thm}

\begin{No}
\tn{By the Verlinda formulas dim $D E\ell(Y)$, $Y$ a closed
surface, is asymptotically $\f{2}{\sqrt{\ell +2}} \left(\sin
\f{\pi}{\ell+2}\right)^{\chi(Y)}$ (the fraction of error converges
to zero). This maybe compared to the much larger dim of $G_{\c,
\ell}$ (proposition 3.8).}
\end{No}

\noindent{\bf Proof: } For each $\mathcal{O}_i$ there are
(orthonormal) vectors $f_{i, j}$ spanning the $k$ tensor factors
on which $\mathcal{O}_i$ acts nontrivially so that $E_{\c, i}$ the
lowest eigenspace of $\mathcal{O}_i$,  has the form span$(f_{i,1},
f_{i, 2}, \ldots f_{i, \delta_i} )\otimes$ (the remaining $n-k$
factors) where $\delta_i = \dim(E_{\c, i})$.  The $f_{i , j}$ are
assumed to be chosen coherently with respect to the natural
isomorphism $\mathcal{O}_i \cong \mathcal{O}_i^{'}$. For each $i
\, \us{k =1}{\overset{\delta_i}{\oplus}}| f_{i , k}\rangle \langle
f_{i,k}|$ constitute a skein relations $s_i$ as explained above.
Thus $X \cap G_{\c , \ell}$ consists of vectors orthogonal to the
equivalence relation spanned by both $d-$isotopy and the skein
relations $\{s_i\}$.

The easiest example is for the level$=1$ theory where $A= i e^{\pi
i/6}$, $d=-A^2 - A^{-2} =1$. In this case the Jones Wenzl
projector $p_2$ reads: $p_2 = \,)(\, - \,{{\smile}\atop{\frown}}$
or in combinatorial model:


\vskip.2in \epsfxsize=3.5in \centerline{\epsfbox{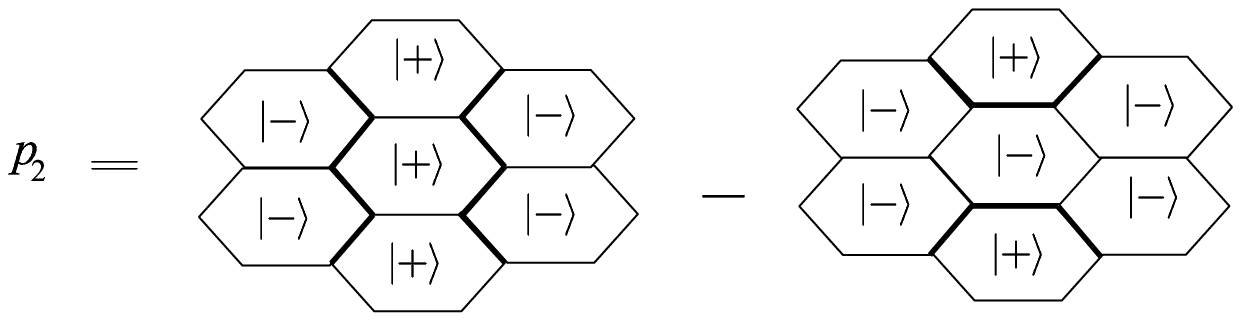 }}
{\centerline{Figure 3.3}} \vskip.2in

Any combinatorial version of the smooth relation will give the
same quotient and same $G_{\c, \ell} \cap X$ provided the
triangulation $\tr$ is sufficiently fine. When $d=1$ generalized
isotopy simply means isotopy and deletion of circles bounding
disks; adding $s_i = p_2^i$ yields unoriented $Z_2 -$homology $-$
as shown in Figure 3.3 $-$ as the quotient equivalence relation.
So, for example on a closed surface $Y$, the space $X\cap G_{\c ,
\ell}$ has dimension $2^{b_1(Y)}$, $b_1$ being the first Betti
number, and is identified with functions $H_1 (Y; Z_2) \la \C$.

For any skein relation $s$ the subset $ <s_{i}> \cap G_{\c,
\ell}\subset \mathcal{H}$, each $s_i$ specializing  $s$ at
locations $U_i$, is an ideal of the surface algebra $G_{\c, \ell}$
as defined in section 2.  The uniqueness Thm
 2.5 implies $<s_{i}> = G_{\c , \ell} , \{0\}$, or $\langle p_{\ell +1}\rangle$ (the latter occurs when
$s_{i}$ represents $p_{\ell +1}$ on $U_i$, i.e, $s_i =
p^i_{\ell+1})$. These correspond respectively to the three
alternatives in the theorem. \qed

\begin{obspro}\tn{
Although $\e V = \e \,{\SS} \sigma_x^v$, is promising perturbation
to find $DE\ell$, its ground state vector $\theta_\c
=\f{1}{2^{V/2}} \underset{k=1}{\overset{2^V}{\SS}} (-1)^{\#
|-\rangle} \Psi_k$ has exponentially large $H_{\c , \ell} -$
expectation, $\ell >1$,
\begin{equation}
 \langle\theta_\c | H_{\c , \ell} |\theta_\c
\rangle \,\, \geq e^{L^\alpha}
\end{equation}
some $\a> 0$, and $L$ the refinement scale.}
\end{obspro}

\noindent{\bf Proof: } If a spin configuration is chosen uniformly
at random from the $2^v$ possibilities it follows from an easy
independence argument that the mean number of circles $\#$ of that
configuration is $\mathcal{O}(v)$ and the standard deviation is
s.d. $(\#) \geq \mathcal{O}(v^{1/2})$, $v$ the number of vertices
of $\tr (Y)$. For $d>1$, line 3.15 is deduced using the
$\l$circle" term of $H_{\c , \ell}$ together with the above
inequality on standard deviation. \qed

To consider the possibility of frustration, which is outside our
algebraic ansatz (but quite possible in a
\underline{fundamental} description of an anyonic system, see $\S$ 4, (2) Uniqueness) we
should think about how a general local operator could act on
$G_{\c, \ell}$. Such (non scalar) actions are possible precisely
because $G_{\c , \ell}$ is \underline{not} a topological modular
functor\footnote{For $\ell \leq 2\,\, G_{\c, \ell}$ presumably has
gapless modes which are un - topological. For $\ell >2$ string
tension allows local measurements to yield a global inference.}
(there is no disk axiom here) so a local operators such as
$\mathcal{O}$ may detect statistical information on the topology
of a state $\Phi \i G_{\c , \ell}$. For example the presence of a
bond in the domain wall $\gamma$ may have polynomial influence on
a global topological event.  This phenomena is familiar from
percolation: The state of a single bond in the middle of an $\ell
\times \ell$ piece of lattice has influence, at $p_c$, on the
existence of a percolating cluster joining two opposite boundary
segments which decays as $\ell^{-5/4}$ (for the triangular
lattice)[LSW] and [SW]. $\mathcal{O}$ may determine an effective
reduction $E_{\ld_\c} \subset G_{\c , \ell}$ based on some local
statistical difference between the random components of different
$\Phi_k \i \mathcal{H}$ which the local operators
$|s_{i}\rangle\langle s_{i} |$ detect. The sites of these
operators should be located and oriented randomly w.r.t. the
domain wall $\Pa_k$ of $\Phi_k$ so it is reasonable to believed
that the deviation of $\mathcal{O}$ from scalar $\times$ identity
will be attributable to the topologically essential part $\Pa^+_k$
of $\Pa_k$ since other features may be lost in averaging over the
sites. For example, $\Pa^+_k$ may appear $\l$straighter" than the
$\l$foam" $\Pa_k \diagdown \Pa^+_k$. That is, $\mathcal{O}$ may
favor or disfavor a topologically complex essential-domain-wall
$\Pa^+_1$ over, say, a simpler essential-domain-wall $\Pa^+_2$.

If complexity is favored the ansatz is still capable of describing
the final reduced ground state space, though for an indirect
reason: A reduction $E_{\ld_\c} \subset G_{\c, \ell}$, or a series
of reductions, which preserves the complex topological
representatives will retain many representatives for each class in
the quotient $G_{\c, \ell}/\langle p_{\ell +1} \rangle$ modular
functor so topological information will not be lost, $E_{\ld_\c}
\cap \langle p_{\ell +1} \rangle^\ast = G_{\c, \ell} \cap \langle
p_{\ell +1} \rangle^\ast$. (But note that the intermediate
subspace $E_{\ld_\c}$ fails to respect the multiplicative/tensor
structure on $G_{\c , \ell}$.).

If topological complexity is disfavored in $E_{\lambda_\c}$, then
the action of $\mathcal{O}$ could destroy the topological
information of the modular functor by killing all classes except
for the simplest, $\Pa^+ =\emptyset$, corresponding to no
essential domain wall and foam covering all of $Y$.  In this case
the ansatz is not applicable. But fortunately the sign of $\e$ in
the perturbation $\e\mathrm{V}$ may be possible to control,
placing the system $(\mathcal{H}, \tr , H_\e )$ into the favorable
regime, where a further reduction, perhaps at, higher order, could
still find the modular functor.

Because of the topological character of skein relation, the final
quotient $G_{\c, \ell}/\langle p_{\ell +1} \rangle$ is the same
regardless of which sites and with what coefficient norm the
various $p_{\ell+1}^i$ relation is enforced. Applying $p_{\ell
+1}$ requires no homogeniety of input and in fact $\l$smooths"
local disturbances.

\section{The evidence for a Chern-Simons phase}
There is no proof or $\l$derivation " of a stable (gapped) phase
\linebreak $G_{\e , \ell}\cong G_{\c , \ell}/<p_{\ell +1}>\cong DE
\ell$ but there is evidence in the form of analogy and internal
consistencies under the headings: $\l$UTMF", $\l$uniqueness",  and
$\l$positivity". The first two have been discussed and are only
now summarized, the third is presented in more detail.

{\bf 1)} \underline{\bf UTMF.} The quotient algebra $G_{\c, \ell}
/<p_{\ell +1}> \cong DE\ell$ is an anyonic system, in mathematical
terms a UTMF.  This in itself is consistency check: The presence
of an anyonic quotient system. The spatial correlation scale for
topological information is zero so we expect a gap protecting the
topological degrees of freedom. The system is the quantum double
of a Chern-Simons theory and experience with the FQHE as a
Chern-Simons theory has prepared us to believe that these
beautiful structures can self-organize in nature from the simplest
underlying Hamiltonians - e.g. Coulomb repulsion in a Landau
level.

Mathematically the double has the form of the algebra of operators
on some (fictitious) FQHE - like system $\l X$": domain walls
realize the Wilson loop operators on $X$.  The double is freed
from chiral asymmetry and the extreme physical conditions required
to break time reversal symmetry.  The double is a better place to
look for a realization protected by a large spectral gap.

{\bf 2)} \underline{\bf Uniqueness.} There is a unique candidate
model $G_{\c , \ell}/<p_{\ell +1}> $ respecting the local or
$\l$multiplicative" structure of $G_{\c , \ell}$ (Thm 3.9).
Uniqueness suggests that there is a sharp boundary: there are no
slightly larger quotients which could include low frequency
excitations. Thus the simplest expectation, that the reduction
$G_{\e , \ell}, \ell \geq 2$, be one dimensional (non-degenerate),
cannot be achieved as a joint ground state jgs of local projectors
but requires \underline{frustration}.  A subtle point is involved
here.  The most interesting candidates in solid state physics for
topological states are highly frustrated systems when written out
in their \underline{fundamental} degrees of freedom.  This does
\underline{not} mean that they cannot have \underline{effective}
descriptions as a jgs $(= \tn{unfrustrated})$.  In fact a very
general topological argument suggest that a phase with the
structure of a TQFT always will.  In local models (e.g. [TV]),
product structures $Y\times I$ yield projectors $\mathcal{H}(Y)
\stackrel{p}{\la} \mathcal{H}(Y)$ whose image is the underlying
UTMF, $V(Y)$. However by building $Y \times I$ from overlapping
local product structures, $p$ can be factored into a commuting
family of local projectors $\{p_i\}$ so $V(Y)=$ image $p=jgs
\{p_i\}$.

Finally, uniqueness creates an aesthetic bias: Could nature really
turn down such a possibility?

We now turn to the final consistency check.
\\
\indent{\bf 3)} \underline{\bf Positivity of the Markov trace
pairing on $G_{\c , \ell} /\langle p_{\ell +1}\rangle $.}

On a surface $Y$ with or without boundary, the Hilbert space
$G_{\c , \ell} /\langle p_{\ell +1}\rangle \cong G_{\c, \ell}\cap
\langle p_{\ell +1}\rangle^\ast$ inherits a Hermitian inner
product $\langle\, ,\,\rangle_{\tn{geom.}}$ by inclusion in
$\mathcal{H}$, the space of all spin configurations. (Beginning
with $\{| + \rangle, |-\rangle \}$ as an orthogonal basis for
$\C^2$, the Hilbert space $\mathcal{H}$ acquires the tensor
product pairing which may be restricted to $G^+_{\c, \ell}\cap
\langle p_{\ell +1}\rangle^\ast$.) On the other hand, $G^+_{\c ,
\ell} /\langle p_{\ell +1}\rangle \cong DE_\ell (Y)$ has a
topologically defined $\l$Markov trace" Hermitian inner product
$\langle\, , \,\rangle_{\tn{top.}}$ corresponding to its structure
as a UTMF (see Definition 4.2.).

\begin{pro}
\tn{ Suppose that the perturbed Hamiltonian $H_{\e , \ell}$ has a
spectral gap above its ground state $G_{\c, \ell}\cap \langle
p_{\ell +1}\rangle^\ast$ , then up to a correction which is
exponentially small in the refinement scale of the triangulation
$\tr$,  $\langle\, , \,\rangle_{\tn{geom.}}$ and  $\langle\, ,
\,\rangle_{\tn{top.}}$ are proportional:  $\langle\, ,
\,\rangle_{\tn{geom.}} \cong c \langle\, , \,\rangle_{\tn{top.}}$
for some real number $c\neq 0$. }
\end{pro}

The definition of $\langle\, , \,\rangle_{\tn{top.}}$ is recalled
below.
\begin{defi}\tn{
 Extending the definition given in section 2: If $\Pa_1$ and
$\Pa_2$ are domain walls in $Y$ with identical boundary data then
$\Pa = \Pa_1 \cup \Pa_2$ defines a link in $\widehat{Y} \times
S^1$ where $\widehat{Y}$ is $Y$ with its boundary capped by disks.
(Place $\Pa_1$ and $\Pa_2$ on disjoint $\theta-$levels $\theta_1$
and $\theta_2$, then bend paired endpoints to meet at the
intermediate level $(\theta_1 +\theta_2)/2$ w.r.t. the $S^1$
orientation.) Regarding $\Pa$ as labelled by the $2-$dimensional
representation, $\langle \Pa_1 , \Pa_2 \rangle :=$ Witten
invariant $(\widehat{Y} \times S^1, \Pa)$ at level $\ell$, see
[Wi], [RT], and [BHMV]. We refer to this pairing, also, as the
Markov trace pairing. }
\end{defi}

The combinatorial properties the code space $C:= G_{\c, \ell} \cap
\langle p_{\ell +1}\rangle^\ast$ are such that local operators (in
fact any operator supported on some topological disk $D \subset
Y$) cannot extract or modify information in $C$.  (It is true that
a local operator can rotate $C$ to $C'$, $C' \perp C$, but
nondestructive (of details within $C$) measurements allow the
error to be corrected by a physical operator $\mathcal{F}$ acting
on $\mathcal{H}$ so that the composition $C\,\overset{ \mathcal{E}
}{\la} C' \overset{\mathcal{F}|}{\la} C$ is the identity id$_C$.)
The code property is a kind of combinatorial/topological rigidity
and it is quite natural that, if achieved in a ground state space,
that space should be protected by a spectral gap.

As we argue for proposition 4.1, a final consistency check
emerges:
\begin{equation}
\tn{gap $+$ code $\Rightarrow$ positivity of Markov trace
pairing.}
\end{equation}

This explains how the Markov pairing is picked out and why its
(indefinite) Galois conjugates are unrealizable as stable phases.
The Markov trace pairing is known to be positive [J], [FNWW]
precisely for our choice of $A$, $A= ie^{\pi i/2r}$, $d= -A^2
-A^{-2}$, and it being topologically defined is automatically
invariant under the mapping-class-group of $(Y,\partial Y)$.  For
other roots of unity $A$ the resulting Hermitian pairings are of
mixed sign so cannot, by positivity of $\langle \,,\,
\rangle_{\tn{geom.}}$ on $\mathcal{H}$ and proposition 4.1,
correspond to stable physical phases. For example, at level $\ell
=3$, Galois conjugate choice $d'= \f{1 -\sqrt{5}}{2}$ is not a
physical. No ground state space modelling $G_\c^{d'} / <p_4^{d'}>$
could have a gap. The corresponding $\l$UTMFs" are only
$\l$unitary" with respect to mixed $(p,q)$ Lorentz form [BHMV],
constructed for each labelled surface of the theory.  These
structures cannot be induced by restricting the standard Hermitian
pairing $\langle\, ,\,\rangle_{\tn{geom.}}$ on $\mathcal{H}$.

By choosing $d= 2 \cos\f{\pi}{\ell +2}$ as we have ensured tr $(a,
\overline{a})$ is positive.  If $G_{\e , \ell}$ has a gap and is
naturally identified with $G_{\c, \ell} /\langle p_{\ell
+1}\rangle$ then 4.1 \underline{implies} positively of the Markov
trace. Positivity is demonstrated by showing that, up to an error
exponentially small in the refinement scale $L$, that the Markov
trace is in the similarity class of the Hermitian form induced
from the standard inner product on $\mathcal{H}$. This is the
argument; it is not mathematically rigorous as the $\l$imaginary -
time $=$ space ansatz" is employed, but we hope that is convincing
physically.

\noindent{\bf Argument, Proposition 4.1. } A surface $(Y, \tr)$
can be gradually changed by bringing bonds in and out of the
triangulation (and perhaps adding or deleting vertices).  With
patience, a Dehn twist can be effected. This takes
$\mathcal{O}(n^2)$ moves on an $n \times n$ square grid torus
$T^2$. Similarly a braid generator for quasiparticle excitations
on a disk takes $\mathcal{O}(n^2)$ such moves where $n$ is the
number of bonds in a loop surrounding the two quasiparticles.
These changes can eventually return $\tr$ to a homeomorphic,
though now twisted, image of itself.

\vskip.2in \epsfxsize=3.5in \centerline{\epsfbox{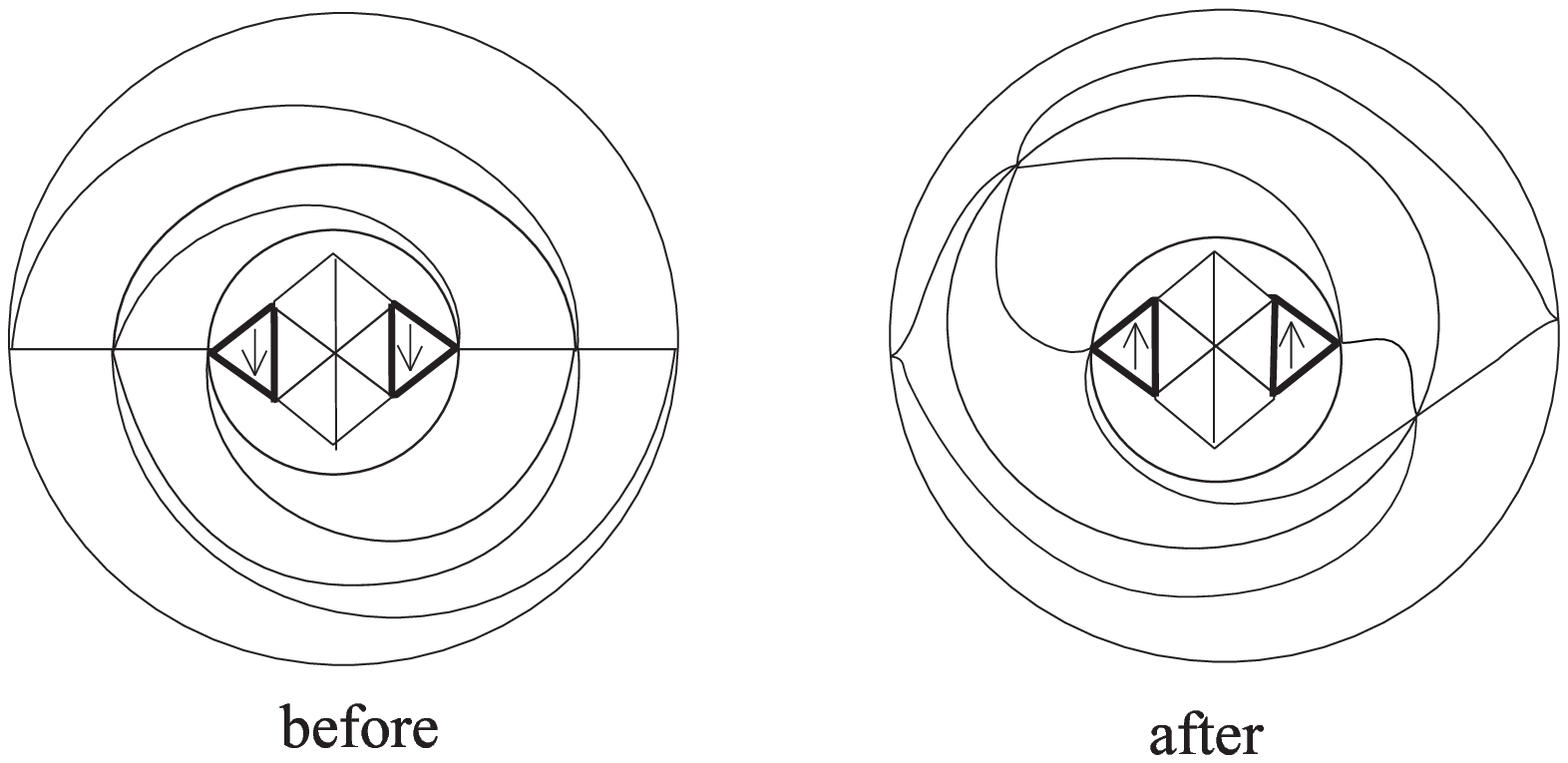 }}
{\centerline{Figure 4.1}} \vskip.2in


If $H_\e$ has a gap, bounded as we change $\tr$ (on which $H_\e$
depends), the adiabatic theorem will define, in the slow
deformation limit: deformation speed $<<$ gap, a time evolution of
vectors in $G_{\e , t} \subset \mathcal{H}$, $t \e [0,1]$.  At
time $t=0, G_{\e , \c} = G_\e$ and finally at time $t=1$, $G_{\e ,
1} = G_{\e}$ again.  This evolution is (incidentally) identical
with the one induced by the canonical connection on the universal
topological bundle of $\{k$ - plane, vector in $k$ - plane $\}\la$
$\{ k $ - planes$\}$.  From the assumption of a gap$=\delta$, one
can argue that this monodromy for a Dehn or braid twist is
accomplished by a composition of $\mathcal{O}(\delta^{-1}n^2)$
local operators, or more precisely operators $A_t$ which have only
an exponentially small nonlocal part. This means that for Pauli
matrices at sites $i$ and $j$, the commutator satisfies:
$\parallel [\sigma_x^i A_t , \sigma_y^j ]\parallel < c_\c
\,e^{-c_1 \parallel i-j
\parallel}$ for all indices $x,y, i, j, t$, and some positive constants $c_\c$ and $c_1$.
We call such operators \underline{quasi-local}. The essential
point is that the local disturbance caused by modifying $H_t$ near
a bond to $H_{t+1}$ dies away exponentially in imaginary time and
hence in space. Let us ignore the exponential tail (which will
lead to a manageable error term), and think of the monodromy as a
composition of $\mathcal{O}(\delta^{-1}n^2)$ local operators:
monodromy $= \us{t}{\prod} A_t^{1 \c c}$. For each $t$ in a
discretized unit interval with $\mathcal{O}(\delta^{-1}n^2)$
points, $\,A_t^{1 \c c}$ is a local unitary operator $\in$ Hom
$(\mathcal{H}, \mathcal{H})$ which carries the code subspace
$G_{\e , t}$ of $\mathcal{H}$ at time $=t$ to the code subspace
$G_{\e , t+1}$ at time $=t+1$.

Adiabatic evolution has provided us with one (local)
representation, $\,\rho_{\tn{geom.}}$: $\pi_1$ (moduli space) $\la
PU_{\tn{geom.}} (G_{\e , t})$, from the fundamental group of
moduli space $(Y)$ (in our discrete context moduli space is the
space of triangulations of $Y$) to the projective unitary
transformations of the perturbed ground state space. The subscript
geom. signifies that $PU$ is defined with respect to $\langle\, ,
\,\rangle_{\tn{geom.}}$. On the other hand, assuming a spectral
gap above $G_{\e , t}$, there is a physical argument that a
second, topologically defined, representation is \underline{also}
\underline{local}.   This representation: $\rho_{\tn{top.}} :
\pi_1$ (moduli space) $\la PU_{\tn{top.}}(G_{\e, t})$ is defined
into the projective unitaries w.r.t. $\langle\, ,
\,\rangle_{\tn{top.}}$ by deforming the triangulation $\tr_t$
while leaving the formal picture ($:=$ superposition of domain
walls) topologically invariant.  This representation can be
defined by choosing a local rotation which interpolates between
the conditions (that define $G_{\c , t} \cap \langle p_{\ell
+1}\rangle^\ast$) in force at time $=t$ but not $t+1$ and those in
force at time $=t+1$ but not $t$. What is not immediate is whether
the effect of this local rotation on the $jgs$ can be achieved by
an operator $A_t$ on $\mathcal{H}$ which is quasi-local.  But the
existence of a quasi-local $A_t$ can be argued based on the
$\l$imaginary $-$ time $=$ space" ansatz ($\S$ 3 lines (8)-(14)).
Similarly, if we view the ground state $G_{\e, t}$ as a local
excitation of $H_{\e, t+1}$, but one without topological content,
we expect that they can be annihilated by a quasi-local $A_t$.

But if a local operator carries one code space into another, that
operator restricted to the first code space is \underline{unique},
up to a scalar, among all restrictions of such local operators.
This is particularly clear in the present case when the operators
are unitary and all the code spaces have the same dimension.
Suppose both $A$ and $B$ are unitary operators carrying $C_1$ into
$C_2$, then $B^\dag \c A |_{C_1} : C_1 \la C_1$ is also local and
so multiplication by some unit norm scalar $\ld$. Thus $B |_{C_1}
= \overline{\ld} A |_{C_1}$.

So assuming a gap, the proceeding observation shows first that
$\rho_{\tn{geom.}}$ and $\rho_{\tn{top.}}$ are both actually well
defined as maps from the fundamental group (see Figure 4.2) and
second that $\rho_{\tn{geom.}}$ will be projectively the same
$\rho_{\tn{top.}}$ up to an error exponentially small in the
refinement scale $L$ (when measured in the operator norm). The
latter, $\rho_{\tn{top}}$ is simply parallel transport in Witten's
[Wi] projectively flat connection on the modular functor bundle
$V(Y_t )$ over the moduli space of surfaces $\{Y_t \}$ ($t$ now an
arbitrary parameter).  Projective flatness as well as uniqueness
of this connection follow formally from locality properties: As
the surface is gradually changed (discretely this is done by moves
on the triangulation $\tr$) the two surfaces $Y_t$ and $Y_{t +1}$
can be canonically identified in the complement of a disk $D$
supporting the changing bonds, and the identification can be
extended arbitrarily over $D$. From the disk axiom and the gluing
axiom of section 2, we have a unique canonical projective
isomorphism of modular functors $V(Y_t ) \la V(Y_{t+1})$. This
determines, via differentiation, a unique connection. Projective
flatness follows by applying this uniqueness to a loop of
identifications representing a small cycle of changes to $\tr$
collectively supported in a disk $D \subset Y$.   Similar loops span
the relations in $\pi_1$ (moduli spaces).


\vskip.2in \epsfxsize=5in \centerline{\epsfbox{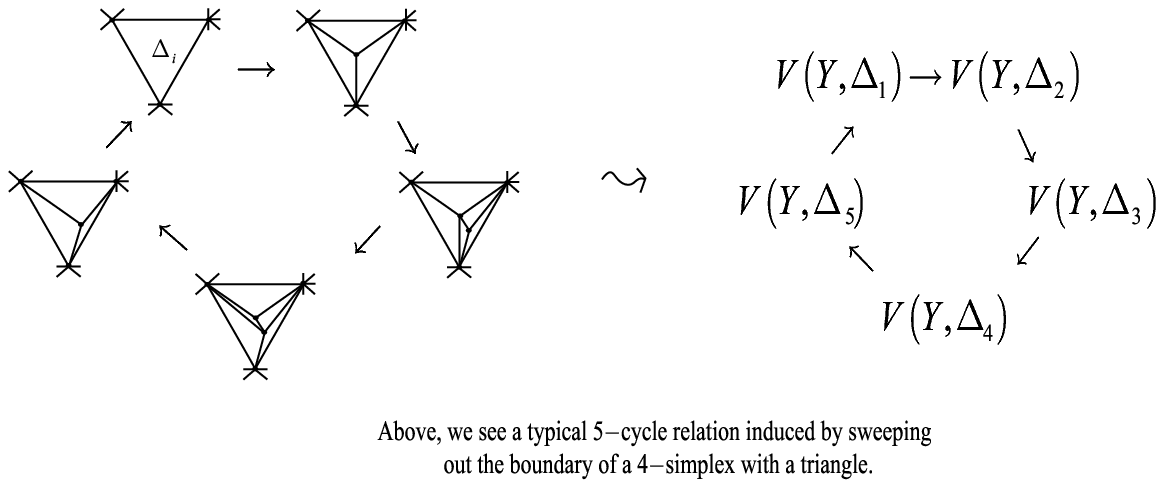 }}
{\centerline{Figure 4.2}} \vskip.2in

Let $V=DE\ell \,\, \ell\neq1,2,$ or $4$. It is known [FLW2] that
for a sphere with $4$ or more punctures (or a higher genus
surface) the braid (or mapping class group) acts densely in the
projective unitary transformations of each label sector $P
U_{\tn{top.}} (V(Y , \overrightarrow{t}))$.  But identifying $V(Y
, \overrightarrow{t})$ with a ground state space $G_{\e, \ell}$
(followed by the idempotent $\overrightarrow{t}$ defined in $\S
3$), we have on the one hand the adiabatic evolution which must be
unitary w.r.t. the Hermitian pairing \underline{induced} from the
\underline{standard} Hermitian pairing on $\mathcal{H}$, and on
the other hand, exponentially close to this, transport in Witten's
connection. Both define (nearly) the same dense homomorphism from
$\pi_1 :=$ the fundamental group of moduli space:
\begin{equation}
 \rho_{\tn{geom.}} \,\,\, \rho_{\tn{top.}} : \pi_1 \la {\tn{End}}\big(V(Y,\overrightarrow{t})\big).
\end{equation}
In the case $Y$ is planar and all boundary labels equal, $\pi_1$
is a familiar braid group.

It follows from the rapid approximation algorithm [KSV],[So],[K2]
of elements of $PU_{\tn{geom.}}, PU_{\tn{top.}} \subset {\tn{End}} \big(V (Y ,
\overrightarrow{t})\big)$ by words in $\pi_1$, that the
\underline{induced} Hermitian metric on $V$ must be exponentially
close (in $L$) to the intrinsic UTMF metric on $V$ up to the
overall scalar $c$. The mathematical fact that we are using here
is that Hermitian pairings can be recovered from their symmetries:

\begin{lemma}
Suppose a vector space $V$ has Hermitian (but not necessarily
positive definite) inner products $\langle \, , \,\rangle_1$ and
$\langle \, , \,\rangle_2$ with symmetry groups $U_1 , U_2 \subset
{\tn{End}}(V)$ respectively, if $U_1 = U_2$ then
 $\langle \, , \,\rangle_1 = c \langle \, , \,\rangle_2$ for some
 real constant $c \neq 0$.  Furthermore if $U_1 \neq U_2$ but instead if
 for all $A_1 \i U_1$ there exists an $A_2 \i U_2$ with $|| A_1
 -A_2 || < \e >0$, and for all $A_2 \i U_2$ there exists an $A_1
 \i U_1$ with $|| A_2 -A_1 || < \e$, then for all $v \i V,\, \,
 \langle v,v\rangle^2_1 / \langle v,v\rangle^2_2 =$ const. $+ \mathcal{O} (\e)$
\end{lemma}

\noindent{\bf Proof:} Up to linear conjugacy the type of the form
is determined by dimension, signature and nullity.  If $U_1 = U_2$
(even approximately) these invariants agree.  Let $M \,\e\,
{\tn{End}} (V)$ transform $\langle \, , \,\rangle_1$ and
$\langle\, , \,\rangle_2 \break \big(\langle M^\dagger v , M w
\rangle_1 = \langle v, w\rangle_2\big)$. Then $U_1 M =MU_2$ so if
$ U_1 =U_2 =:U$ then $M$ normalizes $U$.  Let $s+t = \dim (V)$. In
$PGL (s,t; \C), \, PU (s,t)$ is its own normalizer establishing
the lemma when $U$ is nonsingular.  If the forms have radicals,
these must agree and the preceding argument applies modulo the
radical.  Finally, in the case $U_1$ and $U_2$ are not identical
but have Hausdorff distance $=\e$, a counter example to the lemma
would yield a Lie algebra element $\a \in \,pg\ell(s, t)\diagdown
pu (s,t)$ with $ad_\a (pu(s,t)) \subset pu (s,t)$, but $pu (s,t)
\subset pg \ell(s,t)$ is a maximal proper sub Lie algebra. \qed

Consequently, $G_{\e, \ell}$ is not just linearly $V$ but
metrically $V$, provided $H_{\e , \ell}$ has a gap.

The spectral gap assumption implies that the combinatorically
defined Markov trace pairing is induced from the standard inner
product of $\mathcal{H}$.  The Markov trace pairing is a rather
intricate structure in its relation to gluing (see axiom 2).  That
it arises from a simple assumption can be viewed as a valuable
$\l$consistency check" on that assumption - the existence of a
spectral gap above $G_{\e , \ell}$.

\section{The $H_{\e, \ell,t}$ medium as a quantum computer}
In the literature one finds at least three polynomially equivalent
models of quantum computation defined: $q-$Turing machine [D],
$q-$circuit model [Y], $q-$cellular automata [Ll]. Nearly all
proposed architectures ([NC] is an excellent survey) presume
localization of the fundamental degrees of freedom.  This may be
called the $\l$qubit approach" although {\it qunit} might be more
precise since there is nothing special about two state systems,
the number $n$ of of states per site may even be infinite, as in
optical cavity models $-$ what is important in these
architectures is the tensorial structure of the computational
degrees of freedom.

However, there is another approach [FKLW] in which the global
tensor structure becomes redundant. The physical degrees of
freedom still have a local tensor structure $-$ as is universal in
quantum  mechanics $-$ but these are never touched directly.
Instead a system is engineered so that these local degrees
interact through a Hamiltonian $H$ whose eigenstates $E_\ld$ are
highly degenerate codes spaces capable of storing, protecting and
processing quantum information. For us $E_\ld$ will be the
internal symmetries of an anyonic system in which position
coordinates have been frozen out. The processing will consist of
braiding anyons in $(2+1)-$ dimensional space time.

To make sense of this, consider a definition of $\l$ universal
quantum computer" which does not presuppose  any tensor
decomposition.  We need:
\begin{enumerate}
  \item $h$: A Hilbert $h$ space on which to act.  (Its dimension
  should scale exponentially in a physical parameter.)
  \item $\Psi_\c \i h$: We need to be able to initialize the
  system.
  \item $\rho$: Operations $\la U(h)$, a representation of some
  group (or at least semigroup) of operation on the unitary
  transformations  of $h$ which can be physically implemented - preferably with error
  scaling like $e^{-\tn{constant } L}$ for some physical parameter $L$. (Lack of such scaling is the Achilles heel of qubit
  models.) The representation $\rho$ should have dense image in $SU(h)$: This, together with the rapid convergence
  property of dense subgroups of $U(h)$, ensures universality.
  \item Compiler:  This is a classical computer which takes a
  $q-$algorithm and an instance, e.g. Shor's poly time factoring
  algorithm [S2] and a thousand bit integer, and maps the pair into a
  string $s$ of operations as in (3.).
  \item $\Psi_f$:  The result of the quantum portion of the
  calculation is a final state $\Psi_f = \rho (s) \Psi_\c$.
  \item Observation: There must be a Hermitian operator which
  serves as the observation: projecting $\Psi_f$ into an eignstate
  $\Psi_\ld$ with probability $|a_\ld |^2$, $\Psi_f =\SS
  a_\ld \Psi_\ld$.  The eigenvalue $\ld$ is what is
  actually observed.
  \item Answer:  Another poly-time classical computation is now
  made to convert the observed eigenvalue, perhaps for many
  executions of $1) \la 6)$, into a probabilistic output. The class of
  problems that can be answered in polynomial time by $1) \la 7)$
  with bounded error probability (say error $< \f{1}{4}$) is
  called $B\mathcal{Q}P$.  For example, factoring [S2] is in $B\mathcal{Q}P$.  Computer scientists believe, and
  cryptographers hope, that factoring is not in the corresponding classical computational class
  $BPP$.
\end{enumerate}

The reader can easily take any qubit architecture (see [NC] for
details of these) and fit it into the proceeding format.  Let us
now do this for our anyonic system with Hamiltonian $H_{\e,\ell}$.
As explained in the introduction, the system was chosen to have
two spatial dimensions topologically, i.e. to live on a
triangulated surface $(Y, \tr)$, so that exotic statistics become
a possibility. By mathematical excising neighborhoods of the
excitations we reduce to the case of the studying the ground state
space $G_t$ of a time dependent $H_{\e , \ell, t}$, on a highly
punctured surface $Y^-$ with labelled boundary. The subscript $t$
reminds us of the time dependence of the surface $(Y^- , \tr)$ as
the position of the punctures evolves. The ground state space
$G_t$ describes the internal symmetries of a collection of
quasiparticle (anyon) excitations whose spatial locations are $t$
dependent. The space $G_{t_\c}$ is a representation space for the
braid group (or generalized braid group) which describes the
motion of these quasiparticles on $Y$. Because of the presumed
spectral gap, the quasiparticles are expected to have
exponentially decaying tails. Chopping off and ignoring these
tails amounts to the puncturing $Y$ at the quasiparticles. This
identifies the excited state $\Psi'_{\c, t} \i\, E_{\ld,t}$
containing anyons on $Y$ with a ground state $\Psi_{\c, t} \i
\,G_t$ on a multiply punctured $Y^-$ with labeled boundary. So by
puncturing and labeling the surface $Y$, a ground state in $G_t$
can be used to represent the anyonic state $\i \,E_{\ld, t}$, so
the discussion of $\S$ 3 applies to $E_{\ld , t}$.

Let us walk
through steps 1 through 7 for our anyonic model, though it is not
efficient to do this in strict order.
\begin{itemize}
\item[{\bf 1. \& 2.}] {\bf $h =E_{\ld,t} \cong G_t$:} In [FLW1] and [FKLW] abstract anyonic
  models for (but with no known Hamiltonian $H$) were analyzed algebraically.  In [F2] an explicit but
  artificial Hamiltonian was given as an existence theorem. The UTMFs of [FLW1],[F2] and [FKLW] required a two (with care 1.5) quasiparticle
  pairs per qubit simulated. $DE \ell$ is a closely related UTMF and for $\ell=3$ a
  similar encryption yields one qubit per 1.5 pairs of $(0,2)$ type excitatitons.
  Physically one imagines a disk of quantum media
  governed by $H_t$ and trivial outer boundary condition, lying in its (nondegenerate) ground state =
  $\l$the  vacuum". The steps required to build $h$ are a subset of those discussed in [BK] in connection with their
  CS2 model. The disk is struck in some way (with a hammer?) at a
  point to create a pair of excitations.  Already in building
  $E_{\ld, t}$ we need measurement to tell if the newly created pair is type $\big((0,2);(0,2)\big)$.
  If the pair is of this type, we keep it, if not it is returned to the vacuum.  Repeat (perhaps thousands of times)
  until a sufficiently large Hilbert space
  $E_{\ld, t} (Y)\cong G_t (Y^- ) \cong h$, and initial vector, $\Psi_\c \i \, G_t$ is realized.
  The initial state $\Psi_\c$ is determined by the condition that
all circles surrounding (not separating) the created pairs
acquire label $(0,0)$.  How many pairs are required depends on the
problem instance.  For example, for factoring problem, it is a
small multiple of the number of bits of the number to be factored.
\item[{\bf 6.}] {\bf Measurement:} The creation process is probabilistic.
So even at the start, there must
be a local observation which tells us which anyon pairs have been
created $((0,0), (0,0));((0,2), (0,2)); (2,0),(2,0)$; or
$((2,2),(2,2))$.  One hopes that will not require a
$\l$topological microscope".  Because \underline{quasi} particles
are arrangements of elementary degrees of freedom spins, charges,
etc$\ldots$ of the system, one expects each quasiparticle when
examined electromagnetically to have its own unique signature:
e.g. quadruple moment etc$\ldots$. In this view, localized
quasiparticles would always be $\l$measured" by their environment
and never lie is superpositions.  However, it is essential for
quantum computation that a well separated pair of quasiparticles
$-$ before being fused $-$ could be in a super position of
collective states. Another idea [SF], discussed in the
introduction, is that a phase transition be employed for
measurement.

\item[{\bf 3.}] {\bf Braiding:} The group of operations is the braid group of
quasiparticles moving on the disk.  In order to implement this
mathematically  known representation on $h$  we need to be able to
grab hold of the quasiparticle and, within some allowable
dispersion corridor, nudge it along to execute the braid $s$
dictated by the output of the classical compiler, step {\bf 4}.
This, like observation, should in principle be possible, using the
characteristic electric and magnetic attributes of the nontrivial
quasiparticles (whatever they are eventually measured to be). It
may be possible to design wells that trap, and when desired, move
specific quasiparticles.

\item[{\bf 5.}] $\Psi_f$ is the internal state after braiding
$\Psi_f=\rho(s)\Psi_\c$.

\item[{\bf 6.}] {\bf Observation:} We already discussed the necessity to observe
halves of newly created pairs.  To read out quantum information
after braiding take two quasiparticles in the system $\Psi_f$ and
fuse them.  Although they will retain their individual identities
during braiding and still be both of type $(0,2)$, after fusing,
two outcomes are possible:  $(0,0)$ or $(0,2)$. The probabilities
attached to these outcomes is the classical distillation of
quantum information equivalent to measuring a qubit in the usual
architecture (see [FKLW] for details of the read out and its
relation to quantum topology and the Jones polynomial.) The
braiding has rearranged, in an exponentially intricate fashion the
structure of the composite pairs of $(0,2) -$ quasiparticles. This
recoupling is the heart of the computation.

\item[{\bf 7. \& 4.}] The final conversion of eigenvalues observed to a probabilistic output
is the same as for the qubit architecture.  The structure of the
compiler is also similar but must include a rapid approximation
algorithm [K2][So] subroutine.
\end{itemize}

We have presented $H_{\e,3, t}$ as a theoretical candidate for an
anyonic medium capable of universal quantum computation. Its
experimental realization would a landmark.

\section{Appendix. Ideals in Temperley-Lieb Catergory}

\centerline{\small{Frederick M. Goodman and Hans Wenzl}}
\appendix{ \input{AppendFinal11-02.tex}}

\end{document}

%% file: AppendFinal11-02.tex

\def\ignore#1{\relax}

\font\fig=cmr8
\def\twlrm{}
\newdimen\ybox\ybox=5pt
\setbox249=\vbox{\hbox{\vrule{\vbox to\ybox{}}\kern\ybox}\hrule}
\setbox248=\hbox{\copy249}

\newcount\n
\def\R#1,{%
\n=#1%
\nointerlineskip
\hbox{%
\loop\ifnum\n>0\copy248\advance\n by-1\repeat%
\vrule}%
}

\def\YD#1.{\,\vcenter{\hrule#1}\,}

\def\g{\mathfrak g}
\def\wt{{\rm{wt}}}
\def\r{{\bf r}}

\def\Z{{\mathbb Z}}
\def\nat{{\mathbb N}}
\def\Q{{\mathbb Q}}
\def\C{{\mathbb C}}
\def\F{{\bf F}}
\def\la{\lambda}
\def\La{\Lambda}
\def\ka{\kappa}
\def\T{{\bf T}}
\def\Pb{{\bf P}}
\def\dlm{d_{\la, \mu }}
\def\alm{\alpha_{\la, \mu }}
\def\W{\mathcal W}
\def\S{\mathcal S}
\def\H{\mathcal H}
\def\N{\mathcal N}
\def\A{\mathcal A}
\def\Ca{\mathcal C}
\def\L{\mathcal L}
\def\one{\mathbf 1}
\def\Lm{{\bf F}_m}
\def\Tm{{\bf T}_m}
\def\Ll{{\bf F}_\La}
\def\H{\mathcal H}
\def\P{\mathcal P}
\def\Se{\mathcal S}
\def\A{\mathcal A}
\def\E{\mathcal E}
\def\T{{\bf T}}
\def\a{{\bf a}}
\def\b{{\bf b}}
\def\W{\mathcal W}
\def\M{\mathcal M}
\def\B{{\mathcal B}}
\def\Nu{\underline{K}}
\def\Au{\underline{A}}
\def\U{\mathbb U}
\def\Xl{X_+^{(l)}}
\def\Xpr{X_+^{(p^r)}}
\def\K{\overline{K}}
\def\ni{\noindent}
\def\ident{1\!\!1}
\def\v{\vskip 2.5mm}
\def\Lbar{\Z[v,v^{-1}]_0}
\def\inv{^{-1}}
\def\ignore#1{\relax}
\def\one{{\bf 1}}
\def\Nbar{\underline{N}}
\def\Mbar{\underline{M}}

\def\Tl{{\bf T}_\Lambda}
\def\om{\omega}
\def\eps{\epsilon}
\def\TL{{\rm{TL}}}
\def\Tr{{\rm{Tr}}}
\def\Hom{{\rm{Hom}}}
\def\End{{\rm{End}}}
\def\id{{\rm{id}}}
\def\tr{{\rm{tr}}}
\def\Ne{{\rm Neg}}
\def\Neg{{\rm Neg}}
\def\ve{\varepsilon}
\def\glm{g_{\la, \mu}}
\def\al{\alpha}



{\theoremstyle{plain}
\newtheorem{theorem}{Theorem}[section]
}

{\theoremstyle{plain}
\newtheorem{proposition}[theorem]{Proposition}
}

{\theoremstyle{plain}
\newtheorem{corollary}[theorem]{Corollary}
}

{\theoremstyle{plain}\newtheorem{lemma2}[theorem]{Lemma}}

{\theoremstyle{plain}
\newtheorem{conjecture}[theorem]{Conjecture}
}

{\theoremstyle{definition}
\newtheorem{definition}[theorem]{Definition}
}

{\theoremstyle{definition}
\newtheorem{example}[theorem]{Example}
}

{\theoremstyle{remark}
\newtheorem{remark}[theorem]{Remark}
}

{\theoremstyle{remark}
\newtheorem{notation}[theorem]{Notation}
}




\renewcommand\thesection       {\arabic{section}}
\renewcommand{\thetable}{\thesection.\arabic{table}}
\numberwithin{equation}{section}

\renewcommand{\labelenumi}{{ \theenumi.}}
\renewcommand{\labelenumii}{{(\alph{enumii})}}

\newcommand\aappendix{\appendix
\renewcommand{\thefigure}{\thechapter.\arabic{figure}}
\renewcommand{\thetable}{\thechapter.\arabic{table}}
\renewcommand\theequation {\thechapter.\arabic{equation}}
}


\ignore{
\input BoxedEPS
\SetRokickiEPSFSpecial
\HideDisplacementBoxes
\SetEPSFDirectory{./graphics/} }

\ignore{
\input BoxedEPS
\SetTexturesEPSFSpecial
\HideDisplacementBoxes
\SetEPSFDirectory{:graphics:}
}








\bigskip
 \maketitle

This appendix contains a proof of the following result, which is
used in the paper of Michael Freedman, {\em A magnetic model with
a possible Chern-Simons phase}.

\begin{theorem}
When the parameter $d$ is equal to
$2\cos(j\pi/n)$ with  $n \ge 3$ and $j$ coprime to $n$,
then the Temperley-Lieb category has
exactly one non-zero, proper ideal, namely the ideal of negligible
morphisms. For all other values of $d$, the Temperley-Lieb
category has no non-zero, proper tensor ideal.

\end{theorem}

We are grateful to Michael Freedman for bringing the question of tensor
ideals in the Temperley-Lieb category  to our attention and for allowing
us to present the proof as an appendix to his paper.

Our notation in the appendix differs slightly from that in the main text.
We write $t$ instead of $-A^2$,  $T_n$ for the Temperley-Lieb
algebra with $n$ strands, and $TL$ for the Temperley-Lieb category.
We trust that this notational variance will not cause the reader any
difficulty.

This appendix can be read independently of the main text.

\section{The Temperley-Lieb Category}

\subsection{The Generic Temperley Lieb Category}

Let $t$ be an indeterminant over $\C$, and let $d = (t + t\inv)$.    The
{\em
generic Temperley Lieb category}  TL is a strict tensor categor whose
{\em
objects}  are elements of
$\nat_0 =
\{0, 1, 2, \dots\}$.
The set of {\em morphisms} $\Hom(m,n)$ from $m$ to $n$ is a $\C(t)$
vector
space described as follows:

\smallskip
If $n-m$ is odd, then $\Hom(m,n)$ is the zero vector space.

For $n-m$ even, we first define
$(m,n)$--TL diagrams,  consisting of:
\smallskip\noindent
\begin{enumerate}
\item  a closed rectangle $R$  in the plane with two opposite edges
designated as top and bottom.
\item $m$  marked points (vertices) on  the top edge and $n$ marked
points on the bottom edges.
\item $(n+m)/2$ smooth curves (or ``strands") in $R$ such that for each
curve
$\gamma$,
$\partial
\gamma = \gamma \cap \partial R$ consists of two of the $n+m$  marked
points, and such that the curves are pairwise non-intersecting.
\end{enumerate}

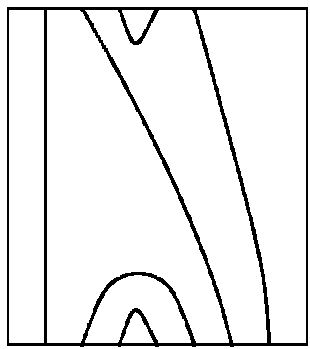
\begin{figure}[ht]
\centering
\setlength{\unitlength}{0.00087489in}
\begingroup\makeatletter\ifx\SetFigFont\undefined%
\gdef\SetFigFont#1#2#3#4#5{%
  \reset@font\fontsize{#1}{#2pt}%
  \fontfamily{#3}\fontseries{#4}\fontshape{#5}%
  \selectfont}%
\fi\endgroup%
{\renewcommand{\dashlinestretch}{30}
\begin{picture}(1824,2084)(0,-10)
\path(12,2047)(1812,2047)(1812,22)
    (12,22)(12,2047)
\thicklines
\path(462,2047)(463,2046)(464,2044)
    (466,2040)(470,2034)(475,2025)
    (482,2013)(490,1998)(500,1981)
    (512,1960)(526,1936)(541,1909)
    (557,1880)(575,1849)(594,1816)
    (613,1781)(634,1745)(654,1707)
    (676,1668)(698,1628)(720,1587)
    (742,1545)(765,1502)(788,1457)
    (812,1411)(836,1364)(861,1315)
    (886,1264)(912,1211)(939,1156)
    (966,1099)(993,1041)(1020,982)
    (1047,922)(1076,855)(1104,791)
    (1129,730)(1153,672)(1174,618)
    (1194,567)(1212,520)(1228,476)
    (1242,434)(1256,395)(1268,358)
    (1280,322)(1290,289)(1300,256)
    (1309,226)(1317,197)(1324,170)
    (1331,145)(1337,121)(1343,101)
    (1347,82)(1351,66)(1355,53)
    (1357,42)(1359,34)(1360,28)
    (1361,25)(1362,23)(1362,22)
\path(237,2047)(237,22)
\path(462,22)(463,25)(465,32)
    (470,45)(476,63)(484,86)
    (494,114)(505,144)(517,175)
    (529,205)(541,234)(552,261)
    (564,286)(575,308)(586,328)
    (596,345)(607,360)(618,374)
    (630,386)(642,397)(655,407)
    (668,416)(682,424)(697,431)
    (713,437)(729,442)(746,446)
    (764,449)(781,451)(800,451)
    (818,451)(835,449)(853,446)
    (870,442)(886,437)(902,431)
    (917,424)(931,416)(944,407)
    (957,397)(969,386)(981,374)
    (992,360)(1003,345)(1013,328)
    (1024,308)(1035,286)(1047,261)
    (1058,234)(1070,205)(1082,175)
    (1094,144)(1105,114)(1115,86)
    (1123,63)(1129,45)(1134,32)
    (1136,25)(1137,22)
\path(1137,2047)(1138,2045)(1139,2041)
    (1141,2033)(1144,2021)(1149,2004)
    (1155,1981)(1163,1952)(1173,1917)
    (1184,1877)(1196,1832)(1210,1783)
    (1224,1729)(1240,1673)(1255,1614)
    (1272,1555)(1288,1495)(1304,1435)
    (1320,1376)(1336,1318)(1351,1262)
    (1365,1208)(1379,1157)(1392,1107)
    (1405,1060)(1416,1016)(1427,974)
    (1438,934)(1448,896)(1457,860)
    (1465,826)(1473,794)(1481,764)
    (1488,734)(1494,707)(1501,680)
    (1506,654)(1512,629)(1520,592)
    (1528,556)(1535,521)(1541,488)
    (1546,456)(1552,425)(1556,393)
    (1561,362)(1564,330)(1568,298)
    (1571,266)(1574,233)(1577,201)
    (1579,170)(1581,139)(1583,111)
    (1584,87)(1585,66)(1586,49)
    (1586,36)(1587,28)(1587,24)(1587,22)
\path(687,2047)(687,2046)(689,2041)
    (693,2029)(699,2009)(708,1984)
    (717,1956)(728,1928)(737,1903)
    (746,1881)(755,1865)(763,1852)
    (770,1843)(777,1839)(785,1837)
    (792,1839)(800,1843)(809,1852)
    (818,1865)(829,1881)(841,1903)
    (854,1928)(868,1956)(882,1984)
    (894,2009)(904,2029)(909,2041)
    (912,2046)(912,2047)
\path(687,22)(687,23)(689,28)
    (693,40)(699,60)(708,85)
    (717,113)(728,141)(737,166)
    (746,188)(755,204)(763,217)
    (770,226)(777,230)(785,232)
    (792,230)(800,226)(809,217)
    (818,204)(829,188)(841,166)
    (854,141)(868,113)(882,85)
    (894,60)(904,40)(909,28)
    (912,23)(912,22)
\end{picture}
}
\caption{A (5,7)--Temperley Lieb Diagram}
\label{TL diagram}
\end{figure}

Two such diagrams are {\em equivalent} if they induce the same pairing
of
the
$n+m$ marked points.
 $\Hom(m,n)$ is defined to be the $\C(t)$  vector space
with basis the set of equivalence classes of $(m,n)$--TL diagrams; we
will refer to equivalence classes of diagrams simply as diagrams.

The composition of morphisms is defined first on the level of diagrams.
The
composition
$ba$ of an
$(m,n)$--diagram
$b$ and an
$(\ell,m)$--diagram
$a$  is defined by the following steps:
\smallskip\noindent

\begin{enumerate}
\item
 Juxtapose the rectangles of $a$ and $b$, identifying the
bottom edge of $a$ (with its $m$  marked points) with the top edge of
$b$
(with its $m$ marked points).
\item Remove from the resulting rectangle any closed loops in its
interior.  The result is a $(n,\ell)$--diagram $c$.
\item The product $ba$ is $d^r c$, where $r$ is the number of
closed loops removed.
\end{enumerate}

The composition product evidently respects equivalence of diagrams, and
extends
uniquely to a bilinear product
$$\Hom(m,n)
\times
\Hom(\ell,m)
\longrightarrow  \Hom(\ell,n),$$ hence to a linear map
$$\Hom(m,n) \otimes
\Hom(\ell,m)
\longrightarrow  \Hom(\ell,n).$$

The {\em tensor product of objects} in $\TL$ is given by $n\otimes n' =
n+n'$. The {\em tensor product of morphisms}  is defined by horizontal
juxtposition.  More exactly,  the tensor $a \otimes b$ of
an
$(n,m)$--TL diagram
$a$ and an $(n',m')$--diagram $b$ is defined by horizontal juxtposition
of
the diagrams, the result being an $(n+n',m+m')$--TL diagram.

The tensor product extends uniquely to a bilinear product $$\Hom(m,n)
\times
\Hom(m',n')
\longrightarrow  \Hom(m+m',n+n'),$$ hence to a linear map
$$\Hom(m,n) \otimes
\Hom(m',n')
\longrightarrow  \Hom(m+m',n+n').$$

For each $n \in \nat_0$, $T_n := \End(n)$ is a $\C(t)$--algebra,
with the composition product.  The identity $1_n$ of $T(n)$ is the
diagram with $n$ vertical (non-crossing) strands.  We have canonical
embeddings of $T_n$ into $T_{n+m}$ given by $x \mapsto x \otimes 1_m$.
If
$m>n$ with $m-n$ even, there also exist obvious embeddings of
$\Hom(n,m)$ and $\Hom(m,n)$ into $T_m$ as follows: If $\cap$ and $\cup$
denote the morphisms in $\Hom(0,2)$ and $\Hom(2,0)$, then
we have linear embeddings
$$a\in\Hom(n,m)\mapsto a\otimes \cup^{\otimes (m-n)/2}\in T_m$$
and
$$b\in\Hom(m,n)\mapsto b\otimes \cap^{\otimes (m-n)/2}\in T_m.$$

Note that these maps have left inverses which are given by
premultiplication
by an
element of $\Hom(n, m)$ in the first case, and postmultiplication by an
element of
$\Hom(m,n)$ in the second.  Namely,
$$
a = d^{-(m-n)/2} (a\otimes \cup^{\otimes (m-n)/2}) \circ
(\one_n \otimes \cap^{\otimes
(m-n)/2})
$$
and
$$
b = d^{-(m-n)/2}
(\one_n \otimes \cup^{\otimes
(m-n)/2})   \circ (b\otimes \cap^{\otimes (m-n)/2})
$$

By an {\em ideal} $J$ in TL  we shall mean a vector
subspace of  $\bigoplus_{n,m} \Hom(n,m)$ which is closed under
composition
and tensor product with arbitrary morphisms.  That is, if $a, b$ are
composible morphisms, and one of them is in $J$, then the composition
$ab$ is in $J$; and if $a, b$ are any morphisms, and one of them is in
$J$,
then the tensor product $a \otimes b$ is in $J$.

Note that any ideal is closed under the embeddings described just above,
and
under
their left inverses.

\subsection{Specializations and evaluable morphisms.}  For any $\tau \in
\C$,
we define the specialization $\TL(\tau)$ of the Temperley Lieb category
at
$\tau$, which is obtained by replacing the indeterminant $t$ by $\tau$.
More
exactly,  the objects of $\TL(\tau)$ are again elements of $\nat_0$, the
set
of morphisms
$\Hom(m,n)(\tau)$ is the $\C$--vector space with basis the set of
$(m,n)$--TL diagrams, and the composition rule is as before, except that
$d$ is replaced by $d(\tau) = (\tau + \tau\inv)$.  Tensor products
are defined as before.
$T_n(\tau) := \End(n)$ is
a complex algebra, and $x \mapsto x \otimes 1_m$ defines a canonical
embedding of $T_n(\tau)$ into $T_{n+m}(\tau)$.  One also has embeddings
$\Hom(m,n) \rightarrow T_n$ and $\Hom(n,m) \rightarrow T_n$, when $m <
n$, as before.   An ideal in $\TL(\tau)$ again means a subspace of
$\bigoplus_{n,m} \Hom(n,m)$ which is closed under composition
and tensor product with arbitrary morphisms.

Let $\C(t)_\tau$ be the ring of rational functions without pole at
$\tau$.
The set of {\em evaluable} morphisms in $\Hom(m,n)$ is the
$\C(t)_\tau$--span
of the basis of $(n,m)$--TL diagrams.  Note that the composition and
tensor product of evaluable morphisms are evaluable.  We have an {\em
evaluation map} from the set of evaluable morphisms to morphisms of
$\TL(\tau)$ defined by
$$a=\sum s_j(t)a_j \mapsto a(\tau)=\sum s_j(\tau)a_j,$$
where the $s_j$ are in $\C(t)_\tau$, and the $a_j$ are TL-diagrams.
We write $x \mapsto x(\tau)$ for the evaluation map.
The evaluation map is a homomorphism for the composition and tensor
products.
In particular, one has a $\C$--algebra homomorphism from the algebra
$T_n^\tau$ of evaluable endomorphisms of $n$ to the algebra $T_n(\tau)$
of
endomorphisms of $n$ in $\TL(\tau)$.

The principle of constancy of dimension  is an
important tool for analyzing the specialized categories $\TL(\tau)$.  We
state it
in the form which we need here:

\begin{proposition}  Let $e \in T_n$ and $f \in T_m$ be evaluable
idempotents
in the generic Temperley Lieb category.  Let $A$ be the $\C(t)$--span in
 $\Hom(m,n)$ of a certain set of $(m,n)$--TL diagrams, and let $A(\tau)$
be
the $\C$--span in  $\Hom(m,n)(\tau)$ of the same set of diagrams.
Then
$$
\dim_{\C(t)} e A f = \dim_\C e(\tau) A(\tau) f(\tau).
$$
\end{proposition}

\begin{proof}  Let $X$ denote the set of TL  diagrams spanning $A$.
Clearly
$$\dim_{\C(t)} A = \dim_\C A(\tau) = |X|.$$
  Choose a basis
of $e(\tau) A(\tau) f(\tau)$  of the form
$
\{e(\tau)
x f(\tau) : x \in X_0\},
$
where $X_0$ is some subset of $X$.  If the set
$
\{ exf  : x \in X_0\},
$
 were linearly dependent over
$\C(t)$,
then it would be linearly dependent over $\C[t]$, and evaluating at
$\tau$
would give a linear dependence of
$
\{e(\tau)
x f(\tau) : x \in X_0\}
$
over $\C$.
  It
follows
that
$$
\dim_{\C(t)} e A f  \ge \dim_\C e(\tau) A(\tau) f(\tau).
$$
But one has similar inequalities with $e$ replaced by $\one - e$ and/or
$f$
replaced by $\one - f$.  If any of the inequalities were strict, then
adding
them
would give $\dim_{\C(t)} A > \dim_\C A(\tau)$, a contradiction.
\end{proof}

\subsection{The Markov trace} The Markov trace $\Tr=\Tr_n$  is defined
 on $T_n$  (or on $T_n(\tau)$) by the following picture, which
represents an
element in $\End(0) \cong \C(t)$  (resp. $\End(0) \cong \C$).

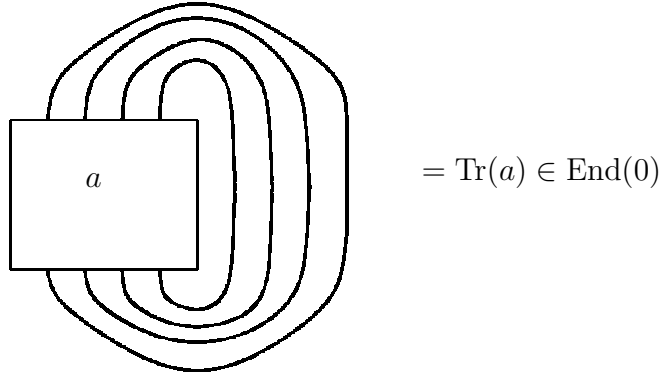
\begin{figure}[h]
\centering
\setlength{\unitlength}{0.00087489in}
\begingroup\makeatletter\ifx\SetFigFont\undefined%
\gdef\SetFigFont#1#2#3#4#5{%
  \reset@font\fontsize{#1}{#2pt}%
  \fontfamily{#3}\fontseries{#4}\fontshape{#5}%
  \selectfont}%
\fi\endgroup%
{\renewcommand{\dashlinestretch}{30}
\begin{picture}(4271,2267)(0,-10)
\path(12,1531)(1137,1531)(1137,631)
    (12,631)(12,1531)
\drawline(1362,1801)(1362,1801)
\thicklines
\path(912,1531)(912,1532)(912,1537)
    (913,1548)(913,1566)(915,1590)
    (916,1617)(918,1645)(921,1671)
    (924,1695)(928,1716)(932,1734)
    (937,1750)(943,1765)(950,1778)
    (956,1790)(964,1801)(973,1812)
    (982,1822)(993,1832)(1004,1842)
    (1017,1851)(1030,1859)(1043,1867)
    (1057,1873)(1070,1879)(1084,1883)
    (1098,1887)(1111,1889)(1124,1891)
    (1137,1891)(1147,1891)(1158,1890)
    (1168,1888)(1179,1885)(1190,1882)
    (1201,1878)(1212,1872)(1223,1866)
    (1234,1858)(1244,1850)(1255,1840)
    (1265,1829)(1274,1818)(1283,1805)
    (1292,1792)(1299,1777)(1307,1762)
    (1313,1746)(1319,1729)(1325,1711)
    (1329,1695)(1332,1678)(1336,1660)
    (1339,1640)(1342,1619)(1344,1595)
    (1347,1570)(1349,1541)(1351,1511)
    (1352,1478)(1354,1442)(1356,1405)
    (1357,1366)(1358,1326)(1359,1287)
    (1360,1251)(1361,1217)(1361,1188)
    (1361,1164)(1362,1147)(1362,1135)
    (1362,1129)(1362,1126)
\path(912,631)(912,630)(912,625)
    (913,614)(914,597)(916,577)
    (918,555)(922,535)(925,517)
    (930,502)(935,488)(942,477)
    (950,466)(957,457)(966,449)
    (977,441)(988,433)(1001,426)
    (1015,419)(1030,412)(1045,407)
    (1061,402)(1076,398)(1092,395)
    (1107,393)(1122,391)(1137,391)
    (1148,391)(1160,392)(1172,394)
    (1184,396)(1196,399)(1208,404)
    (1220,409)(1232,416)(1244,424)
    (1256,433)(1267,443)(1277,455)
    (1287,467)(1296,481)(1304,496)
    (1312,512)(1318,530)(1325,548)
    (1329,564)(1332,580)(1336,597)
    (1339,616)(1342,637)(1344,660)
    (1347,686)(1349,713)(1351,743)
    (1352,776)(1354,811)(1356,849)
    (1357,887)(1358,927)(1359,965)
    (1360,1002)(1361,1035)(1361,1064)
    (1361,1088)(1362,1105)(1362,1117)
    (1362,1123)(1362,1126)
\path(687,631)(687,630)(687,625)
    (688,614)(689,597)(691,576)
    (693,554)(697,533)(700,514)
    (705,498)(710,484)(717,471)
    (725,458)(732,448)(741,438)
    (752,428)(763,417)(776,407)
    (790,397)(805,387)(820,377)
    (836,368)(851,359)(867,351)
    (882,344)(897,337)(912,331)
    (927,325)(942,320)(957,315)
    (974,310)(990,305)(1007,301)
    (1025,298)(1042,294)(1059,292)
    (1075,290)(1092,288)(1107,287)
    (1122,286)(1137,286)(1152,286)
    (1167,287)(1182,288)(1199,290)
    (1215,292)(1232,294)(1250,298)
    (1267,301)(1284,305)(1300,310)
    (1317,315)(1332,320)(1347,325)
    (1362,331)(1373,336)(1385,341)
    (1397,347)(1409,353)(1421,360)
    (1433,368)(1445,377)(1457,387)
    (1469,398)(1481,410)(1492,423)
    (1502,437)(1512,452)(1521,468)
    (1529,485)(1537,502)(1543,521)
    (1550,541)(1554,557)(1557,574)
    (1561,592)(1564,612)(1567,633)
    (1569,657)(1572,682)(1574,711)
    (1576,741)(1577,774)(1579,810)
    (1581,847)(1582,886)(1583,926)
    (1584,965)(1585,1001)(1586,1035)
    (1586,1064)(1586,1088)(1587,1105)
    (1587,1117)(1587,1123)(1587,1126)
\path(462,631)(462,630)(462,627)
    (463,619)(465,607)(467,590)
    (471,570)(475,550)(480,530)
    (486,512)(493,495)(502,480)
    (512,465)(523,451)(537,436)
    (548,426)(560,415)(573,404)
    (587,393)(603,381)(620,369)
    (638,357)(657,345)(677,332)
    (697,320)(719,308)(741,297)
    (762,286)(784,275)(806,265)
    (828,256)(849,247)(871,240)
    (891,232)(912,226)(933,220)
    (954,215)(975,210)(997,206)
    (1019,202)(1042,199)(1065,197)
    (1089,195)(1113,194)(1137,193)
    (1161,194)(1185,195)(1209,197)
    (1232,199)(1255,202)(1277,206)
    (1299,210)(1320,215)(1341,220)
    (1362,226)(1383,233)(1403,240)
    (1425,248)(1446,257)(1468,267)
    (1490,278)(1512,290)(1533,303)
    (1555,317)(1577,332)(1597,347)
    (1617,364)(1636,382)(1654,400)
    (1671,418)(1687,437)(1701,457)
    (1714,477)(1726,497)(1737,518)
    (1745,537)(1753,556)(1760,576)
    (1766,598)(1771,621)(1777,646)
    (1781,673)(1786,703)(1789,734)
    (1793,769)(1796,805)(1799,843)
    (1802,883)(1804,923)(1806,963)
    (1808,1000)(1809,1034)(1810,1064)
    (1811,1087)(1811,1105)(1812,1117)
    (1812,1123)(1812,1126)
\path(237,631)(237,630)(238,627)
    (239,619)(241,605)(245,587)
    (250,567)(256,545)(264,523)
    (273,503)(284,483)(297,464)
    (311,445)(329,426)(350,406)
    (363,394)(377,381)(393,368)
    (410,354)(428,340)(448,325)
    (468,310)(490,294)(512,278)
    (536,262)(560,246)(585,230)
    (610,214)(636,199)(661,183)
    (687,169)(712,155)(737,141)
    (761,128)(785,116)(808,105)
    (831,95)(853,85)(875,76)
    (898,67)(921,59)(944,51)
    (967,45)(991,39)(1014,34)
    (1039,30)(1063,26)(1087,24)
    (1112,22)(1137,22)(1162,22)
    (1187,24)(1211,26)(1235,30)
    (1260,34)(1283,39)(1307,45)
    (1330,51)(1353,59)(1376,67)
    (1400,76)(1421,85)(1443,95)
    (1466,105)(1489,116)(1513,129)
    (1537,142)(1562,156)(1587,170)
    (1613,186)(1638,202)(1664,218)
    (1689,235)(1714,253)(1738,271)
    (1762,288)(1784,306)(1806,324)
    (1826,342)(1846,360)(1864,377)
    (1881,394)(1897,411)(1911,427)
    (1925,443)(1939,463)(1952,483)
    (1964,503)(1975,524)(1984,545)
    (1992,569)(2000,593)(2006,620)
    (2012,648)(2018,678)(2022,709)
    (2026,739)(2030,769)(2032,795)
    (2034,817)(2036,835)(2036,846)
    (2037,853)(2037,856)
\path(687,1531)(687,1532)(687,1535)
    (688,1543)(688,1556)(690,1575)
    (692,1598)(694,1622)(697,1647)
    (701,1671)(705,1694)(710,1714)
    (715,1732)(722,1749)(729,1764)
    (737,1779)(747,1793)(757,1806)
    (767,1819)(779,1831)(792,1844)
    (806,1857)(822,1870)(838,1882)
    (855,1895)(872,1907)(890,1919)
    (908,1930)(926,1941)(944,1951)
    (961,1960)(978,1968)(994,1976)
    (1009,1982)(1025,1988)(1041,1995)
    (1057,2000)(1074,2005)(1091,2008)
    (1107,2012)(1124,2014)(1141,2015)
    (1158,2016)(1175,2015)(1192,2014)
    (1209,2012)(1225,2008)(1241,2005)
    (1256,2000)(1272,1995)(1287,1988)
    (1301,1982)(1315,1976)(1330,1968)
    (1345,1959)(1361,1949)(1376,1938)
    (1392,1926)(1408,1913)(1424,1898)
    (1440,1882)(1455,1865)(1469,1848)
    (1482,1829)(1495,1810)(1506,1790)
    (1517,1770)(1526,1748)(1535,1726)
    (1540,1708)(1546,1690)(1550,1670)
    (1555,1649)(1559,1627)(1562,1602)
    (1565,1576)(1568,1547)(1571,1515)
    (1574,1481)(1576,1445)(1578,1407)
    (1580,1368)(1581,1328)(1583,1289)
    (1584,1251)(1585,1218)(1586,1188)
    (1586,1165)(1587,1147)(1587,1135)
    (1587,1129)(1587,1126)
\path(462,1531)(462,1532)(462,1535)
    (463,1543)(464,1557)(466,1576)
    (468,1600)(471,1625)(475,1651)
    (479,1677)(485,1700)(491,1722)
    (497,1743)(505,1762)(514,1780)
    (525,1798)(537,1816)(548,1831)
    (560,1845)(573,1861)(587,1876)
    (603,1892)(620,1909)(638,1925)
    (657,1942)(677,1958)(697,1975)
    (719,1991)(741,2006)(762,2021)
    (784,2035)(806,2049)(828,2061)
    (849,2072)(871,2083)(891,2092)
    (912,2101)(933,2109)(954,2116)
    (975,2122)(997,2128)(1019,2133)
    (1042,2137)(1065,2140)(1089,2142)
    (1113,2144)(1137,2144)(1161,2144)
    (1185,2142)(1209,2140)(1232,2137)
    (1255,2133)(1277,2128)(1299,2122)
    (1320,2116)(1341,2109)(1362,2101)
    (1381,2093)(1400,2085)(1419,2075)
    (1438,2065)(1458,2053)(1478,2041)
    (1498,2028)(1518,2014)(1537,1998)
    (1557,1982)(1577,1965)(1596,1948)
    (1614,1929)(1631,1911)(1648,1891)
    (1664,1872)(1678,1852)(1692,1832)
    (1705,1811)(1716,1791)(1727,1770)
    (1737,1748)(1745,1728)(1753,1708)
    (1760,1686)(1766,1663)(1771,1639)
    (1777,1613)(1781,1585)(1786,1555)
    (1789,1522)(1793,1487)(1796,1450)
    (1799,1411)(1802,1371)(1804,1330)
    (1806,1290)(1808,1253)(1809,1218)
    (1810,1189)(1811,1165)(1811,1147)
    (1812,1135)(1812,1129)(1812,1126)
\path(237,1531)(237,1532)(237,1535)
    (238,1543)(240,1557)(242,1577)
    (246,1601)(251,1627)(256,1654)
    (263,1680)(271,1705)(280,1728)
    (290,1750)(302,1771)(316,1791)
    (331,1811)(350,1831)(363,1845)
    (377,1859)(393,1874)(410,1889)
    (428,1904)(448,1920)(468,1936)
    (490,1953)(512,1970)(536,1986)
    (560,2003)(585,2020)(610,2036)
    (636,2052)(661,2068)(687,2083)
    (712,2097)(737,2111)(761,2124)
    (785,2136)(808,2147)(831,2157)
    (853,2167)(875,2176)(898,2185)
    (921,2193)(944,2201)(967,2207)
    (991,2213)(1014,2218)(1039,2222)
    (1063,2226)(1087,2228)(1112,2230)
    (1137,2230)(1162,2230)(1187,2228)
    (1211,2226)(1235,2222)(1260,2218)
    (1283,2213)(1307,2207)(1330,2201)
    (1353,2193)(1376,2185)(1400,2176)
    (1421,2167)(1443,2157)(1466,2147)
    (1489,2136)(1513,2124)(1537,2111)
    (1562,2097)(1587,2083)(1613,2068)
    (1638,2053)(1664,2037)(1689,2021)
    (1714,2004)(1738,1988)(1762,1972)
    (1784,1955)(1806,1939)(1826,1924)
    (1846,1908)(1864,1893)(1881,1879)
    (1897,1865)(1911,1852)(1925,1838)
    (1939,1823)(1953,1808)(1965,1793)
    (1976,1778)(1986,1763)(1995,1748)
    (2003,1732)(2010,1716)(2016,1699)
    (2021,1682)(2026,1665)(2029,1647)
    (2032,1629)(2034,1611)(2035,1593)
    (2036,1574)(2037,1555)(2037,1535)
    (2037,1515)(2037,1493)(2037,1475)
    (2037,1456)(2037,1435)(2037,1413)
    (2037,1389)(2037,1363)(2037,1335)
    (2037,1304)(2037,1271)(2037,1235)
    (2037,1196)(2037,1156)(2037,1114)
    (2037,1071)(2037,1029)(2037,990)
    (2037,954)(2037,922)(2037,897)
    (2037,878)(2037,866)(2037,859)(2037,856)
\put(462,1126){\makebox(0,0)[lb]
{\smash{{{\SetFigFont{12}{14.4}{\rmdefault}{\mddefault}{\updefault}$a$}}
}}}
\put(2487,1171){\makebox(0,0)[lb]
{\smash{{{\SetFigFont{12}{14.4}{\rmdefault}{\mddefault}{\updefault}$ =
\Tr(a) \in
 \End(0) $}}}}}
\end{picture}
}

\caption{The categorical trace of an element $a \in T_n$.}

\end{figure}

 On an $(n,n)$--TL diagram
$a\in T_n$, the trace is evaluated
 geometrically by closing  up the diagram as in the figure, and counting
the
number
$c(a)$ of components (closed loops); then $\Tr(a)=d^{c(a)}$.

It will be useful to give the following inductive description
of closing up a diagram. We define a map $\ve_n:T_{n+1}\to T_n$
(known as a conditional expectation in operator algebras) by
only closing up the last strand; algebraically it can be defined by
$$a\in T_{n+1}\quad \mapsto \quad (1_n\otimes \cup)\circ(a\otimes 1)
\circ (1_n\otimes \cap).$$
\ignore{where $1_i$ indicates $i$ vertical strands, and composition of
our morphism from right to left corresponds to putting the graphs
on top of each other.} If $k>n$, the map $\ve_{n,k}$ is defined by
$\ve_{n,k}=\ve_{n}\circ\ve_{n+1}\ ...\ \circ\ve_{k-1}$.
It follows from the definitions that $\Tr(a)=\ve_{0,n}$ for $a\in T_n$.

It is well-known that $\Tr$ is indeed a functional
satisfying $\Tr(ab)=\Tr(ba)$; one easily checks that this equality is
even
true if $a\in \Hom(n,m)$ and $b\in \Hom(m,n)$. We need the following
well-known fact:

\begin{lemma2}\label{condexp}
Let $f\in T_{n+m}$ and let $p\in T_n$ such that $(p\otimes
1_m)f(p\otimes
1_m)
=f$, where $p$ is a minimal idempotent in $T_n$.
Then $\ve_{n,n+m}(f)=\gamma p$, where $\gamma = Tr_{n+m}(f)/Tr_n(p)$
\end{lemma2}

\begin{proof}
It follows from the definitions that
$$p\ve_{n,n+m}(f)p=\ve_{n,n+m}((p\otimes 1_m)f(p\otimes
1_m))=\ve_{n,n+m}(f).$$
As $p$ is a minimal idempotent in $T_n$, $\ve_{n,n+m}(f)=\gamma p,$  for
some scalar $\gamma$. Moreover, by our definition of trace, we have
$Tr_{n+m}(f) =Tr_n(\ve_{n,n+m}(f))=\gamma Tr_n(p)$. This determines the
value of $\gamma$.
\end{proof}

The {\em negligible morphisms}
$\Ne(n,m)$ are defined to be all elements $a\in \Hom(n,m)$
for which $Tr(ab)=0$ for all $b\in \Hom(m,n)$. It is well-known
that the set of all negligible morphisms form an ideal in $\TL$.

\section{The structure of the Temperley Lieb algebras}

\subsection{The generic Temperley Lieb algebras}

Recall that a {\em Young  diagram} $\lambda = [\lambda_1,\lambda_2,\
....\
\lambda_k]$ is a left justified array of boxes with $\lambda_i$ boxes in
the $i$-th row and $\lambda_i\geq \lambda_{i+1}$ for all $i$.
For example,
$$
[5,3] = {\YD \R5,\R3,.}.
$$
{\em All Young diagrams in this paper will have at most two rows.}
For $\la$ a Young diagram with $n$ boxes, a
{\em  Young tableau  of shape $\la$} is
a filling of
$\lambda$ with the numbers 1 through $n$  so that
the numbers increase in each row and column.   The number of Young
tableax
of
shape $\la$ is denoted by $f_\la$.

The generic Temperley Lieb algebras $T_n$ are known (\cite{J1})
to decompose as direct sums of full matrix algebras over the field
$\C(t)$, $T_n = \bigoplus_\la T_\la$, where the sum is over all
Young diagrams $\lambda$ with $n$ boxes (and with no more than
two rows), and $T_\lambda$ is isomorphic to an $f_\la$-by-$f_\la$
matrix algebra.

When $\lambda$ and $\mu$  are Young diagrams of size $n$ and $n+1$,
one has a
(non-unital) homomorphism of $T_\lambda$ into $T_\mu$ given by
$x \mapsto (x \otimes 1)z_\mu$, where $z_\mu$ denotes the minimal
central
idempotent in $T_{n+1}$ such that $T_\mu = T_{n+1} z_\mu$.  Let
$\glm$ denote the rank of
$ (e \otimes 1)z_\mu$, where $e$ is any minimal idempotent in
$T_\lambda$.  It is known that $\glm = 1$ in case $\mu$ is obtained from
$\la$ by adding one box, and $\glm = 0$ otherwise.

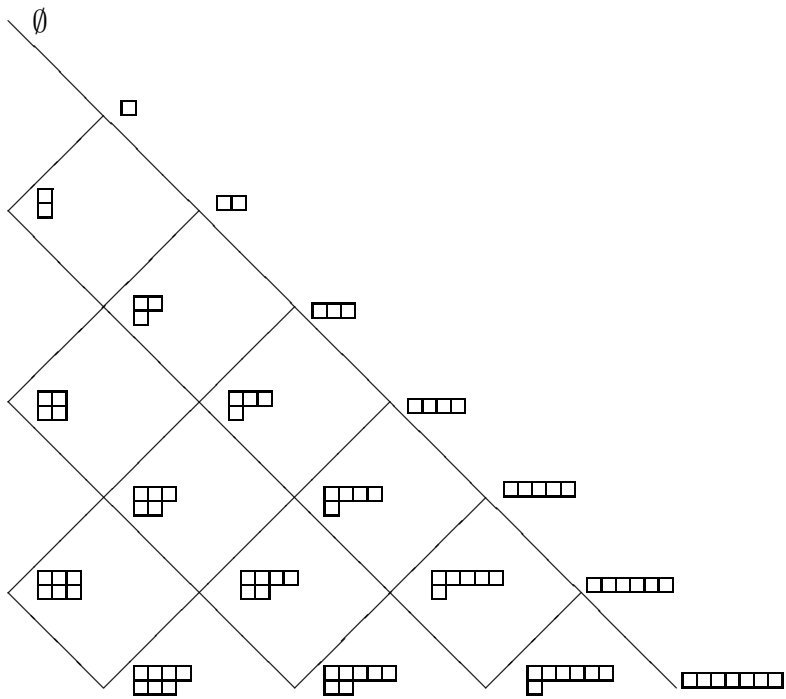
\begin{figure}
\begin{center}
\setlength{\unitlength}{0.0125in}
\begin{picture}(348,304)(0,-10)
\drawline(0,280)(280,0)
\drawline(120,160)(0,40)(40,0)(160,120)
\drawline(80,200)(0,120)(120,0)(200,80)
\drawline(40,240)(0,200)(200,0)(240,40)
\put(240,40){\makebox(0,0)[lb]{\raisebox{0pt}[0pt][0pt]
{\shortstack[l]{{\twlrm $\YD \R6,.$}}}}}
\put(50,0){\makebox(0,0)[lb]{\raisebox{0pt}[0pt][0pt]
{\shortstack[l]{{\twlrm $\YD \R4,\R3,.$}}}}}
\put(280,0){\makebox(0,0)[lb]{\raisebox{0pt}[0pt][0pt]
{\shortstack[l]{{\twlrm $\YD \R7,.$}}}}}
\put(215,0){\makebox(0,0)[lb]{\raisebox{0pt}[0pt][0pt]
{\shortstack[l]{{\twlrm $\YD \R6,\R1,.$}}}}}
\put(130,0){\makebox(0,0)[lb]{\raisebox{0pt}[0pt][0pt]
{\shortstack[l]{{\twlrm $\YD \R5,\R2,.$}}}}}
\put(10,40){\makebox(0,0)[lb]{\raisebox{0pt}[0pt][0pt]
{\shortstack[l]{{\twlrm $\YD \R3,\R3,.$}}}}}
\put(175,40){\makebox(0,0)[lb]{\raisebox{0pt}[0pt][0pt]
{\shortstack[l]{{\twlrm $\YD \R5,\R1,.$}}}}}
\put(95,40){\makebox(0,0)[lb]{\raisebox{0pt}[0pt][0pt]
{\shortstack[l]{{\twlrm $\YD \R4,\R2,.$}}}}}
\put(130,75){\makebox(0,0)[lb]{\raisebox{0pt}[0pt][0pt]
{\shortstack[l]{{\twlrm $\YD \R4,\R1,.$}}}}}
\put(50,75){\makebox(0,0)[lb]{\raisebox{0pt}[0pt][0pt]
{\shortstack[l]{{\twlrm $\YD \R3,\R2,.$}}}}}
\put(10,115){\makebox(0,0)[lb]{\raisebox{0pt}[0pt][0pt]
{\shortstack[l]{{\twlrm $\YD \R2,\R2,.$}}}}}
\put(90,115){\makebox(0,0)[lb]{\raisebox{0pt}[0pt][0pt]
{\shortstack[l]{{\twlrm $\YD \R3,\R1,.$}}}}}
\put(50,155){\makebox(0,0)[lb]{\raisebox{0pt}[0pt][0pt]
{\shortstack[l]{{\twlrm $\YD \R2,\R1,.$}}}}}
\put(205,80){\makebox(0,0)[lb]{\raisebox{0pt}[0pt][0pt]
{\shortstack[l]{{\twlrm $\YD \R5,.$}}}}}
\put(165,115){\makebox(0,0)[lb]{\raisebox{0pt}[0pt][0pt]
{\shortstack[l]{{\twlrm $\YD \R4,.$}}}}}
\put(125,155){\makebox(0,0)[lb]{\raisebox{0pt}[0pt][0pt]
{\shortstack[l]{{\twlrm $\YD \R3,.$}}}}}
\put(10,275){\makebox(0,0)[lb]{\raisebox{0pt}[0pt][0pt]
{\shortstack[l]{{\twlrm $\emptyset$}}}}}
\put(45,240){\makebox(0,0)[lb]{\raisebox{0pt}[0pt][0pt]
{\shortstack[l]{{\twlrm $\YD \R1,.$}}}}}
\put(85,200){\makebox(0,0)[lb]{\raisebox{0pt}[0pt][0pt]
{\shortstack[l]{{\twlrm $\YD \R2,.$}}}}}
\put(10,200){\makebox(0,0)[lb]{\raisebox{0pt}[0pt][0pt]
{\shortstack[l]{{\twlrm $\YD \R1,\R1,.$}}}}}
\end{picture}

\caption{Bratteli diagram for the sequence $(T_n)$}
\label{bratteli}
\end{center}
\end{figure}

One can describe the embedding of $T_n$ into $T_{n+1}$ by a {\em
Bratteli
diagram}  (or induction-restriction diagram), which is a bipartite graph
with
vertices labelled by two-row Young diagrams of size $n$ and $n+1$
(corresponding to the simple components of
$T_n$ and $T_{n+1}$) and with $\glm$ edges joining the vertices labelled
by
$\la$ and $\mu$.  That is $\lambda$ and $\mu$ are  joined by an edge
precisely when $\mu$ is obtained from $\la$ by adding one box.  The
sequence
of embeddings $T_0 \rightarrow T_1 \rightarrow T_2 \rightarrow \dots$ is
described by a multilevel Bratteli diagram, as shown in Figure
\ref{bratteli}.

 A  tableau of shape $\la$
may be identified with an
increasing sequence of Young diagrams beginning with the empty diagram
and
ending at $\la$; namely  the $j$-th  diagram in the sequence is the
subdiagram of $\la$ containing the numbers 1, 2, \dots, $j$. Such a
sequence
may
also be interpreted as a {\em path} on the Bratteli diagram of Figure
2.5, beginning at  the empty diagram and ending at
$\la$.

\subsection{Path idempotents}
One can define a familiy of minimal idempotents $p_t$ in $T_n$, labelled
by
paths $t$  of length $n$ on the Bratteli diagram (or equivalently, by
Young
tableaux of size $n$), with the following
properties:

\begin{enumerate}
\item $p_t p_s = 0$ if $t, s$ are different paths both of length $n$.
\item $z_\la = \sum \{p_t : t \text{ ends at } \la\}$.
\item $p_t \otimes 1 = \sum \{p_s : s  \text{ has length $n+1$ and
extends } t \}$
\end{enumerate}

Let $t$ be a path of length $n$ and shape $\lambda$ and let
$\mu$  be a Young diagram of size $n + m$.  It follows that $(p_t
\otimes 1_m) z_\mu \ne 0$ precisely when there is a path on the
Bratteli diagram from $\la$ to $\mu$. It has been shown in
\cite{J1} that (in our notations) $\Tr(p_t)=[\la_1-\la_2+1]$,
where $[m]=(t^m-t^{-m})/(t-t^{-1})$ for any integer $m$, and
where $\la$ is the endpoint of the path $t$. Observe that we get
the same value for diagrams $\la$ and $\mu$ (of different sizes) that are
in the same column in the Bratteli diagram.

The idempotents $p_t$ were defined by recursive formulas in [W2],
generalizing the formulas for the Jones-Wenzl idempotents in
[W1].

\subsection{Specializations at  non-roots of unity}

When $\tau$ is not a proper root of unity,  the Temperley Lieb algebras
$T_n(\tau)$ are semi-simple complex algebras with the ``same" structure
as
generic Temperley Lieb algebras.  That is,
$T_n(\tau) = \bigoplus_\la T_\la(\tau)$, where  $T_\lambda(\tau)$ is
isomorphic to an
$f_\la$-by-$f_\la$ matrix algebra over $\C$.  The embeddings
$T_n(\tau) \rightarrow T_{n+1}(\tau)$ are described by the Bratteli
diagram
as before.  The idempotents $p_t$, and the minimal central idempotents
$z_\lambda$, in the generic algebras $T_n$,  are
evaluable at
$\tau$, and the  evaluations $p_t(\tau)$, resp. $z_\lambda(\tau)$,
satisfy
analogous properties.

\subsection{Specializations at roots of unity and evaluable idempotents}
We require some terminology for discussing the case where $\tau$ is a
root of unity. Let $q = \tau^2$, and suppose that $q$ is a primitive
$\ell$-th root of unity.  We say that a Young diagram $\la$  is
{\em critical} if $w(\la) :=\la_1 - \la_2 +1$ is divisible by $\ell$.
The $m$-th {\em critical line} on the Bratteli diagram for the generic
Temperly
Lieb algebra is the line containing the diagrams $\la$ with $w(\la) =
ml$.
See
Figure 2.6.

Say that two non-critical diagrams $\la$ and
$\mu$
with the same number of boxes are {\em reflections of one another in the
$m$-th
critical line} if
$\la \ne \mu$ and $|w(\la) - m\ell| = |w(\mu) - m\ell| < \ell$.
(For example, with $\ell = 3$, $[2,2]$ and $[4]$ are reflections in the
first
critical line $w(\la) = 3$.)

\begin{figure}[ht]
\centering
\setlength{\unitlength}{0.0115in}
\begin{picture}(360,395)(0,-10)
\dashline{4.000}(280,380)(280,0)
\dashline{4.000}(180,380)(180,0)
\dashline{4.000}(80,380)(80,0)
\drawline(340,40)(320,20)
\drawline(320,60)(280,20)
\drawline(300,80)(240,20)
\drawline(280,100)(200,20)
\drawline(260,120)(160,20)
\drawline(240,140)(120,20)
\drawline(220,160)(80,20)
\drawline(200,180)(40,20)
\drawline(180,200)(0,20)
\drawline(160,220)(0,60)(40,20)
\drawline(140,240)(0,100)(80,20)
\drawline(120,260)(0,140)(120,20)
\drawline(100,280)(0,180)(160,20)
\drawline(80,300)(0,220)(200,20)
\drawline(60,320)(0,260)(240,20)
\drawline(60,320)(60,320)
\drawline(40,340)(0,300)(280,20)
\drawline(20,360)(0,340)(320,20)
\drawline(0,380)(360,20)
\end{picture}
\caption{Critical lines}
\label{critical lines}
\end{figure}
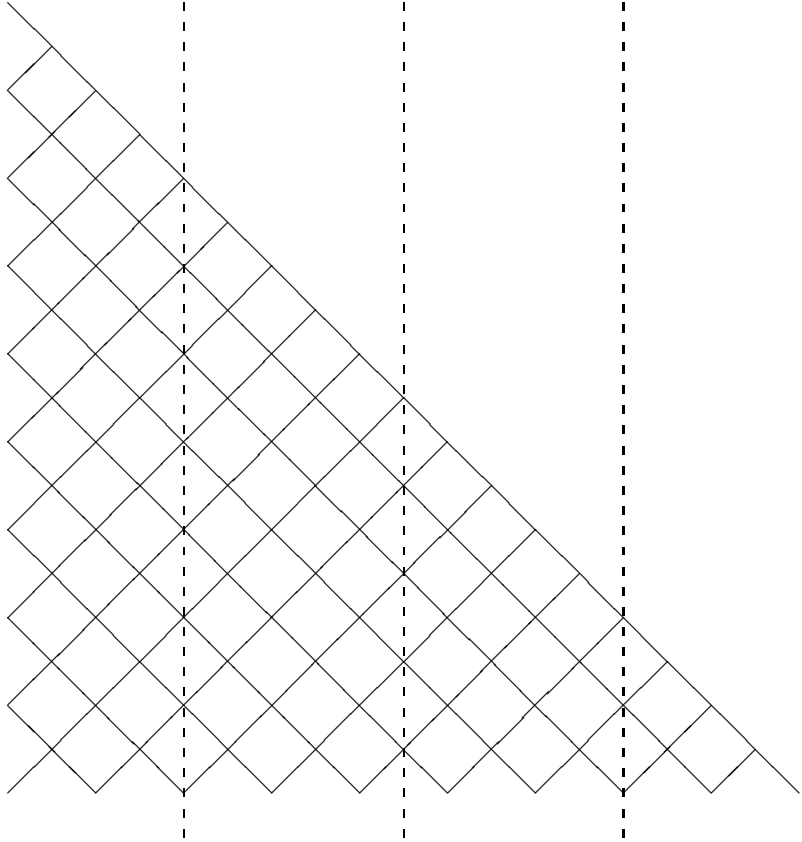

For $\tau$ a proper root of unity, the formulas for path idempotents in
[W1] and [W2] generally contain
poles
at $\tau$, i.e. the idempotents are not evaluable. However, suitable
sums
of path idempotents are evaluable.  We will review some facts from
~\cite{pacific} about such evaluable sums .

Suppose $w(\la) \le \ell$ and $t$ is a path of shape $\la$
which stays strictly
to the left of the first critical line (in case $w(\la) < \ell$), or
hits
the
first critical line for the first time at $\la$ (in case $w(\la) =
\ell$);
then
$p_t$ is evaluable at
$\tau$, and  furthermore $\Tr(p_t) = [w(\la)]_\tau =
(\tau^{w(\la)}-\tau^{-{w(\la)}})/(\tau-\tau^{-1})$.

\begin{figure}[ht]
\centering
\setlength{\unitlength}{0.0117in}
\begin{picture}(360,395)(0,-10)
\dottedline{5}(40,340)(0,300)(280,20)
\thicklines
\drawline(0,380)(80,300)(100,280)
    (80,260)(180,160)(140,120)
    (160,100)(140,80)
\drawline(180,160)(220,120)(200,100)(220,80)
\thinlines
\dottedline{5}(320,20)(0,340)(20,360)
\dottedline{5}(360,20)(0,380)
\dottedline{5}(320,60)(280,20)
\dottedline{5}(340,40)(320,20)
\dottedline{5}(280,100)(200,20)
\dottedline{5}(260,120)(160,20)
\dottedline{5}(300,80)(240,20)
\dottedline{5}(220,160)(80,20)
\dottedline{5}(200,180)(40,20)
\dottedline{5}(240,140)(120,20)
\dottedline{5}(160,220)(0,60)(40,20)
\dottedline{5}(140,240)(0,100)(80,20)
\dottedline{5}(120,260)(0,140)(120,20)
\dottedline{5}(100,280)(0,180)(160,20)
\dottedline{5}(80,300)(0,220)(200,20)
\dottedline{5}(60,320)(0,260)(240,20)
\dottedline{5}(180,200)(0,20)
\dashline{4.000}(80,380)(80,0)
\dashline{4.000}(180,380)(180,0)
\dashline{4.000}(280,380)(280,0)
\drawline(60,320)(60,320)
\end{picture}
\caption{A path and its reflected path.}
\label{conjugate paths}
\end{figure}
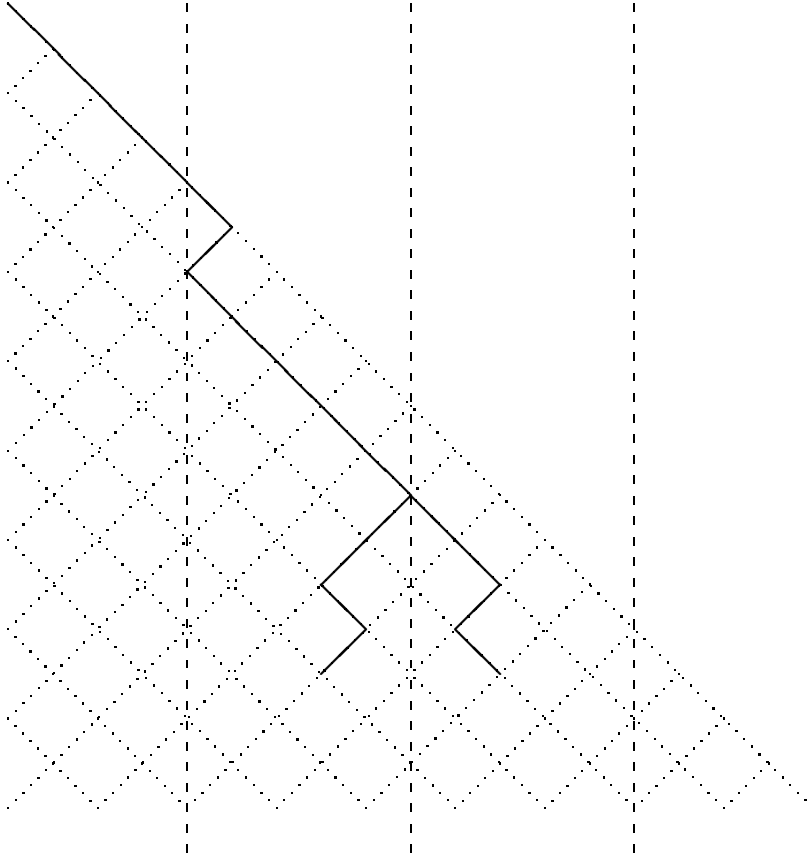

 For each
critical diagram $\la$  of size $n$, the minimal central idempotent
$z_\la$
in
$T_n$ is evaluable at $\tau$.  Furthermore, for each non-critical
diagram
$\la$ of size $n$, an evaluable idempotent $z_\la^L =\sum p_{t}\in T_n$
was defined in ~\cite{pacific} as follows: The summation goes over all
paths $t$ ending in $\la$ for which the last critical line hit by $t$ is
the
one nearest to $\la$ to the left {\it and} over the paths obtained from
such
$t$ by reflecting its part after the last critical line in the critical
line
(see Figure \ref{conjugate paths}).

These idempotents have the
following properties (which were shown in  [GW]:
\begin{enumerate}
\item $\{z_\la(\tau) : \la \text{ critical }\} \cup
\{z_\mu^L(\tau) : \mu \text{ non-critical }\}$ is a partition
of unity in
$T_n(\tau)$;  that is, the idempotents are mutually orthogonal and sum
to
the identity.
\item   $z_\la(\tau)$  is a  minimal central idempotent in $T_n(\tau)$
if $\la$ is critical, and
$z_\la^L(\tau)$ is minimal
central modulo the nilradical of $T_n$ if $\la$ is not critical
  (see
~\cite{pacific}, Theorem 2.2 and Theorem 2.3).
\item  For $\la$ and $\mu$ non-critical, $z_\la^L(\tau) T_n(\tau)
z_\mu^L(\tau) \ne 0$ only if $\la =\mu$, or if there is exactly
one critical line between
$\la$ and  $\mu$ which reflects $\la$ to $\mu$.
If in this case $\nu$ denotes the leftmost of the two diagrams $\la$ and
$\mu$, then
$z_\la^L T_n z_{\mu}^L \subseteq T_\nu$
 (in the generic Temperley Lieb algebra).
\item Let $z_n^{reg}=\sum p_t$, where the summation goes over all paths
$t$ which stay strictly to the left of the first critical line,
and let $z_n^{nil}=\one-z_n^{reg}$.
Then both $z_n^{reg}$ and $z_n^{nil}$ are evaluable; this is a direct
consequence of the fact that $z_n^{reg}=\sum_\la z_\la^L$,
where the summation goes over diagrams $\la$ with $n$ boxes with
width $w(\la)<\ell$.
\end{enumerate}
\ignore{
Finally, we also mention that the restriction rule in the root of unity
case for the semisimple quotients of $T_n$ is very similar to the
generic case. It coincides in all cases except if one of the
involved diagrams lies on a critical line. If a predecessor of
$\la$ lies in a critical line
a simple $T_{n,\la}$-module $W$ decomposes into the direct sum of
3 simple $T_{n-1}$ modules.}

\begin{proposition}
The ideal of negligible morphisms in $\TL(\tau)$ is generated by the
idempotent $p_{[\ell-1]}(\tau) \in T_{\ell-1}(\tau)$.
\end{proposition}

\begin{proof} Let us first show
that $z_n^{nil}(\tau)$ is in the ideal generated by
 $p_{[\ell-1]}(\tau)$ for all $n$. This is clear for $n<\ell$,
as $z_{\ell -1}^{nil}=p_{[\ell-1]}$ and $z_n^{nil}=0$ for $n<\ell-1$.

Moreover, $z^{nil}_n$ is a central idempotent in the maximum semisimple
quotient of $T_n$, whose minimal central idempotents are the
$z_\la^L$ with $w(\la)\geq \ell$. One checks pictorially that
$p_{[\ell-1]}z_\la^L\neq 0$ for any such $\lambda$ (i.e.
the path to $[\ell-1]$ can be extended to a path $t$ for which $p_t$
is a summand of $z_\la^L$).  This proves our assertion
in the maximum semisimple quotient of $T_n$; it is well-known that
in this case also the idempotent itself must be in the ideal generated
by $p_{[\ell-1]}$. In particular,  $\Hom(n,m)z_m^{nil}(\tau)
+z_n^{nil}(\tau)\Hom(n,m)$ is also contained in this ideal.

By [GW], Theorem 2.2 (c), for $\la$ a Young diagram of size $n$,
with $w(\la) < \ell$,
$z_\la^L T_n z_\la^L(\tau)$ is a full matrix algebra, which moreover
contains a minimal idempotent $p_t$ of trace $\Tr(p_t) =
[w(\la)]_\tau \ne 0$.  Therefore $$z_\la^L T_n z_\la^L(\tau) \cap
\Neg(n,n) = (0).$$   Furthermore, $z_n^{reg} T_n z_n^{reg}(\tau) =
\sum z_\la^L T_n  z_\la^L(\tau) $, by Fact 4 above, so
$$z_n^{reg} T_n z_n^{reg}(\tau) \cap \Neg(n,n) = (0)$$
as well.  Now for $x \in \Neg(n,n)$,
one has $z_n^{reg}(\tau) x z_n^{reg}(\tau) = 0$, so
$$x \in T_n(\tau) z_n^{nil}(\tau)+z_n^{nil}(\tau)T_n (\tau).$$

We have shown that $\Neg(n,n)$ is contained in the ideal of $\TL(\tau)$
generated
by $p_{[l-1]}$, for all $n$.    That the same is true for $\Neg(m,n)$
with $n\neq m$ follows
from using the embeddings, and their left inverses, described at the end
of
Section 1.1.

\ignore{
On the other hand, if $n=m$, its complement $z_n^{reg}T_nz_n^{reg}$
(as a linear direct summand) has the same dimension as the quotient
of $T_n/\Ne(n,n)$: this can be read off inductively from the
Bratteli diagram of the quotient and the definition of $z_n^{reg}$.
Hence $T_nz_n^{nil}+z_n^{nil}T_n$ must contain all negligible morphisms
in $T_n$. The claim for the general case follows from this and
the embedding described at the end of Section 1.1
}

\end{proof}

\section{Ideals}

\begin{proposition} Any proper ideal in $\TL$ (or in $\TL(\tau)$) is
contained in the ideal of negligible morphisms.
\end{proposition}

\begin{proof} Let $a \in \Hom(m, n)$.  For all $b \in \Hom(n,m)$,
$\tr(ba)$ is in the intersection of the ideal generated by $a$ with the
scalars $\End(0)$.  If $a$ is not negligible, then
the ideal generated by $a$ contains an non-zero scalar, and therefore
contains all morphisms.
\end{proof}

\begin{corollary}  The categories $\TL$ and $\TL(\tau)$ for $\tau$ not a
proper root of unity have no non-zero proper ideals.
\end{corollary}

\begin{proof}  There are no non-zero negligible morphisms in
$\TL$ and in $\TL(\tau)$ for $\tau$ not a
proper root of unity.
\end{proof}

\ignore{
\begin{lemma2} \label{embedding lemma}
For  $m, n \in \nat_0$,
there is an injective linear map $$\iota : \Hom(m,n)(\tau) \rightarrow
T_{\max(m,n)}(\tau)$$ such that for any ideal
$J$ in $\TL(\tau)$,  $$\iota(J \cap \Hom(m,n)) \subseteq J \cap
T_{\max(m,n)}(\tau).$$
\end{lemma2}
\begin{proof}  blah, blah
\end{proof}
}

\begin{theorem}  Suppose that $\tau$ is a proper root of unity.  Then
the negligible morphisms form the unique non-zero proper ideal in
$\TL(\tau)$.
\end{theorem}

\begin{proof}  Let $J$ be a non-zero proper ideal in $\TL(\tau)$.
By the embeddings discussed at the end of Section 1.1, we can assume
$J\cap T_n\neq 0$ for some $n$.

Now let $a$ be a non-zero element of $J \cap T_n(\tau)$.  Since
$\{z_\la(\tau)  \}
\cup
\{z_\mu^L(\tau) \}$  is a partition
of unity in $T_n(\tau)$, one of the following conditions hold:
{\renewcommand{\labelenumi}{{(\alph{enumi})}}
\begin{enumerate}
\item $b = a z_\mu(\tau) \ne 0$ for some
critical diagram
$\mu$.
\item
$b = z_\mu^L(\tau) a z_\mu^L(\tau) \ne 0$ for some non-critical diagram
$\mu$.
\item
$b = z_\la^L(\tau) a z_{\la'}^L(\tau) \ne 0$
 for some pair $\la, \la'$ of non-critical diagrams which are
reflections of
one another in a critical line.  In this  case, let $\mu$ denote the
leftmost
of the two diagrams $\la, \la'$.
\end{enumerate}
}

In each of the three cases, one has $b \in e(\tau) T_n(\tau) f(\tau)$,
where
$e, f$ are evaluable idempotents in $T_n$ such that $e T_n f  \subseteq
T_\mu$.
Let $\alpha$ be a Young diagram on the
first critical line  of size $n+m$, such that there exists a path on the
generic
Bratteli diagram connecting $\mu$ and $\alpha$.  Then one has
\begin{align*}
\dim_\C\ &z_\alpha(\tau)(e(\tau) \otimes 1_m) ( T_n(\tau)  \otimes \C\,
1_m)
(f(\tau)
\otimes 1_m) \\ &=
\dim_{\C(t)}z_\alpha (e \otimes \id_m )( T_n  \otimes  \C(t) 1_m)
(f \otimes 1_m )
\\
&= \dim_{\C(t)} e T_n f =
\dim_\C e(\tau) T_n(\tau) f(\tau)
\end{align*}
where the first and last equalities result from the principle of
constancy
of
dimension, and the second equality is because $x \mapsto  z_\alpha(x
\otimes
1_m) $ is injective from $T_\mu$ to $T_\alpha$.
But then it follows that $ x \mapsto z_\alpha(\tau) (x \otimes 1_m)$ is
injective
on $e(\tau) T_n(\tau) f(\tau)$.  In particular
$(b \otimes 1_m)z_\alpha $ is a non-zero element of $J \cap T_\alpha$.
Hence there exists $c\in T_\al$ such that $g=c(b\otimes 1_m)z_\al$
is an idempotent. After conjugating (and multiplying with $p_{[\ell-1]}
\otimes 1_m$, if necessary), we can assume $g$ to be a subidempotent
of  $p_{[\ell-1]} \otimes 1_m$. But then $\ve_{\ell-1+m,\ell-1}(g)$
is a multiple of  $p_{[\ell-1]}$, by Lemma 1.2,
with the multiple equal to the rank of $g$ in $T_\al$.
This, together with Proposition 2.1, finishes the proof.
\end{proof}
It is easily seen that $\TL$ has a subcategory $\TL^{ev}$
whose objects consist of even numbers of points, and with the same
morphisms between sets of even points as for $\TL$. The evaluation
$\TL^{ev}(\tau)$ is defined in complete analogy to $\TL(\tau)$.

\begin{corollary} If $\tau^2$ is a proper root of unity
of degree $\ell$ with $\ell$ odd, the negligible morphisms
form the unique non-zero proper ideal in
$\TL^{ev}$.
\end{corollary}

\begin{proof}
If $\ell$ is odd, $p_{[\ell-1]}$ is a morphism in $\TL^{ev}$.
The proof of the last theorem goes through word for word (one only
needs to make sure that one stays within $\TL^{ev}$, which is easy
to check).
\end{proof}
